\documentclass[12pt]{article}
\usepackage{float}
\usepackage{authblk}
\usepackage{amssymb,amsmath}
\usepackage{graphicx}
\usepackage{caption}
\usepackage{subcaption}
\usepackage{cite}
\usepackage{xcolor}
\usepackage{setspace}
\usepackage{appendix}
\usepackage{makecell}
\usepackage{array}
\newcolumntype{M}[1]{>{\centering\arraybackslash}m{#1}}
\usepackage[colorlinks=true,urlcolor=blue,citecolor=blue,linkcolor=red,pagecolor=blue,linktocpage=true,pdfproducer=medialab]{hyperref}
\usepackage[a4paper,width=17cm,height=25cm]{geometry}
\usepackage{morefloats}
\usepackage{siunitx}
\makeatletter \renewcommand{\@dotsep}{10000} \makeatother
\title{ Footprints of New Physics in the angular distribution of $B_{c}\to D_{s}^{\ast}(\to D_{s}\gamma,(D_{s}\pi))\ell^{+}\ell^{-}$ decays}
\author{Marwah Zaki%
	\thanks{\texttt mzaki.msphy20sns@student.nust.edu.pk}}
\author {M. Ali Paracha%
	\thanks{\texttt aliparacha@sns.nust.edu.pk}}
\author{Faisal Munir Bhutta%
	\thanks{\texttt faisal.munir@sns.nust.edu.pk}}
\affil{Department of Physics, School of Natural Sciences, National University of Sciences and Technology, H-12, Islamabad, Pakistan}
\date{\today}
\begin{document}
\maketitle
\begin{abstract}
We investigate the angular decay distribution of the four-fold $B_{c}\to D^{\ast}_{s}(\to D_{s}\gamma)\mu^{+}\mu^{-}$, and  $B_{c}\to D^{\ast}_{s}(\to D_{s}\pi)\mu^{+}\mu^{-}$ decays that proceed through $b\to s\mu^{+}\mu^{-}$ quark level transition. We use the model independent effective Hamiltonian with vector and axial vector new physics operators to formulate the angular observables and study the implications of different latest new physics scenarios, taken from the global fits to all the $b\to s$ data, on these observables. We also give Standard Model and new physics predictions of several observables such as differential branching ratios, forward backward asymmetry, longitudinal
polarization fraction of $D_s^{\ast}$, and the unpolarized and polarized lepton flavor universality violating ratios.
Future measurements of the predicted angular observables, both at current and future high energy colliders, will add to the useful complementary data required to clarify the structure of new physics in $b\to s\ell\ell$ neutral current decays.
\end{abstract}
\section{Introduction}\label{intro}
High Energy Physics community has put a lot of effort over the past decade in searching the new physics (NP) via exclusive decays of $B$ meson based on flavor changing neutral (FCNC) transitions, in particular $b\to s\ell^{+}\ell^{-}$ mode. These FCNC transitions occur only at loop level in the Standard Model (SM) and hence provide a fertile ground to investigate NP as well as the SM parameters. $B\to K^{\ast}(\to K\pi)\mu\mu$, and $B\to K\mu\mu$ decays and their angular distributions have been studied in great detail at the LHCb experiments \cite{LHCb:2015svh,LHCb:2020lmf,LHCb:2020gog,CMS:2020oqb,LHCb:2014auh,LHCb:2016due,CMS:2018qih}. From the angular distributions of such decays, new set of observables have been constructed which are free from the Cabibbo-Kobayashi-Maskawa (CKM) uncertainties, and therefore furnish a complementary way to diagnose the status of NP \cite{Becirevic:2011bp}. However the main hindrance to chalk out the status of NP via angular observables are the hadronic uncertainties. There is an improvement in controlling the uncertainties in the hadronic matrix elements of local quark operators and in few cases the uncertainty is about $10\%$. On the other hand, the matrix element of the non-local quark operators appearing from the coupling of charmonium states, remains a daunting task to handle \cite{Ciuchini:2015qxb}. Both of the above mentioned hadronic uncertainties are almost negligibly small in the ratios $R_{K^{(\ast)}}=\frac{\mathcal{B}(B\to K^{(\ast)}\mu^{+}\mu^{-})}{\mathcal{B}(B\to K^{(\ast)}e^{+}e^{-})}$ \cite{Hiller:2003js}. Recently the updated measurements of these observables $R_{K}$, and $R_{K^{\ast}}$ \cite{LHCb:2022qnv,LHCb:2022zom}, have put stringent constraints on the NP couplings and the NP models.

Among several $b\to s\mu\mu$ observables, showing deviations from the SM predictions, there are branching fractions of $B\to K\mu^{+}\mu^{-}$ \cite{LHCb:2014cxe}, $B\to K^{\ast}\mu^{+}\mu^{-}$ \cite{LHCb:2014cxe,LHCb:2013zuf,LHCb:2016ykl}, and $B_{s}\to\phi\mu^{+}\mu^{-}$ \cite{LHCb:2013tgx,LHCb:2015wdu} decays. The values of these branching fractions are found to be on the lower side as compared to their SM predictions. Also, in angular observables, $P_{5}^{\prime}$ observable in the $B^{0}\to K^{\ast 0}\mu^{+}\mu^{-}$ decay, \cite{Descotes-Genon:2012isb,Descotes-Genon:2013vna}, has shown mismatch from the SM values. For instance, ATLAS \cite{ATLAS:2018gqc}, and LHCb \cite{LHCb:2015svh,LHCb:2020lmf}, measured the value of $P_{5}^{\prime}$ in the kinematical region $4.0<q^{2}<6.0$ $\text{GeV}^{2}$ and found departure from the SM value to be more than $3\sigma$ \cite{Aebischer:2018iyb}. Furthermore Belle \cite{Belle:2016xuo,Belle:2016fev} and CMS \cite{CMS:2017rzx} measured the value of $P_{5}^{\prime}$ for the same decay mode in $q^{2}$ bin $4.0<q^{2}<8.0$ $\text{GeV}^{2}$ and $6.0<q^{2}<8.68$ $\text{GeV}^{2}$ respectively. Belle measurement shows the deviation of $2.6\sigma$ from the SM prediction and CMS measurement shows a discrimination of $1\sigma$ from the SM value.

Considering all the $b\to s$ data, including the above mentioned, several model independent global fit analyses \cite{Alguero:2021anc,Descotes-Genon:2015uva,Altmannshofer:2017fio,Alok:2017sui,Altmannshofer:2017yso,
Geng:2017svp,Ciuchini:2017mik,Capdevila:2017bsm,Alguero:2019ptt,Alok:2019ufo,Ciuchini:2019usw,Datta:2019zca,Aebischer:2019mlg,
Kowalska:2019ley,Arbey:2019duh,Bhattacharya:2019dot,Biswas:2020uaq,Alok:2022pjb} have been performed with NP present only in the muon sector, that found two simple
one-dimensional (1D) NP scenarios (S1) $C_{9\mu}^{\text{NP}}$ or (S2) $C_{9\mu}^{\text{NP}}=-C_{10\mu}^{\text{NP}}$, which give
better fit to all the data, with preferences reaching $\approx5-6\sigma$ compared to the SM. Interestingly, if global fits predict the NP effects being present in the observables of $B\to K^{(\ast)}\mu^{+}\mu^{-}$ and $B_{s}\to\phi\mu^{+}\mu^{-}$ decay modes, following $b\to s\mu^{+}\mu^{-}$ transition, then it is worth wondering that similar NP effects should also emerge in the observables of other complementary semileptonic decay modes followed by the same quark level transition. In this context, different complementary decay modes $B\to K_{1}\mu^{+}\mu^{-}$ \cite{Huang:2018rys,MunirBhutta:2020ber}, $B\to K_{2}^{\ast}\mu^{+}\mu^{-}$ \cite{Das:2018orb,Mohapatra:2021izl}, $B_s\to f_{2}^{\prime}\mu^{+}\mu^{-}$ \cite{Rajeev:2020aut,Mohapatra:2021izl}, and $B_{c}\to D_{s}^{(\ast)}\mu^{+}\mu^{-}$ \cite{Dutta:2019wxo,Mohapatra:2021ynn} have been investigated both in model independent approach and the specific NP models. For example, in Ref. \cite{Dutta:2019wxo}, $B_{c}\to D_{s}^{(\ast)}\mu^{+}\mu^{-}$ decay observables have been investigated model independently with various 1D and 2D NP scenarios whereas the authors of Ref. \cite{Mohapatra:2021ynn} analyzed the $B_{c}\to D_{s}^{(\ast)}\mu^{+}\mu^{-}$, and $B_{c}\to D_{s}^{(\ast)}\nu\bar{\nu}$ decays in a $Z^{\prime}$ and leptoquark models.

In this work, we use the model independent effective Hamiltonian in the presence of only vector and axial vector NP operators and perform the four-fold angular analysis of $B_{c}\to D^{\ast}_{s}(\to D_{s}\gamma,D_{s}\pi)\mu\mu$ decays using the relativistic quark model (RQM) form factors in the low energy $q^{2}$ range. For the decay channels $D^{\ast}_{s}\to D_{s}\gamma$ the probability is $93\%$, and the probability of the channel $D^{\ast}_{s}\to D_{s}\pi$ is $5\%$. As our NP extensions cater both new vector and axial vector couplings, therefore for the NP scenarios, we choose the best fit values of NP couplings in different 1D and 2D scenarios, from the recent global fit analysis \cite{Alok:2022pjb}. We give the predictions of  different physical observables such as differential branching fractions, forward-backward asymmetry, longitudinal helicity fraction of $D_{s}^{\ast}$ meson, lepton flavor universality violating (LFUV) ratios, when $D_{s}^{\ast}$ meson is longitudinally and transversely polarized and the individual angular observables within the SM and in different NP scenarios.

 The organization of the paper is as follows. In section \ref{TH}, we start with the general effective Hamiltonian, for $b\to s\mu\mu$ transition, in the presence of vector and axial vector NP operators after which we express the matrix elements in terms of form factors for $B_{c}\to D^{\ast}_{s}\mu^{+}\mu^{-}$ decay. Further, the helicity formalism is followed by the expressions of the helicity amplitudes, angular coefficients and the physical observables for the decay $B_{c}\to D^{\ast}_{s}(\to D_{s}\gamma,D_{s}\pi)\mu^{+}\mu^{-}$. In section \ref{NUM}, we present the phenomenological analysis of all the observables, in the SM and the NP scenarios, and section \ref{concl} concludes our discussion.

 \section{Theoretical Framework}\label{TH}
In this section, we present the effective Hamiltonian which is used to compute the full angular distribution of $B_{c}\to D_{s}^{\ast}\to(D_{s}\gamma,D_{s}\pi)\mu^{+}\mu^{-}$ decays. We give the expressions of the helicity amplitudes and express all the angular coefficients in terms these helicity amplitudes. Using the full form of the four fold angular decay distribution, we can extract the $q^{2}$ dependent angular coefficients, which will be used to analyze the effects of various 1D and 2D NP scenarios.

\subsection{Effective Hamiltonian and Decay Amplitude of $B_{c}\to D_{s}^{\ast}\mu^{+}\mu^{-}$}\label{EffecHam}
The most general low energy effective Hamiltonian for rare $|\Delta B| = |\Delta S| = 1$ transition, in the presence of new vector and axial vector operators is written as \cite{Dutta:2019wxo},
\begin{align}\label{H1}
\mathcal{H}_{\text{eff}}=-\frac{4 G_{F}}{\sqrt{2}}V_{tb}V^{\ast}_{ts}&\Bigg[C_{7}^{\text{eff}}O_{7}+C_{7^{\prime}}O_{7^{\prime}}+\sum_{i=9,10}\Big((C_{i}+C_{i\ell}^{\text{NP}})O_{i}+C_{i^{\prime}\ell}^{\text{NP}}O_{i^{\prime}}\Big)\Bigg],
\end{align}
where $G_F$ is the Fermi coupling constant, $V_{ij}$ are the CKM matrix elements. The expressions of the dipole operators $O_{7^{(\prime)}}$, and the semileptonic operators $O_{9^{(\prime)}, {10}^{(\prime)}}$ are given as,
\begin{align}\label{op1}
O_{7} &=\frac{e}{16\pi ^{2}}m_{b}\left( \bar{s}\sigma _{\mu \nu }P_{R}b\right) F^{\mu \nu },
&  O_{7^{\prime}} &=\frac{e}{16\pi ^{2}}m_{b}\left( \bar{s}\sigma _{\mu \nu }P_{L}b\right) F^{\mu \nu },\notag\\
O_{9} &=\frac{e^{2}}{16\pi ^{2}}(\bar{s}\gamma _{\mu }P_{L}b)(\bar{l}\gamma^{\mu }l),
&  O_{9^{\prime}} &=\frac{e^{2}}{16\pi ^{2}}(\bar{s}\gamma _{\mu }P_{R}b)(\bar{l}\gamma^{\mu }l),\notag\\
O_{10} &=\frac{e^{2}}{16\pi ^{2}}(\bar{s}\gamma _{\mu }P_{L}b)(\bar{l} \gamma ^{\mu }\gamma _{5} l),
&  O_{10^{\prime}} &=\frac{e^{2}}{16\pi ^{2}}(\bar{s}\gamma _{\mu }P_{R}b)(\bar{l} \gamma ^{\mu }\gamma _{5} l),
\end{align}
where $e$ $(g_s)$ is the electromagnetic (strong) coupling constant, and $m_b$ in $O_{7^{(\prime)}}$, is assumed to be the running $b-$quark mass in the $\overline{\text{MS}}$ scheme. $O_{i^{\prime}}$ are the chirality flipped operators. Within the SM, contributions of $O_{7^{\prime}}$ operator are suppressed by $m_s/m_b$, therefore we neglect them and further we do not consider NP scenarios with radiative coefficients $C^{\text{NP}}_{7^{(\prime)}}$ as they are well constrained \cite{Paul:2016urs}.
Moreover, for the present study, we have ignored the non-factorizable contributions such as the long distance charm-loop corrections in the effective Hamiltonain, although they are expected to be significant at large recoil.

In Eq.(\ref{H1}), $C_{i}(\mu)$ are the corresponding Wilson coefficients at the energy scale $\mu$. The expressions of the $C_{7}^{\text{eff}}(q^{2})$ and $C_{9}^{\text{eff}}(q^{2})$ Wilson coefficients \cite{Bobeth:1999mk,Beneke:2001at,Asatrian:2001de,Asatryan:2001zw,Greub:2008cy,Du:2015tda}, that contain the factorizable contributions from current-current, QCD penguins and chromomagnetic dipole operators $O_{1-6,8}$ are explicitly given in appendix \ref{append}. Using the above effective Hamiltonian, the amplitude for the $B_{c}\to D_{s}^{\ast}\ell^{+}\ell^{-}$ decay in the framework of SM as well as NP can be written as,
\begin{eqnarray}
\mathcal{M}\left(B_{c}\to D_{s}^{\ast}\ell^{+}\ell^{-}\right)=\frac{G_{F}\alpha}{2\sqrt{2}\pi}V_{tb}V^{\ast}_{ts}\Big\{T^{1,D_{s}^{\ast}}_{\mu}(\bar{\ell}\gamma^{\mu}\ell)
+T^{2,D_{s}^{\ast}}_{\mu}(\bar{\ell}\gamma^{\mu}\gamma_{5}\ell)\Big\},\label{Amp1}
\end{eqnarray}
where
\begin{eqnarray}
T^{1,D_{s}^{\ast}}_{\mu}&=&(C_{9}^{\text{eff}}+C_{9\ell}^{\text{NP}})\Big\langle D_{s}^{\ast}(k,\varepsilon)|\bar s\gamma_{\mu}(1-\gamma_{5})b|B_{c}(p)\Big\rangle
+C_{9^\prime\ell}^{\text{NP}}\Big\langle D_{s}^{\ast}(k,\varepsilon)|\bar s\gamma_{\mu}(1+\gamma_{5})b|B_{c}(p)\Big\rangle\notag\\
&-&\frac{2m_{b}}{q^{2}}C_{7}^{\text{eff}}
\Big\langle D_{s}^{\ast}(k,\varepsilon)|\bar s i\sigma_{\mu\nu}q^{\nu}(1+\gamma_{5})b|B_{c}(p)\Big\rangle,\label{Amp1a}
\\
T^{2,D_{s}^{\ast}}_{\mu}&=&(C_{10}+C_{10\ell}^{\text{NP}})\Big\langle D_{s}^{\ast}(k,\varepsilon)|\bar s\gamma_{\mu}(1-\gamma_{5})b|B_{c}(p)\Big\rangle
+C_{10^\prime\ell}^{\text{NP}}\Big\langle D_{s}^{\ast}(k,\varepsilon)|\bar s\gamma_{\mu}(1+\gamma_{5})b|B_{c}(p)\Big\rangle.\notag\\
\label{Amp1b}
\end{eqnarray}
where $T^{i,D_{s}^{\ast}}_{\mu}$, $i=(1,2)$, contain the matrix elements of $B_{c}\to D_{s}^{\ast}$.
\subsection{Matrix Elements for $B_{c}\to D_{s}^{\ast}\mu^{+}\mu^{-}$ Decay}
The hadronic matrix element for $B_{c}\to D_{s}^{\ast}\mu^{+}\mu^{-}$ can be parameterized in terms of form factors as follows,
\begin{align}
\left\langle D_{s}^\ast(k,\overline\epsilon)\left\vert \bar{s}\gamma
_{\mu }b\right\vert B_c(p)\right\rangle &=\frac{2\epsilon_{\mu\nu\alpha\beta}}
{m_{B_c}+m_{D_{s}^\ast}}\overline\epsilon^{\,\ast\nu}p^{\alpha}k^{\beta}V(q^{2}),\label{2.13a}
\\
\left\langle D_{s}^\ast(k,\overline\epsilon)\left\vert \bar{s}\gamma_{\mu}\gamma_{5}b\right\vert
B_c(p)\right\rangle &=i\left(m_{B_c}+m_{D_{s}^\ast}\right)g_{\mu\nu}\overline\epsilon^{\,\ast\nu}A_{1}(q^{2})
\notag\\
&-iP_{\mu}(\overline\epsilon^{\ast}\cdot q)\frac{A_{2}(q^{2})}{\left(m_{B_c}+m_{D_{s}^\ast}\right)}\notag\\
&-i\frac{2m_{D_{s}^\ast}}{q^{2}}q_{\mu}(\overline\epsilon^{\,\ast}\cdot q)
\left[A_{3}(q^{2})-A_{0}(q^{2})\right],\label{2.13b}
\end{align}
where $P_{\mu}=p_{\mu}+k_{\mu}$, $q_{\mu}=p_{\mu}-k_{\mu}$, and
\begin{eqnarray}
A_{3}(q^{2})&=&\frac{m_{B_{c}}+m_{D_{s}^\ast}}{2m_{D_{s}^\ast}}A_{1}(q^{2})
-\frac{m_{B_{c}}-m_{D_{s}^\ast}}{2m_{D_{s}^\ast}}A_{2}(q^{2}),\label{A3}
\end{eqnarray}
with $A_3(0)=A_0(0)$. We have used $\epsilon_{0123}=+1$ convention throughout the study. The additional tensor form factors are expressed as,
\begin{align}
\left\langle D_{s}^\ast(k,\overline\epsilon)\left\vert \bar{s}i\sigma
_{\mu \nu }q^{\nu }b\right\vert B_{c}(p)\right\rangle
&=-2\epsilon _{\mu\nu\alpha\beta}\overline\epsilon^{\,\ast\nu}p^{\alpha}k^{\beta}T_{1}(q^{2}),\label{FF11}\\
\left\langle D_{s}^\ast(k,\overline\epsilon )\left\vert \bar{s}i\sigma
_{\mu \nu }q^{\nu}\gamma_{5}b\right\vert B_{c}(p)\right\rangle
&=i\Big[\left(m^2_{B_c}-m^2_{D_{s}^\ast}\right)g_{\mu\nu}\overline\epsilon^{\,\ast\nu}\notag\\
&-(\overline\epsilon^{\,\ast }\cdot q)P_{\mu}\Big]T_{2}(q^{2})+i(\overline\epsilon^{\,\ast}\cdot q)\notag\\
&\times\left[q_{\mu}-\frac{q^{2}}{m^2_{B_{c}}-m^2_{D_{s}^\ast}}P_{\mu}
\right]T_{3}(q^{2}).\label{F3}
\end{align}
%The form factors given in Eqs. (\ref{2.13a})-(\ref{F3}) are calculated in the framework of relativistic quark model, and their explicit parametrization is given in \cite{Ebert:2010dv}.
\subsection{Helicity Formalism of $B_{c}\to D_{s}^{\ast}\mu^{+}\mu^{-}$ Decay}
\begin{figure*}[ht!]
\centering
\includegraphics[scale=0.4]{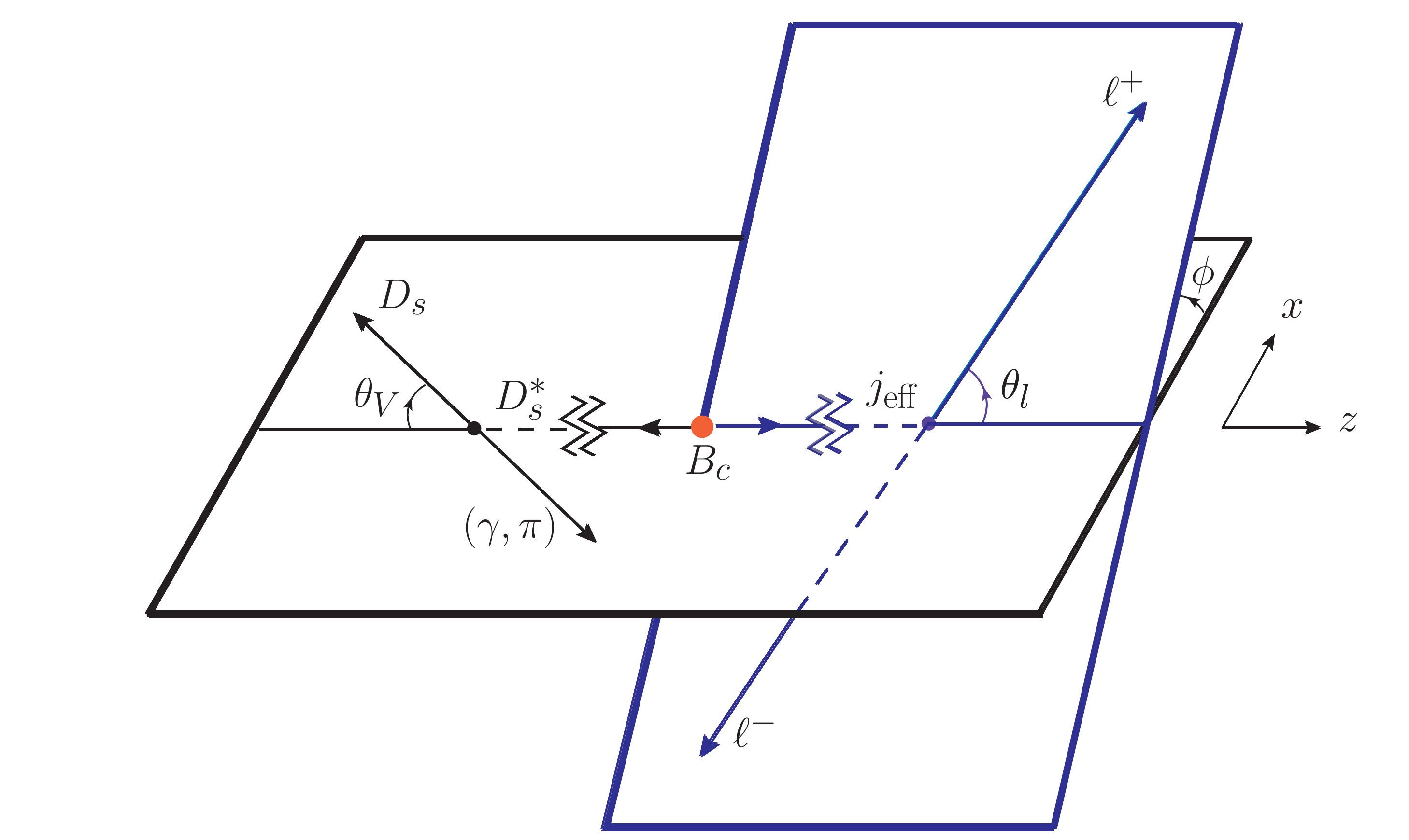}
\caption{Kinematics of the $B_{c}\to D^{\ast}_{s}(\to D_{s}\gamma, (D_{s}\pi) )\ell^{+}\ell^{-}$ decays.}
\label{figfourfold}
\end{figure*}

For $B_{c}\to D_{s}^{\ast}\mu^{+}\mu^{-}$ decay, the amplitude can be expressed in terms of helicity basis. For kinematics of the four-body decay (see Fig. \ref{figfourfold}), we closely follow Ref. \cite{Faessler:2002ut}, where detailed formalism is given. The completeness and orthogonality properties of helicity basis can be expressed as follows,
\begin{eqnarray}
\varepsilon^{\ast\alpha}(n)\varepsilon_{\alpha}(l)=g_{nl}, \qquad\quad \sum_{n, l=t, +, -, 0}\varepsilon^{\ast\alpha}(n)\varepsilon^{\beta}(l)g_{nl}=g^{\alpha\beta},\label{C22}
\end{eqnarray}
with $g_{nl}=\text{diag}(+, -, -, -)$. From the completeness relation given in Eq. (\ref{C22}), the contraction of leptonic tensors $L^{(k)\alpha\beta}$ and hadronic tensors $H^{ij}_{\alpha\beta}=T^{i,D^{\ast}_s}_{\alpha}\overline{T}^{\,j,D^{\ast}_s}_{\beta}$ $(i, j=1, 2)$, can be written as
\begin{eqnarray}
L^{(k)\alpha\beta}H^{ij}_{\alpha\beta}=\sum_{n, n^{\prime}, l, l^{\prime}}L^{(k)}_{nl}g_{nn^{\prime}}g_{ll^{\prime}}H^{ij}_{n^{\prime}l^{\prime}},\label{LH}
\end{eqnarray}
where the leptonic and hadronic tensors can be written in the helicity basis as follows
\begin{eqnarray}
L^{(k)}_{nl}=\varepsilon^{\alpha}(n)\varepsilon^{\ast\beta}(l)L^{(k)}_{\alpha\beta}, &&\qquad H^{ij}_{nl}=\varepsilon^{\ast\alpha}(n)\varepsilon^{\beta}(l)H^{ij}_{\alpha\beta}.\label{LHT}
\end{eqnarray}
Both leptonic and hadronic tensors shown in Eq. (\ref{LHT}), can be evaluated in two different frame of references. The lepton tensor $L^{(k)}_{nl}$ is evaluated in $\mu^{+}\mu^{-}$ centre of mass (CM) frame, and the hadronic tensor $H^{ij}_{nl}$ is evaluated in the rest frame of $B_{c}$ meson. For the said decay one can write the hadronic tensor as follows,
\begin{eqnarray}
H^{ij}_{nl}&=&\big(\varepsilon^{\ast\alpha}(n)T^{i,D^{\ast}_{s}}_{\alpha}\big)\cdot\big(\overline{\varepsilon^{\ast\beta}(l)T^{j,D^{\ast}_{s}}_{\beta}}\big)\notag
\\
&=&\big(\varepsilon^{\ast\alpha}(n)\overline\epsilon^{\ast\mu}(r)T^{i,D^{\ast}_{s}}_{\alpha,\mu}\big)\cdot\big(\overline{\varepsilon^{\ast\beta}(l)
\overline\epsilon^{\ast\nu}(s)T^{j,D^{\ast}_{s}}_{\beta,\nu}}\big)\delta_{rs}\equiv H^{i,D^{\ast}_{s}}_n \,\overline{H}^{\, j,D^{\ast}_{s}}_l,\label{HA5}
\end{eqnarray}
where, from angular momentum conservation, $r=n$ and $s=l$ for $n, l=\pm,0$ and $r, s=0$ for $n, l=t$.
The explicit expressions of the helicity amplitudes for $B_{c}\to D_{s}^{\ast}$, are obtained in terms of the SM and NP Wilson coefficients as,
\begin{align}
H^{1,D_{s}^{\ast}}_t&=-i\sqrt{\frac{\lambda}{q^2}}(C_{9}^{\text{eff}}+C_{9\ell}^{\text{NP}}-C_{9^{\prime}\ell}^{\text{NP}})A_0,\notag
\\
H^{2,D_{s}^{\ast}}_t&=-i\sqrt{\frac{\lambda}{q^2}}(C_{10}+C_{10\ell}^{\text{NP}}-C_{10^{\prime}\ell}^{\text{NP}})A_0,\notag
\\
H^{1,D_{s}^{\ast}}_{\pm}&=-i\left(m^2_{B_c}-m^2_{D_{s}^\ast}\right)\Big[(C_{9}^{\text{eff}}+C_{9\ell}^{\text{NP}}-C_{9^{\prime}\ell}^{\text{NP}})
\frac{A_{1}}{\left(m_{B_c}-m_{D_s^\ast}\right)}\notag
\\
&+\frac{2m_{b}}{q^{2}}C_{7}^{\text{eff}}T_{2}\Big]
\pm i\sqrt{\lambda}\Big[(C_{9}^{\text{eff}}+C_{9\ell}^{\text{NP}}+C_{9^{\prime}\ell}^{\text{NP}})
\frac{V}{\left(m_{B_c}+m_{D_s^\ast}\right)}+\frac{2m_{b}}{q^{2}}C_{7}^{\text{eff}}T_{1}\Big],\notag
\\
H^{2,D_s^\ast}_{\pm}&=-i(C_{10}+C_{10\ell}^{\text{NP}}-C_{10^{\prime}\ell}^{\text{NP}})\left(m_{B_c}+m_{D_{s}^\ast}\right)
A_{1}\notag
\\
&\pm i\sqrt{\lambda}(C_{10}+C_{10\ell}^{\text{NP}}+C_{10^{\prime}\ell}^{\text{NP}})
\frac{V}{\left(m_{B_c}+m_{D_s^\ast}\right)},\notag
\\
H^{1,D_s^\ast}_0&=-\frac{i}{2m_{D_s^\ast}\sqrt{q^2}}\Bigg[(C_{9}^{\text{eff}}+C_{9\ell}^{\text{NP}}-C_{9^{\prime}\ell}^{\text{NP}})
\Big\{(m^2_{B_c}-m^2_{D_s^\ast}-q^2)\left(m_{B_c}+m_{D_s^\ast}\right)A_{1}\notag
\\
&-\frac{\lambda}{m_{B_c}+m_{D_s^\ast}}A_{2}\Big\}+2m_b C_{7}^{\text{eff}}\Big\{(m^2_{B_c}+3m^2_{D_s^\ast}-q^2)T_{2}
-\frac{\lambda}{m^2_{B_c}-m^2_{D_s^\ast}}T_{3}\Big\}
\Bigg],\notag
\end{align}
\begin{align}
H^{2,D_s^\ast}_0&=-\frac{i}{2m_{D_s^\ast}\sqrt{q^2}}(C_{10}+C_{10\ell}^{\text{NP}}-C_{10^{\prime}\ell}^{\text{NP}})
\Bigg[(m^2_{B_c}-m^2_{D_s^\ast}-q^2)\left(m_{B_c}+m_{D_s^\ast}\right)A_{1}\notag
\\
&-\frac{\lambda}{m_{B_c}+m_{D_s^\ast}}A_{2}\Bigg].\label{HA6}
\end{align}
\subsection{Four-fold Angular Distribution of
$B_{c}\to D_{s}^{\ast}(\to D_{s}\gamma (D_{s}\pi))\ell^{+}\ell^{-}$ Decays}
In an effective theory, the NP effects are due to the Wilson coefficients and new operators given in Eq. (\ref{H1}). For the decay modes $B_{c}\to D_{s}^{\ast}(\to D_{s}\gamma (D_{s}\pi))\ell^{+}\ell^{-}$, these effects are contained in the four-dimensional differential decay distribution that depends on the square of the momentum transfer $q^{2}$, angles $\theta_{\ell}$, $\theta_{V}$, and $\phi$ as described in Fig. \ref{figfourfold}. For the decays under consideration, the full differential angular distribution can be written as,
\begin{eqnarray}
 \frac{d^4\Gamma\left(B_{c}\to D_{s}^{\ast}\,(\to D_{s}\gamma(D_s\pi)\ell^+\ell^-\right)}{dq^2 \ d\cos{\theta_{l}} \ d\cos {\theta}_{V} \ d\phi} &=& \frac{9}{32 \pi} \mathcal{B}(D_{s}^{\ast}\to D_{s}\gamma(D_s\pi))\notag
 \\
&\times&\bigg[I^{\gamma}_{1s,\perp}(I_{1s,\|})\sin^2\theta_{V}+I^{\gamma}_{1c,\perp}(I_{1c,\|})\cos^2\theta_{V}\notag\\
&+&\Big(I^{\gamma}_{2s,\perp}(I_{2s,\|})\sin^2\theta_{V}+I^{\gamma}_{2c,\perp}(I_{2c,\|})\cos^2\theta_{V}\Big)\cos{2\theta_{l}}
\notag\\
&+&\Big(I^{\gamma}_{6s,\perp}(I_{6s,\|})\sin^2\theta_{V}+I^{\gamma}_{6c,\perp}\cos^2\theta_{V}\Big)\cos{\theta_{l}}\notag\\
&+&\Big(I^{\gamma}_{3,\perp}(I_{3,\|})\cos{2\phi}
+I^{\gamma}_{9,\perp}(I_{9,\|})\sin{2\phi}\Big)\sin^2\theta_{V}\sin^2\theta_{l}\notag
\\
&+&\Big(I^{\gamma}_{4,\perp}(I_{4,\|})\cos{\phi}+I^{\gamma}_{8,\perp}(I_{8,\|})\sin{\phi}\Big)\sin2\theta_{V}\sin2\theta_{l}\notag
\\
&+&\Big(I^{\gamma}_{5,\perp}(I_{5,\|})\cos{\phi}+I^{\gamma}_{7,\perp}(I_{7,\|})\sin{\phi}\Big)\sin2\theta_{V}\sin\theta_{l}\bigg],\notag\\
\label{fullad}
\end{eqnarray}
where $I_{n\lambda,\perp}^{\gamma}$ and $I_{n\lambda,\|}$ are the angular coefficients. The explicit expressions of $I_{n\lambda,\perp}^{\gamma}$ in terms of the helicity amplitudes are obtained as,
\begin{eqnarray}\label{30df}
I^{\gamma}_{1s,\perp} &=&\frac{(2+\beta_l^2)}{4}N^2\left(|H_+^1|^2+|H_+^2|^2+|H_-^1|^2+|H_-^2|^2\right)+\left(|H_0^1|^2+|H_0^2|^2\right)\notag\\
&+&\frac{2m_l^2}{q^2}N^2\bigg[\left(|H_+^1|^2-|H_+^2|^2+|H_-^1|^2-|H_-^2|^2\right)
+2\left(|H_0^1|^2-|H_0^2|^2+2|H_t^2|^2\right)\bigg],\\
I^{\gamma}_{1c,\perp} &=& \frac{(2+\beta_l^2)}{2}N^2\left(|H_+^1|^2+|H_+^2|^2+|H_-^1|^2+|H_-^2|^2\right)
\notag\\
&+&\frac{4m_l^2}{q^2}N^2\left(|H_+^1|^2-|H_+^2|^2+|H_-^1|^2-|H_-^2|^2\right),\\
I^{\gamma}_{2s,\perp} &=& -\beta_l^2N^2\bigg[\left(|H_0^1|^2+|H_0^2|^2\right)-\frac{1}{4}\left(|H_+^1|^2+|H_+^2|^2+|H_-^1|^2+|H_-^2|^2\right)\bigg]
,\\
I^{\gamma}_{2c,\perp} &=& \frac{\beta_l^2}{2}N^2\left(|H_+^1|^2+|H_+^2|^2+|H_-^1|^2+|H_-^2|^2\right),\\
I^{\gamma}_{3,\perp}&=&\beta_l^2N^2\bigg[\mathcal{R}e\left(H_+^{1}H_-^{1\ast}+H_+^{2}H_-^{2\ast}\right)\bigg],\\
I^{\gamma}_{4,\perp}&=&-\frac{\beta_l^2}{2}N^2\bigg[\mathcal{R}e\left(H_+^{1}H_0^{1\ast}+H_-^{1}H_0^{1\ast}\right)
+\mathcal{R}e\left(H_+^{2}H_0^{2\ast}+H_-^{2}H_0^{2\ast}\right)\bigg],
\end{eqnarray}
\begin{eqnarray}
I^{\gamma}_{5,\perp}&=&\beta_lN^2\bigg[\mathcal{R}e\left(H_+^{1}H_0^{2\ast}-H_-^{1}H_0^{2\ast}\right)
+\mathcal{R}e\left(H_+^{2}H_0^{1\ast}-H_-^{2}H_0^{1\ast}\right)\bigg],\\
I^{\gamma}_{6s,\perp}&=&-2\beta_lN^2\bigg[\mathcal{R}e\left(H_+^{1}H_+^{2\ast}-H_-^{1}H_-^{2\ast}\right)\bigg],\\
I^{\gamma}_{6c,\perp}&=&-4\beta_lN^2\bigg[\mathcal{R}e\left(H_+^{1}H_+^{2\ast}-H_-^{1}H_-^{2\ast}\right)\bigg],\\
I^{\gamma}_{7,\perp}&=&\beta_lN^2\bigg[\mathcal{I}m\left(H_0^{1}H_+^{2\ast}+H_0^{1}H_-^{2\ast}\right)
+\mathcal{I}m\left(H_0^{2}H_+^{1\ast}+H_0^{2}H_-^{1\ast}\right)\bigg],\\
I^{\gamma}_{8,\perp}&=&-\frac{\beta_l^2}{2}N^2\bigg[\mathcal{I}m\left(H_0^{1}H_+^{1\ast}-H_0^{1}H_-^{1\ast}\right)
+\mathcal{I}m\left(H_0^{2}H_+^{2\ast}-H_0^{2}H_-^{2\ast}\right)\bigg],\\
I^{\gamma}_{9,\perp}&=&-\beta_l^2N^2\bigg[\mathcal{I}m\left(H_+^{1}H_-^{1\ast}+H_+^{2}H_-^{2\ast}\right)\bigg],
\end{eqnarray}
whereas the expressions of $I_{n\lambda,\|}$ in terms of the helicity amplitudes are written as,
\begin{eqnarray}\label{IsDpi}
I_{1s,\|} &=& \frac{(2+\beta_l^2)}{2}N^2\left(|H_+^1|^2+|H_+^2|^2+|H_-^1|^2+|H_-^2|^2\right)\notag
\\&+&\frac{4m_l^2}{q^2}N^2\left(|H_+^1|^2-|H_+^2|^2+|H_-^1|^2-|H_-^2|^2\right),\\
I_{1c,\|} &=& 2N^2\left(|H_0^1|^2+|H_0^2|^2\right)+\frac{8m_l^2}{q^2}N^2\left(|H_0^1|^2-|H_0^2|^2+2|H_t^2|^2\right),\\
I_{2s,\|} &=& \frac{\beta_l^2}{2}N^2\left(|H_+^1|^2+|H_+^2|^2+|H_-^1|^2+|H_-^2|^2\right),\\
I_{2c,\|} &=& -2\beta_l^2N^2\left(|H_0^1|^2+|H_0^2|^2\right),\\
I_{3,\|}&=&-2\beta_l^2N^2\bigg[\mathcal{R}e\left(H_+^{1}H_-^{1\ast}+H_+^{2}H_-^{2\ast}\right)\bigg],\\
I_{4,\|}&=&\beta_l^2N^2\bigg[\mathcal{R}e\left(H_+^{1}H_0^{1\ast}+H_-^{1}H_0^{1\ast}\right)
+\mathcal{R}e\left(H_+^{2}H_0^{2\ast}+H_-^{2}H_0^{2\ast}\right)\bigg],\\
I_{5,\|}&=&-2\beta_lN^2\bigg[\mathcal{R}e\left(H_+^{1}H_0^{2\ast}-H_-^{1}H_0^{2\ast}\right)
+\mathcal{R}e\left(H_+^{2}H_0^{1\ast}-H_-^{2}H_0^{1\ast}\right)\bigg],\\
I_{6s,\|}&=&-4\beta_lN^2\bigg[\mathcal{R}e\left(H_+^{1}H_+^{2\ast}-H_-^{1}H_-^{2\ast}\right)\bigg],\\
I_{6c,\|}&=&0,\\
I_{7,\|}&=&-2\beta_lN^2\bigg[\mathcal{I}m\left(H_0^{1}H_+^{2\ast}+H_0^{1}H_-^{2\ast}\right)
+\mathcal{I}m\left(H_0^{2}H_+^{1\ast}+H_0^{2}H_-^{1\ast}\right)\bigg],\\
I_{8,\|}&=&\beta_l^2N^2\bigg[\mathcal{I}m\left(H_0^{1}H_+^{1\ast}-H_0^{1}H_-^{1\ast}\right)
+\mathcal{I}m\left(H_0^{2}H_+^{2\ast}-H_0^{2}H_-^{2\ast}\right)\bigg],\\
I_{9,\|}&=&2\beta_l^2N^2\bigg[\mathcal{I}m\left(H_+^{1}H_-^{1\ast}+H_+^{2}H_-^{2\ast}\right)\bigg],
\end{eqnarray}
where
\begin{eqnarray}\label{24abc}
N=V_{tb}V^{\ast}_{ts}\Bigg[\frac{G_{F}^2\alpha^2}{3.2^{10} \pi^5 m_{B_c}^{3}} q^2\sqrt{\lambda}\beta_l\Bigg]^{1/2},
\end{eqnarray}
with $\lambda\equiv \lambda(m^2_{B_c}, m^2_{D_{s}^{\ast}}, q^2)$ and $\beta_l=\sqrt{1-4m_l^2/q^2}$.
\subsection{Physical Observables for $B_{c}\to D_{s}^{\ast}(\to D_{s}\gamma(D_{s}\pi))\ell^{+}\ell^{-}$ Decays}
In this section, we construct the physical observables for the $B_{c}\to D^{\ast}_{s}(\to D_{s}\gamma(D_{s}\pi))\ell^{+}\ell^{-}$  decays, in terms of the angular coefficients. The observables which we consider are the differential branching ratios $(d\mathcal{B}/dq^2)$, lepton forward-backward asymmetry $(A_{\text{FB}})$, longitudinal polarization fraction of $D^{\ast}_{s}$ $(f_L)$, unpolarized $(R_{D_s^{\ast}})$, and polarized $(R_{D_s^{\ast}}^{L,T})$ LFUV ratios, and the angular coefficients $(\langle I_{n\lambda,\perp}^{\gamma}\rangle, \langle I_{n\lambda,\|}\rangle)$. Other than the differential decay rates and the ratios, all observables are normalized to the corresponding differential decay rate.\\
\textbf{(i) Differential decay rates:}
From the full angular distribution Eq. \eqref{fullad}, $q^{2}$ dependent differential decay rate expressions are obtained in terms of angular coefficients as follows,
\begin{eqnarray}
\frac{d\Gamma (B_{c}\to D^{\ast}_{s}\mu^{+}\mu^{-})}{dq^{2}}&=&\frac{1}{4}(3I^{\gamma}_{1c,\perp}+6I^{\gamma}_{1s,\perp}-I^{\gamma}_{2c,\perp}-2I^{\gamma}_{2s,\perp})\notag\\
&=&\frac{1}{4}(3I_{1c,||}+6I_{1s,||}-I_{2c,||}-2I_{2s,||}).\label{DBR}
\end{eqnarray}
\begin{eqnarray}
\frac{d\Gamma\left(B_{c}\to D^{\ast}_{s}(\to D_{s}\gamma)\mu^{+}\mu^{-}\right)}{dq^{2}}=\mathcal{B}(D_{s}^{\ast}\to D_{s}\gamma)\frac{1}{4}(3I^{\gamma}_{1c,\perp}+6I^{\gamma}_{1s,\perp}-I^{\gamma}_{2c,\perp}-2I^{\gamma}_{2s,\perp}).\label{DBR}
\end{eqnarray}

\begin{eqnarray}
\frac{d\Gamma \left(B_{c}\to D^{\ast}_{s}(\to D_{s}\pi)\mu^{+}\mu^{-}\right)}{dq^{2}}=\mathcal{B}(D_{s}^{\ast}\to D_s\pi)\frac{1}{4}(3I_{1c,||}+6I_{1s,||}-I_{2c,||}-2I_{2s,||}).\label{DBR}
\end{eqnarray}
\textbf{(ii) Lepton forward backward asymmetry:}
The lepton forward backward asymmetry as a function of $q^{2}$ can be expressed as,
\begin{eqnarray}
A_{\text{FB}}(q^{2})=\frac{3\left(I^{\gamma}_{6c,\perp}+2I^{\gamma}_{6s,\perp}\right)}{2(3I^{\gamma}_{1c,\perp}+6I^{\gamma}_{1s,\perp}-I^{\gamma}_{2c,\perp}-2I^{\gamma}_{2s,\perp})}=\frac{6I_{6s,||}}{2(3I_{1c,||}+6I_{1s,||}-I_{2c,||}-2I_{2s,||})}.\label{FB}
\end{eqnarray}
\textbf{(iii) Longitudinal helicity fraction:}
The longitudinal helicity fraction of the decay $B_{c}\to D_{s}^{\ast}(\to D_{s}\gamma,(D_{s}\pi))\mu^{+}\mu^{-}$, when $D_{s}^{\ast}$ meson is longitudinally polarized can be expressed as,
\begin{eqnarray}
f_{L}(q^{2})=\frac{(6I^{\gamma}_{1s,\perp}-2I^{\gamma}_{2s,\perp})-(3I^{\gamma}_{1c,\perp}-I^{\gamma}_{2c,\perp})}{3I^{\gamma}_{1c,\perp}+6I^{\gamma}_{1s,\perp}-I^{\gamma}_{2c,\perp}-2I^{\gamma}_{2s,\perp}}=\frac{3I_{1c,||}-I_{2c,||}}{3I_{1c,||}+6I_{1s,||}-I_{2c,||}-2I_{2s,||}}.
\end{eqnarray}
\textbf{(iv) LFUV ratios for $B_{c}\to D_{s}^{\ast}\ell^{+}\ell^{-}$ Decay:}
The unpolarized and polarized LFUV for the decay $B_{c}\to D_{s}^{\ast}\ell^{+}\ell^{-}$ can be written as,
\begin{eqnarray}
{R_{D_{s}^{\ast}}}_{\left[q^{2}_{\text{min}},\, q^{2}_{\text{max}}\right]}=\frac{\int^{q^{2}_{\text{max}}}_{q^{2}_{\text{min}}}(d\mathcal{B}(B_{c}\to D_{s}^{\ast}\mu^{+}\mu^{-})/dq^{2})dq^{2}}{\int^{q^{2}_{\text{max}}}_{q^{2}_{\text{min}}}(d\mathcal{B}(B_{c}\to D_{s}^{\ast}e^{+}e^{-})/dq^{2})dq^{2}},
\end{eqnarray}
\begin{eqnarray}
{R^{\,L,T}_{D_{s}^{\ast}}}_{\left[q^{2}_{\text{min}},\, q^{2}_{\text{max}}\right]}=\frac{\int^{q^{2}_{\text{max}}}_{q^{2}_{\text{min}}}(d\mathcal{B}(B_{c}\to D_{s}^{\ast{L,T}}\mu^{+}\mu^{-})/dq^{2})dq^{2}}{\int^{q^{2}_{\text{max}}}_{q^{2}_{\text{min}}}(d\mathcal{B}(B_{c}\to D_{s}^{\ast{L,T}}e^{+}e^{-})/dq^{2})dq^{2}}.
\end{eqnarray}
\textbf{(v) Normalized angular observables:}
\begin{eqnarray}
\langle I_{n\lambda,\parallel}\rangle=\frac{I_{n\lambda,\parallel}}{d\Gamma/dq^2},\qquad\quad \langle I_{n\lambda,\perp}^{\gamma}\rangle=\frac{I_{n\lambda,\perp}^{\gamma}}{d\Gamma/dq^2}.
\end{eqnarray}
\textbf{(vi) Binned normalized angular observables:}
\begin{eqnarray}
\langle I_{n\lambda,\parallel}\rangle_{\left[q^{2}_{\text{min}},\, q^{2}_{\text{max}}\right]}=\frac{\int^{q^{2}_{\text{max}}}_{q^{2}_{\text{min}}}I_{n\lambda,\parallel}\,dq^2}{\int^{q^{2}_{\text{max}}}_{q^{2}_{\text{min}}}(d\Gamma/dq^2d)dq^2},\qquad\quad \langle I_{n\lambda,\perp}^{\gamma}\rangle_{\left[q^{2}_{\text{min}},\, q^{2}_{\text{max}}\right]}=\frac{\int^{q^{2}_{\text{max}}}_{q^{2}_{\text{min}}}I_{n\lambda,\perp}^{\gamma}\,dq^2}{\int^{q^{2}_{\text{max}}}_{q^{2}_{\text{min}}}(d\Gamma/dq^2)dq^2}.
\end{eqnarray}
\section{Phenomenological Analysis}\label{NUM}
\subsection{Input Parameters}
To investigate NP effects in the observables of the $B_{c}\to D_{s}^{\ast}(\to D_{s}\gamma,(D_{s}\pi))\ell^{+}\ell^{-}$ decays, we use input parameters such as the transition form factors, which are calculated in the framework of the relativistic quark model (RQM)\cite{Ebert:2010dv}. The RQM, based on the quasipotential approach, reliably determines the form factors in the whole $q^2$ range by incorporating relativistic effects including contributions of intermediate negative energy states and relativistic transformations of the meson wave functions. Furthermore, the form factors obtained in the RQM, satisfy all the model independent symmetry relations arising in the limits of heavy quark mass and large recoil of the final meson \cite{Ebert:2001pc}. The form factors calculated, in the RQM, through the overlap integrals of the initial and final meson relativistic wave functions \cite{Ebert:2010dv}, can be expressed in terms of the expressions involving fitted parameters. Such as, the form factors, $V(q^{2})$, $A_{0}(q^{2})$ and $T_{1}(q^{2})$ given in Eqs. (\ref{2.13a}, \ref{2.13b}), and (\ref{FF11}), are parameterized in the whole kinematical $q^2$ region as,
\begin{align}\label{FF1}
    F(q^{2})=\frac{F(0)}{\big(1-\frac{q^{2}}{{M}^{2}}\big)\big(1-\sigma_{1}\frac{q^{2}}{M^{2}_{B^{\ast}_s}}+\sigma_{2}\frac{q^{4}}{M^{4}_{B^{\ast}_s}}\big)},
\end{align}
where the form factor $A_0(q^2)$ contains a pole at $q^2=M^2\equiv M^2_{B_s}$ and the form factors $V(q^{2})$, $T_{1}(q^{2})$ contain pole at $q^2=M^2\equiv M^2_{B^{\ast}_s}$. The numerical values of these pole masses are $M_{B_s}=5.36692$ GeV, and $M_{B^{\ast}_s}=5.4154$ GeV \cite{ParticleDataGroup:2022pth}. Moreover, the form factors $A_{1}(q^{2})$, $A_{2}(q^{2})$, $T_{2}(q^{2})$, and $T_{3}(q^{2})$, given in Eq. (\ref{2.13b}), and Eq. (\ref{F3}), can be parameterized as follows,
\begin{align}\label{FF2}
    F(q^{2})=\frac{F(0)}{\big(1-\sigma_{1}\frac{q^{2}}{M^{2}_{B^{\ast}_s}}+\sigma_{2}\frac{q^{4}}{M^{4}_{B^{\ast}_s}}\big)}.
\end{align}
For completeness, the numerical values of form factors at $q^{2}=0$, and fitted parameters $\sigma_{1}$ and $\sigma_{2}$, are collected in Table \ref{FF table}. In order to gauge the effects of the form
factor uncertainties on various observables, we allow the parameters in the fitted form factors to deviate by $\pm5\%$.
\begin{table*}[!htbp]
\centering
\captionsetup{margin=0.5cm}
\caption{\small The numerical values of transition form factors for $B_{c}\to D_{s}^{\ast}\mu^{+}\mu^{-}$ decay at $q^{2}=0$, and the fitted parameters $\sigma_{1}$ and $\sigma_{2}$ \cite{Ebert:2010dv}. The reported uncertainties represent the $\pm5\%$ deviations in the parameters of the fitted form factors.}\label{FF table}
 \renewcommand{\arraystretch}{1.5}
    \scalebox{0.90}{
\begin{tabular}{|c|c|c|c|c|c|c|c|}
\hline
&$V$&$A_{0}$&$A_{1}$&$A_{2}$&$T_{1}$&$T_{2}$&$T_{3}$
\\ \hline
$F(0)$ & $0.182^{+0.010}_{-0.010}$ &   $0.070^{+0.004}_{-0.004}$  & $0.089^{+0.005}_{-0.005}$  &   $0.110^{+0.006}_{-0.006}$  & $0.085^{+0.005}_{-0.005}$ & $0.085^{+0.004}_{-0.004}$ & $0.051^{+0.003}_{-0.003}$\\ \hline
$\sigma_{1}$ &  $2.133^{+0.107}_{-0.107}$  &  $1.561^{+0.078}_{-0.078}$   &   $2.479^{+0.124}_{-0.124}$  &  $2.833^{+0.142}_{-0.142}$  & $1.540^{+0.077}_{-0.077}$ & $2.577^{+0.129}_{-0.129}$ & $2.783^{+0.139}_{-0.139}$\\ \hline
 $\sigma_{2}$ &   $1.183^{+0.059}_{-0.059}$ &  $0.192^{+0.010}_{-0.010}$ & $1.686^{+0.084}_{-0.084}$ & $2.167^{+0.108}_{-0.108}$ & $0.248^{+0.013}_{-0.013}$ & $1.859^{+0.093}_{-0.093}$ & $2.170^{+0.109}_{-0.109}$ \\
 \hline
 \end{tabular}}
\end{table*}
The numerical values of Wilson coefficients in the SM, evaluated at the renormalization scale $\mu\sim m_{b}$ \cite{Blake:2016olu}, are presented in Table \ref{wc table}. The other input parameters are listed in Table \ref{input}.
%The other input parameters are the Wilson coefficients, the various masses of quarks and leptons, the branching fractions of $\mathcal{B}(D_{s}^{\ast}\to D_{s}\gamma)$, $\mathcal{B}(D_{s}^{\ast}\to D_{s}\pi)$, the values of CKM matrix and the life time of $B_{c}$ meson. The numerical values of these parmeters are presented in Table \ref{wc table} and Table \ref{input}.
\begin{table*}[!htbp]
\centering
\captionsetup{margin=0.30cm}
\caption{\small The numerical values of the SM Wilson coefficients up to NNLL accuracy, evaluated at the renormalization scale $\mu\sim m_{b}$ \cite{Blake:2016olu}.}
\label{wc table}
\begin{tabular}{|c|c|c|c|c|c|c|c|c|c|}
\hline
$C_{1}$&$C_{2}$&$C_{3}$&$C_{4}$&$C_{5}$&$C_{6}$&$C_{7}$&$C_{8}$&$C_{9}$&$C_{10}$
\\ \hline
  $-0.294$ &   $1.017$  & $-0.0059$  &   $-0.087$  &
  $0.0004$  &  $0.0011$   &   $-0.324$  &  $-0.176$  &
    $4.114$  &  $-4.193$ \\
\hline
\end{tabular}
\end{table*}
\begin{table}[ht]
\centering
\captionsetup{margin=3.0cm}
\caption{\small Values of different input parameters used in the numerical analysis.}
\label{input}
\begin{tabular}{|c|}
\hline
$M_{B_{c}}=6.27$ GeV, $m_{b}=4.28$ GeV, $m_{s}=0.13$ GeV, \\
$m_{\mu}=0.105$ GeV, $m_{\tau}=1.77$ GeV, \\
$\mathcal{B}(D^{\ast}_{s}\to D_{s}\gamma)=93.5\times 10^{-2}$, $\mathcal{B}(D^{\ast}_{s}\to D_{s}\pi)=5.8\times 10^{-2}$\\
$|V_{tb}V_{ts}^{\ast}|=45\times 10^{-3}$, $\alpha^{-1}=137$, $%
G_{F}=1.17\times 10^{-5}$ GeV$^{-2}$, \\
$\tau_{B_{c}}=0.46\times 10^{-12}$ sec, $M_{D^{\ast}_{s}}=2.1123$ GeV. \\ \hline
\end{tabular}\label{Input}
\end{table}
\\
\subsection{NP Scenarios}
In this section, we first specify our choice of the NP scenarios which are used to investigate the effects of NP on various physical observables in the angular distribution of $B_{c}\to D^{\ast}_{s}(\to D_{s}\gamma)\mu^{+}\mu^{-}$ and $B_{c}\to D^{\ast}_{s}(\to D_{s}\pi)\mu^{+}\mu^{-}$ decays in a model independent framework. We choose the best fit values of NP couplings in different scenarios, from the recent global fit analysis \cite{Alok:2022pjb}. The global fit analysis performed by authors of Ref.\cite{Alok:2022pjb}, shows that with the assumption of NP present only in the muon sector, two 1D NP scenarios $C^{\text{NP}}_{9\mu}<0$, and $C^{\text{NP}}_{9\mu}=-C^{\text{NP}}_{10\mu}$, continue to be the most favored scenarios, whereas the 2D scenarios $(C^{\text{NP}}_{9\mu},C^{\text{NP}}_{10^{\prime}\mu}),(C^{\text{NP}}_{9\mu},C^{\text{NP}}_{9^{\prime}\mu})$ and $(C^{\text{NP}}_{9\mu},C^{\text{NP}}_{10\mu})$ provide better fit to data with preference decreasing in the listed order. The best fit values of these 1D and 2D NP scenarios are listed in Table-\ref{tab:bestfitWC}.
\begin{table*}[!htbp]
\begin{center}
\captionsetup{margin=4.0cm}
\caption{\small Best-fit values of the 1D and 2D NP scenarios considering NP in muon sector only \cite{Alok:2022pjb}.}\label{tab:bestfitWC}
			\begin{tabular}{|clc|}
				\hline
				Scenario & & Best-fit value \\
				\hline
                S1 & $C_{9\mu}^{\text{NP}}$                                         &$-0.98$       \\
                S2 & $C_{9\mu}^{\text{NP}}=-C_{10\mu}^{\text{NP}}$                  &$-0.46$        \\
                S3 & $(C_{9\mu}^{\text{NP}}, C_{10^{\prime}\mu}^{\text{NP}})$      &$(-1.15, -0.26)$        \\
                S4 & $(C_{9\mu}^{\text{NP}}, C_{9^{\prime}\mu}^{\text{NP}})$       &$(-1.12, 0.40)$         \\
                S5 & $(C_{9\mu}^{\text{NP}}, C_{10\mu}^{\text{NP}})$                &$(-0.80, 0.24)$        \\
                \hline
            \end{tabular}
	\end{center}
\end{table*}
 %In this context we use these scenarios to analyze the effects of NP on the physical observables such as the branching ratio, forward-backward asymmetry ($A_{FB}$), the longitudinal polarization ($f_{L}$) of $D_{s}^{\ast}$ meson, lepton flavor universality violating ratio (LFUV) ($R_{D^{\ast}_{s}}$), and the angular observables such as $I^{\gamma}_{1s,\perp}(I_{1s,||}),I^{\gamma}_{1c,\perp}(I_{1c,||}),I^{\gamma}_{2s,\perp}(I_{2s,||}),I^{\gamma}_{2c,\perp}(I_{2c,||}),I^{\gamma}_{3,\perp}(I_{3,||}),I^{\gamma}_{4,\perp}(I_{4,||}),I^{\gamma}_{5,\perp}(I_{5,||}),I^{\gamma}_{6s,\perp}(I_{6s,||})$ and $I^{\gamma}_{6c,\perp}(I_{6c,||})$.

\subsection{Analysis of Physical Observables  in $B_{c}\to D^{\ast}_{s}(\to D_{s}\gamma,(D_{s}\pi))\ell^{+}\ell^{-}$ Decays}
\begin{figure*}[t!]
\begin{tabular}{cc}
\hspace{0.6cm}($\mathbf{a}$)&\hspace{1.2cm}($\mathbf{b}$)\\
\includegraphics[scale=0.4]{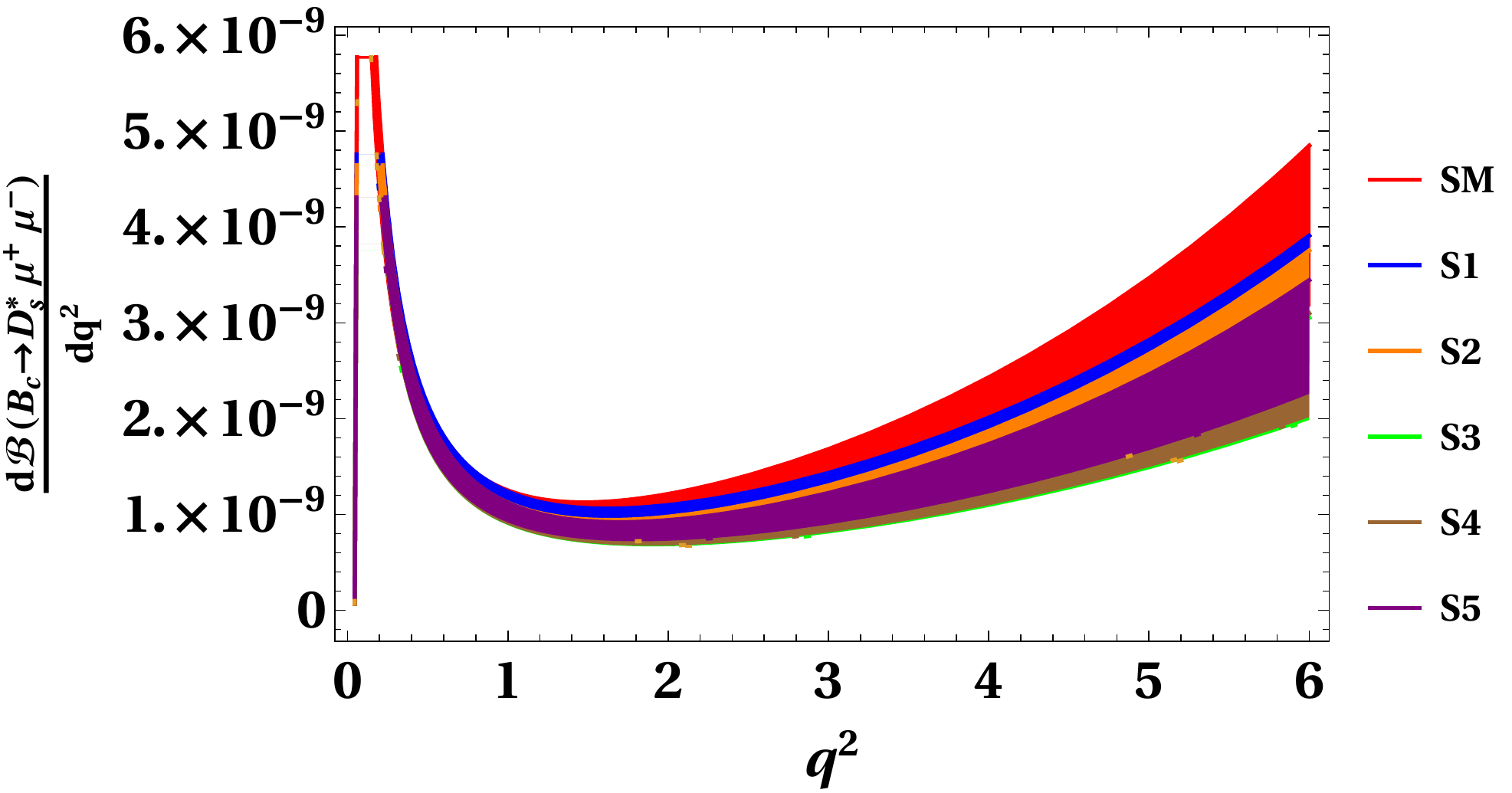}\ \ \
& \ \ \ \includegraphics[scale=0.4]{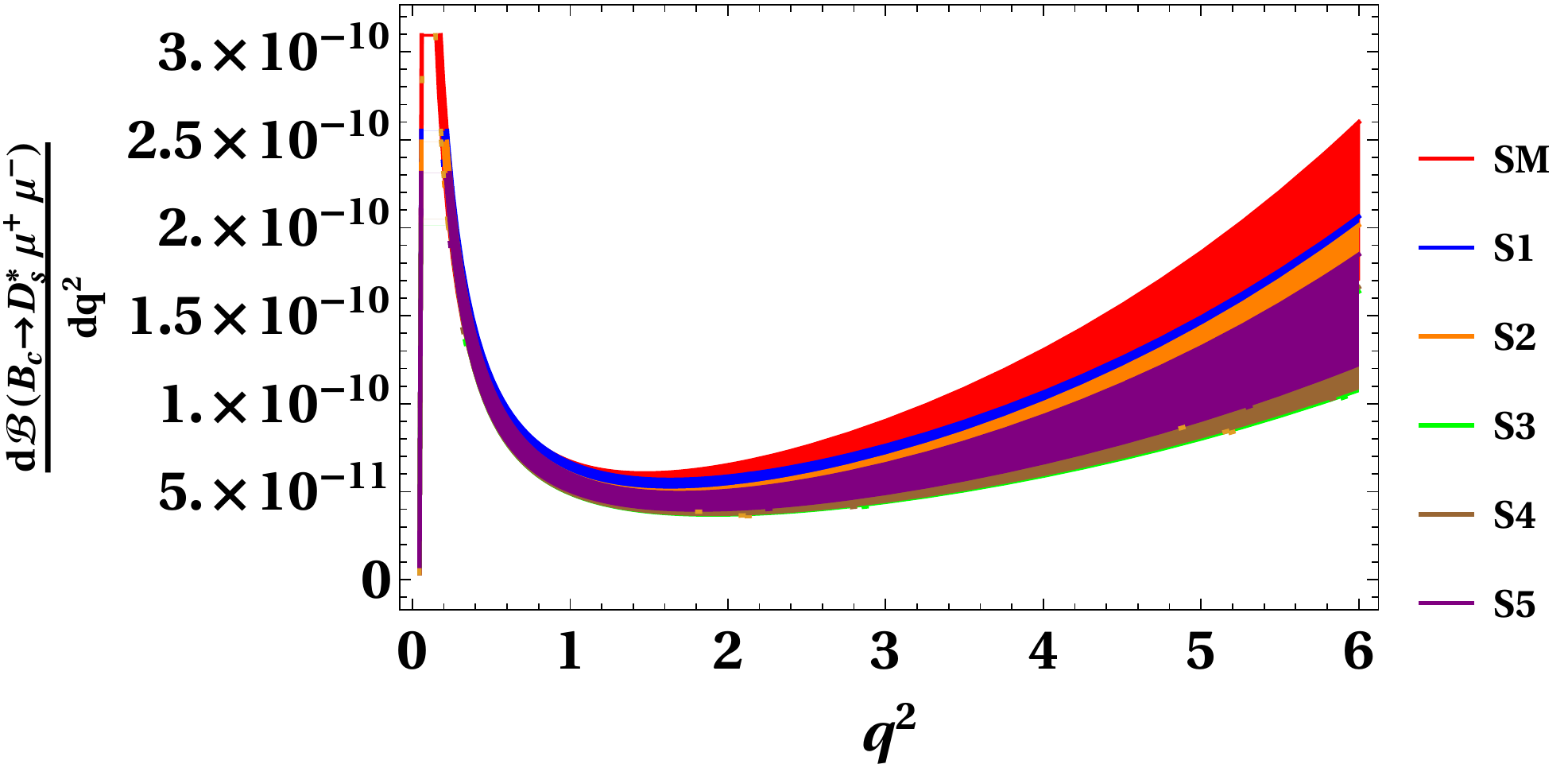}\\\
\end{tabular}
\caption{Differential branching ratio for the decay $B_{c}\to D^{\ast}_{s}\mu^{+}\mu^{-}$ in the SM and the NP scenarios.
%Fig.\ref{figBR}
(a) depicts the differential branching ratio for the cascade decay $B_{c}\to D^{\ast}_{s}(\to D_{s}\gamma)\mu^{+}\mu^{-}$, and
%Fig.\ref{figBR}
(b) depicts the differential branching ratio for the cascade decay $B_{c}\to D^{\ast}_{s}(\to D_{s}\pi)\mu^{+}\mu^{-}$.}\label{figBR}
\end{figure*}
In this section, we now analyze the NP effects via observables which are constructed from the combination of the angular coefficients, such as the differential branching ratios $(d\mathcal{B}/dq^2)$, lepton forward-backward asymmetry $(A_{\text{FB}})$, longitudinal polarization fraction of $D^{\ast}_{s}$ $(f_L)$, unpolarized $(R_{D_s^{\ast}})$, and polarized $(R_{D_s^{\ast}}^{L,T})$ LFUV ratios, in $B_{c}\to D_{s}^{\ast}(\to D_{s}\gamma, (D_{s}\pi))\ell^{+}\ell^{-}$ decays. Binned averaged numerical values of the SM and NP predictions of all these observables, with errors due to the form factors, in different $q^2$ bins, are given in Tables \ref{Obs1}-\ref{Obs7}, of appendix \ref{append1}. In Figs. \ref{figBR}-\ref{FBA}, we plot
differential branching ratios, forward-backward asymmetry, and the longitudinal polarization fraction of $D^{\ast}_{s}$, as a function of $q^{2}$, leading to following observations. Moreover, our results regarding to LFUV ratios, in all NP scenarios, do not show sizable deviations from the SM predictions, therefore, we do not present their $q^2$ plots.
\begin{itemize}
\item In Fig. \ref{figBR}(a) and \ref{figBR}(b), we have plotted the differential branching ratios for  $B_{c}\to D^{\ast}_{s}(\to D_{s}\gamma)\mu^{+}\mu^{-}$ and $B_{c}\to D^{\ast}_{s}(\to D_{s}\pi)\mu^{+}\mu^{-}$ decays as a function of $q^{2}$ in the framework of the SM as well as the NP scenarios under consideration. In SM, the differential branching ratio for $B_{c}\to D^{\ast}_{s}(\to D_{s}\gamma)\mu^{+}\mu^{-}$ decay is of the order $10^{-9}$, whereas for $B_{c}\to D^{\ast}_{s}(\to D_{s}\pi)\mu^{+}\mu^{-}$ decay, it is of the order $10^{-10}$. Both Fig. \ref{figBR}(a) and \ref{figBR}(b), depict
that NP scenarios predictions are compatible with the SM predictions as the error bands emerging due to the uncertainties of the form factors overlap.
However, the central value predictions of all the NP scenarios show trend towards the lesser values of differential branching ratios as compared to the SM expectations.

\item Fig. \ref{FBA}, depicts the forward-backward asymmetry ($A_{\text{FB}}$), and longitudinal helicity fraction ($f_{L}$), of $B_{c}\to D^{\ast}_{s}\mu^{+}\mu^{-}$ decay as a function of $q^{2}$, in the SM framework as well as NP scenarios presented in Table \ref{tab:bestfitWC}. Regarding the zero position of the $A_{FB}$, it is important to mention here that the uncertainty due to the form factors is small and hence the $A_{\text{FB}}$ provides stringent tests to see the NP effects. Fig. \ref{FBA}(a) shows that the zero position of $A_{\text{FB}}$ shifts towards right for all the NP scenarios. The zero crossing in the $A_{\text{FB}}$  at $q^{2}=3$ $\text{GeV}^{2}, q^{2}=2.8$ $\text{GeV}^{2}$, and $q^{2}=3.4$ $\text{GeV}^{2}$ in the presence of NP scenarios (S1, S5), S2, (S3, S4), respectively is quite distinct from the SM prediction at $q^{2}=2.6$ $\text{GeV}^{2}$.
\begin{figure*}[b!]
\begin{tabular}{cc}
\hspace{0.6cm}($\mathbf{a}$)&\hspace{1.2cm}($\mathbf{b}$)\\
\includegraphics[scale=0.4]{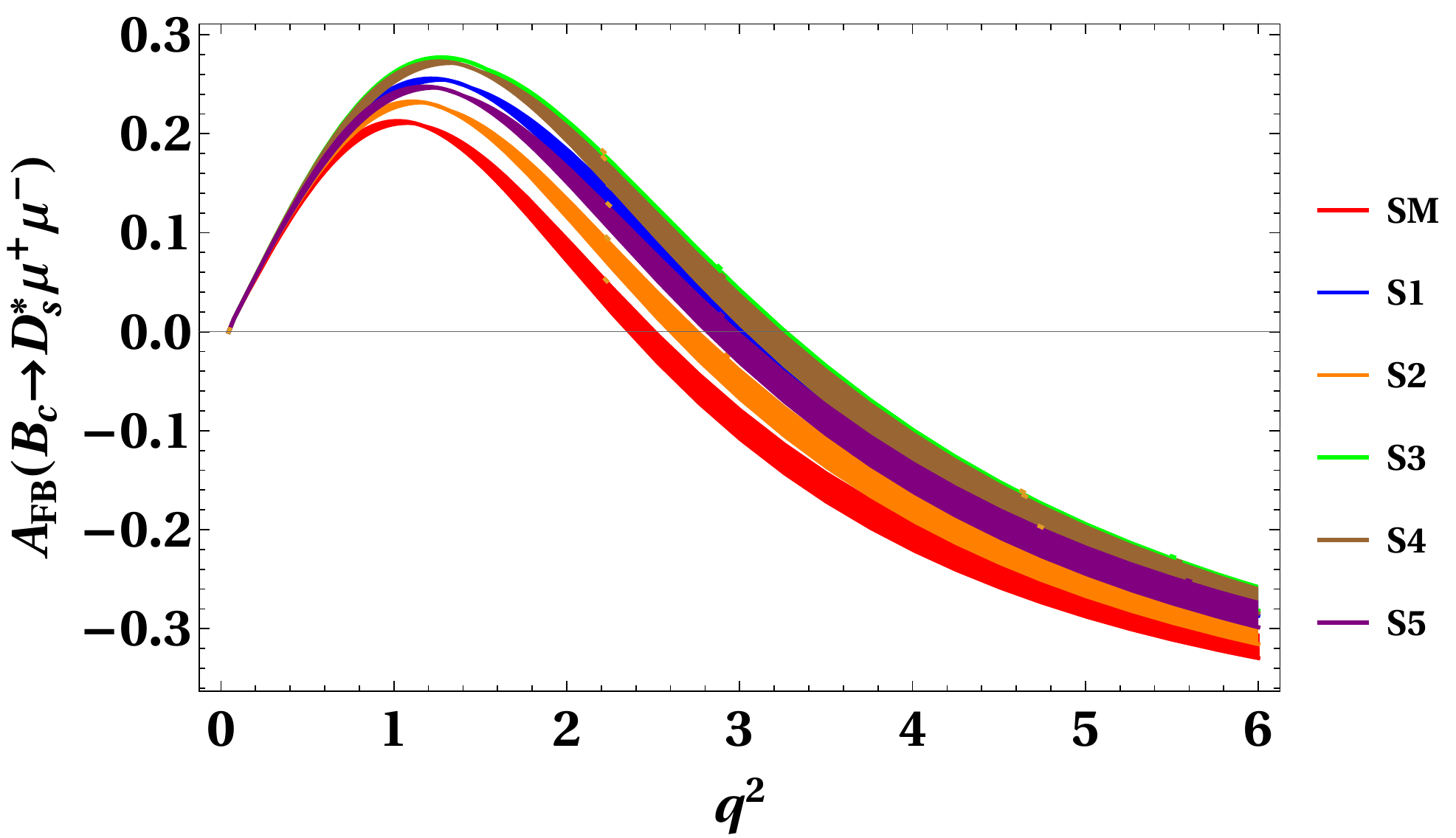}\ \ \
& \ \ \ \includegraphics[scale=0.4]{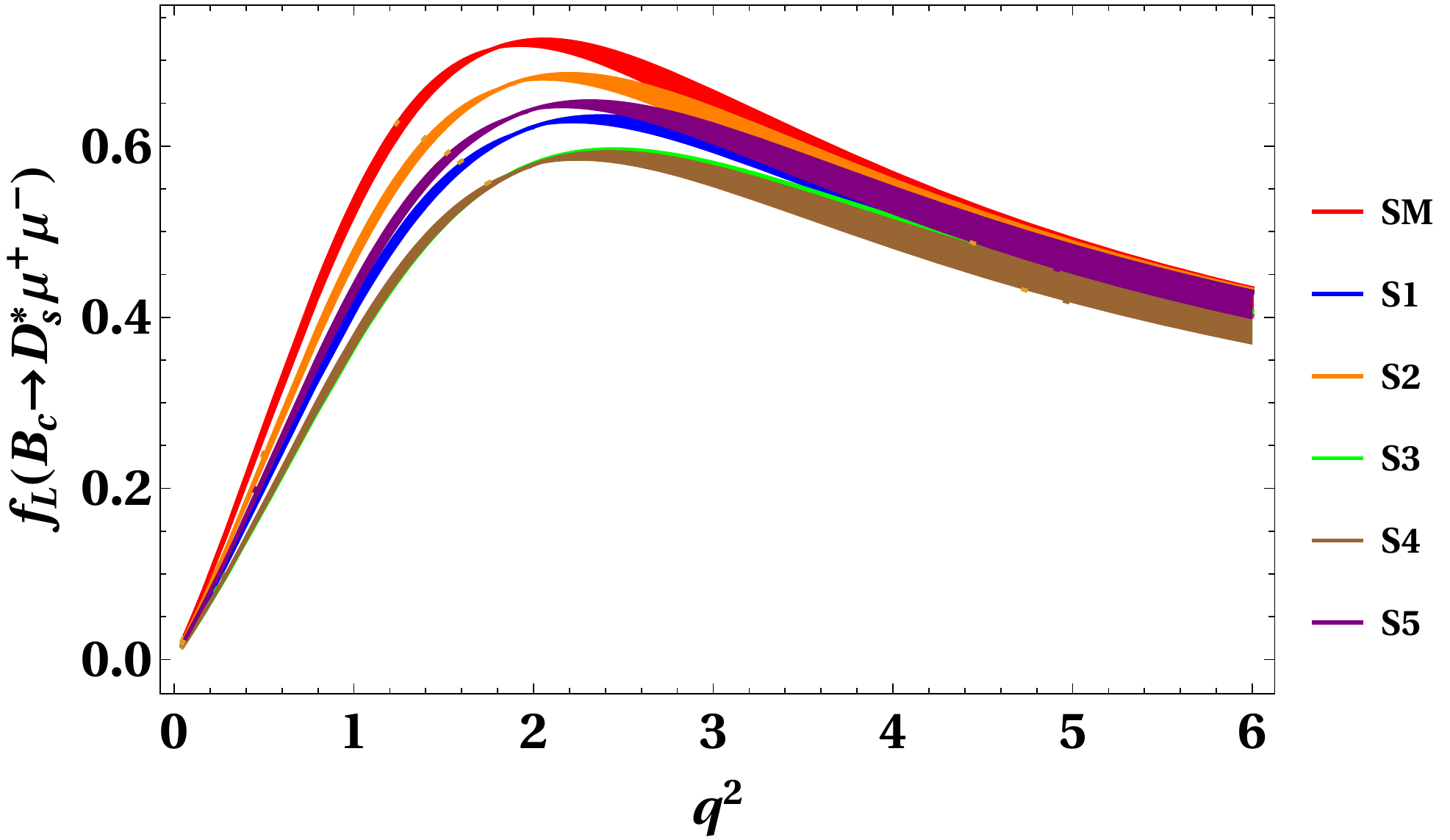}\\\
\end{tabular}
\caption{(a) Lepton forward backward asymmetry $A_{\text{FB}}$, and (b) longitudinal polarization fraction of $D^{\ast}_{s}$ meson $f_{L}$, for the $B_{c}\to D^{\ast}_{s}\mu^{+}\mu^{-}$ decay, in the SM and the NP scenarios.}
\label{FBA}
\end{figure*}
\item Another physical observable useful to investigate the structure of NP is the longitudinal helicity fraction of the final state meson $(f_L)$. In Fig. \ref{FBA}(b), we have shown the longitudinal helicity fraction $f_{L}$ for $B_{c}\to D^{\ast}_{s}\mu^{+}\mu^{-}$ decay as a function of $q^{2}$. We can recognize from Fig. \ref{FBA}(b), that the given NP scenarios in the longitudinal helicity fraction of $D^{\ast}_{s}$ can be distinguished  quite easily in the region $q^{2}=(1-2.5)$ $\text{GeV}^{2}$, and all the NP predictions point out lesser values of $f_L$, compare to the SM. However, for $q^{2}>2.5$ $\text{GeV}^{2}$ the given NP scenarios overlap with each other.
\end{itemize}

\subsection{Analysis of Angular Coefficients in $B_{c}\to D^{\ast}_{s}(\to D_{s}\gamma,(D_{s}\pi))\mu^{+}\mu^{-}$ Decays}
\begin{figure*}[b!]
\begin{tabular}{cc}
\hspace{0.6cm}($\mathbf{a}$)&\hspace{1.2cm}($\mathbf{b}$)\\
\includegraphics[scale=0.40]{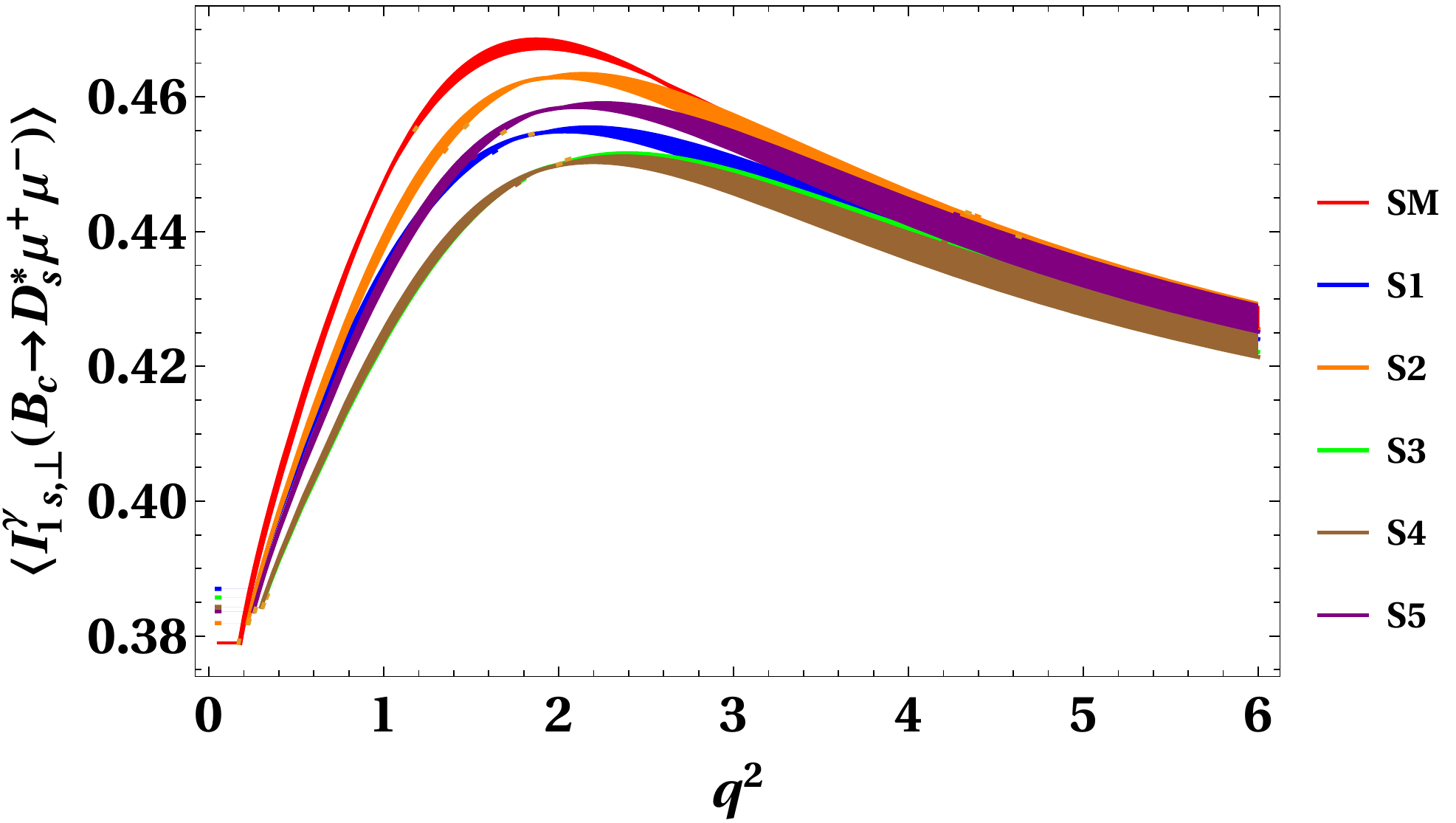}\ \ \
& \ \ \ \includegraphics[scale=0.40]{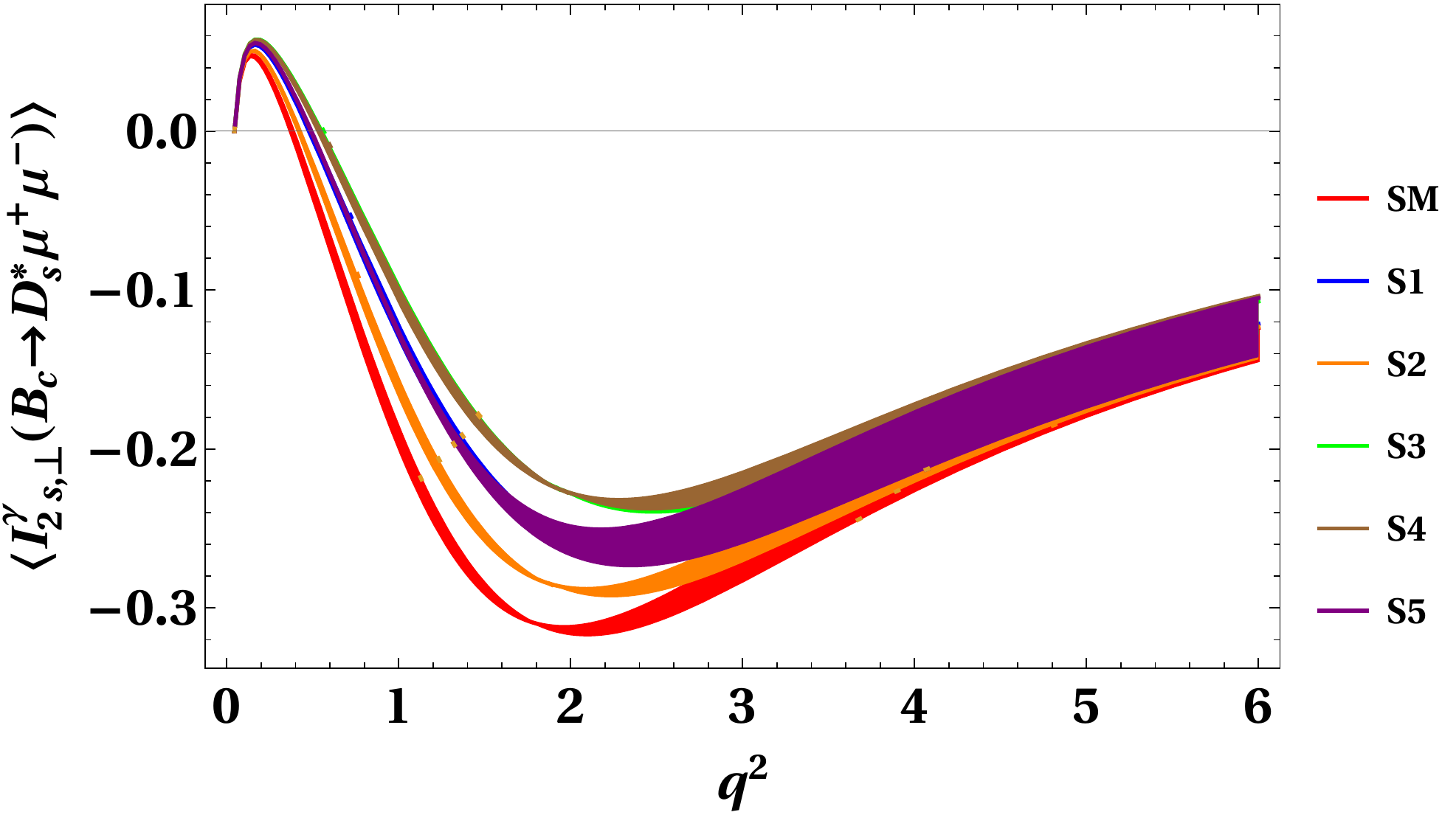}\\\
\hspace{0.6cm}($\mathbf{c}$)&\hspace{1.2cm}($\mathbf{d}$)\\
\includegraphics[scale=0.40]{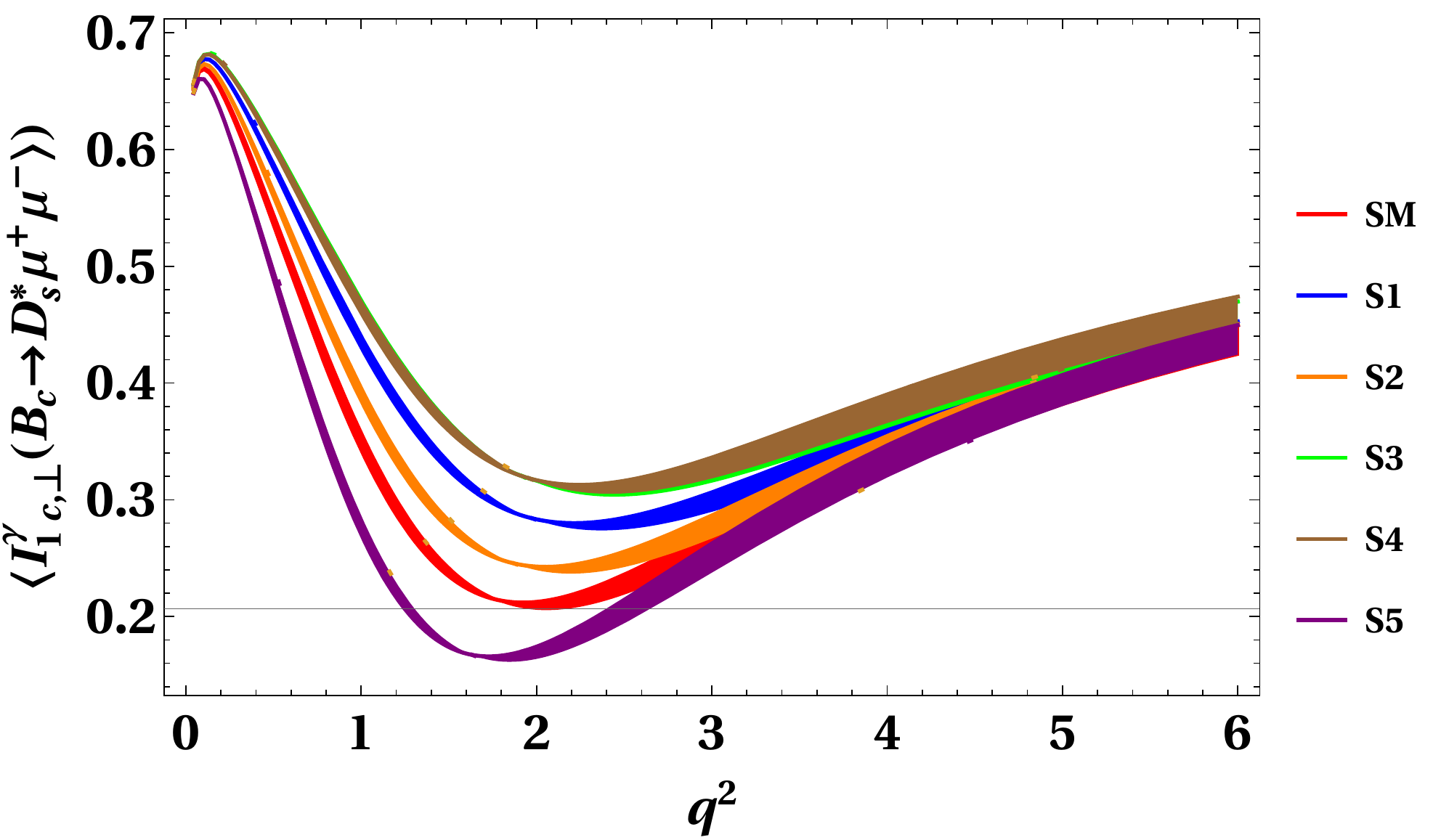}\ \ \
& \ \ \ \includegraphics[scale=0.40]{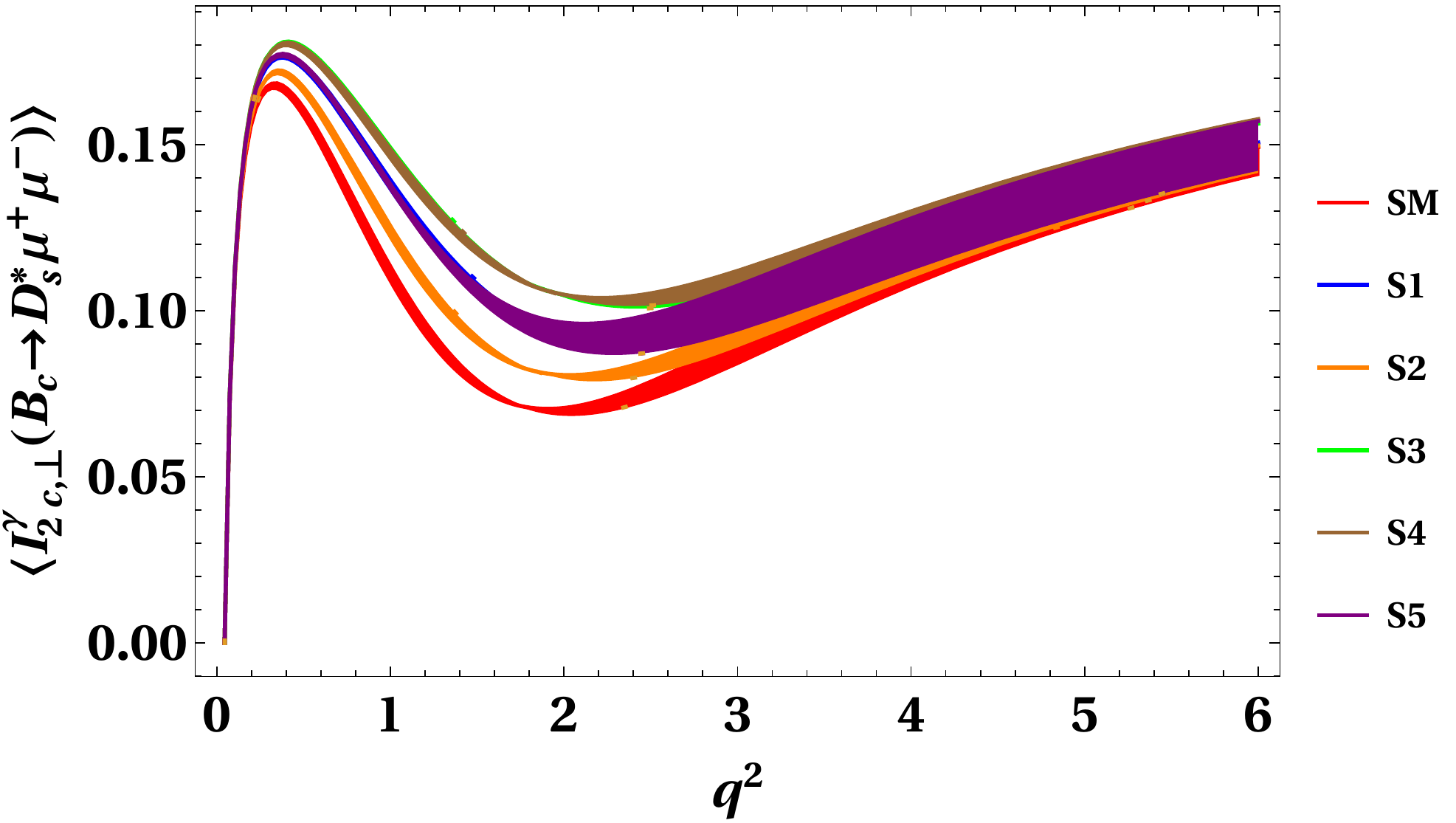}\\\
\hspace{0.6cm}($\mathbf{e}$)&\hspace{1.2cm}($\mathbf{f}$)\\
\includegraphics[scale=0.40]{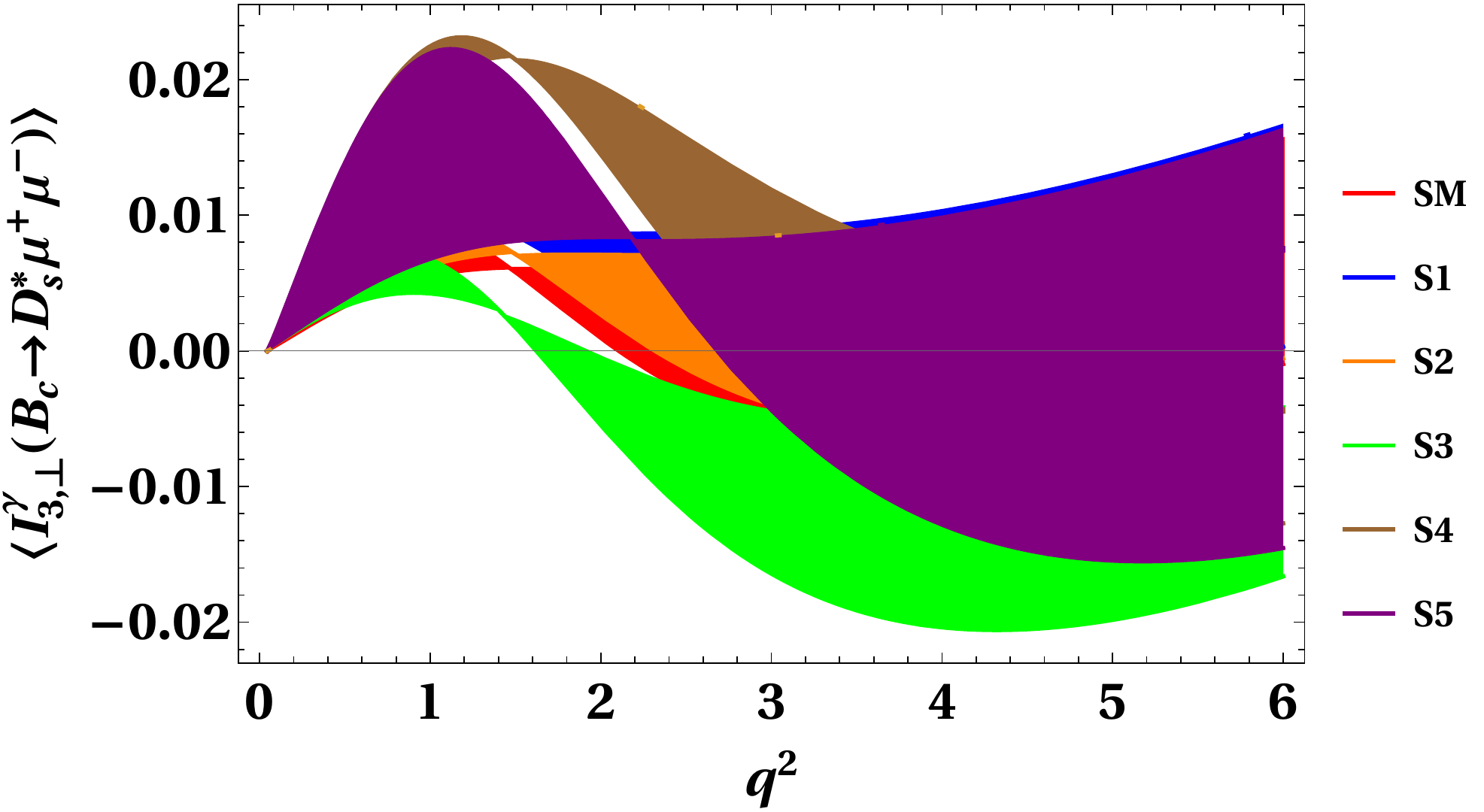}\ \ \
& \ \ \ \includegraphics[scale=0.40]{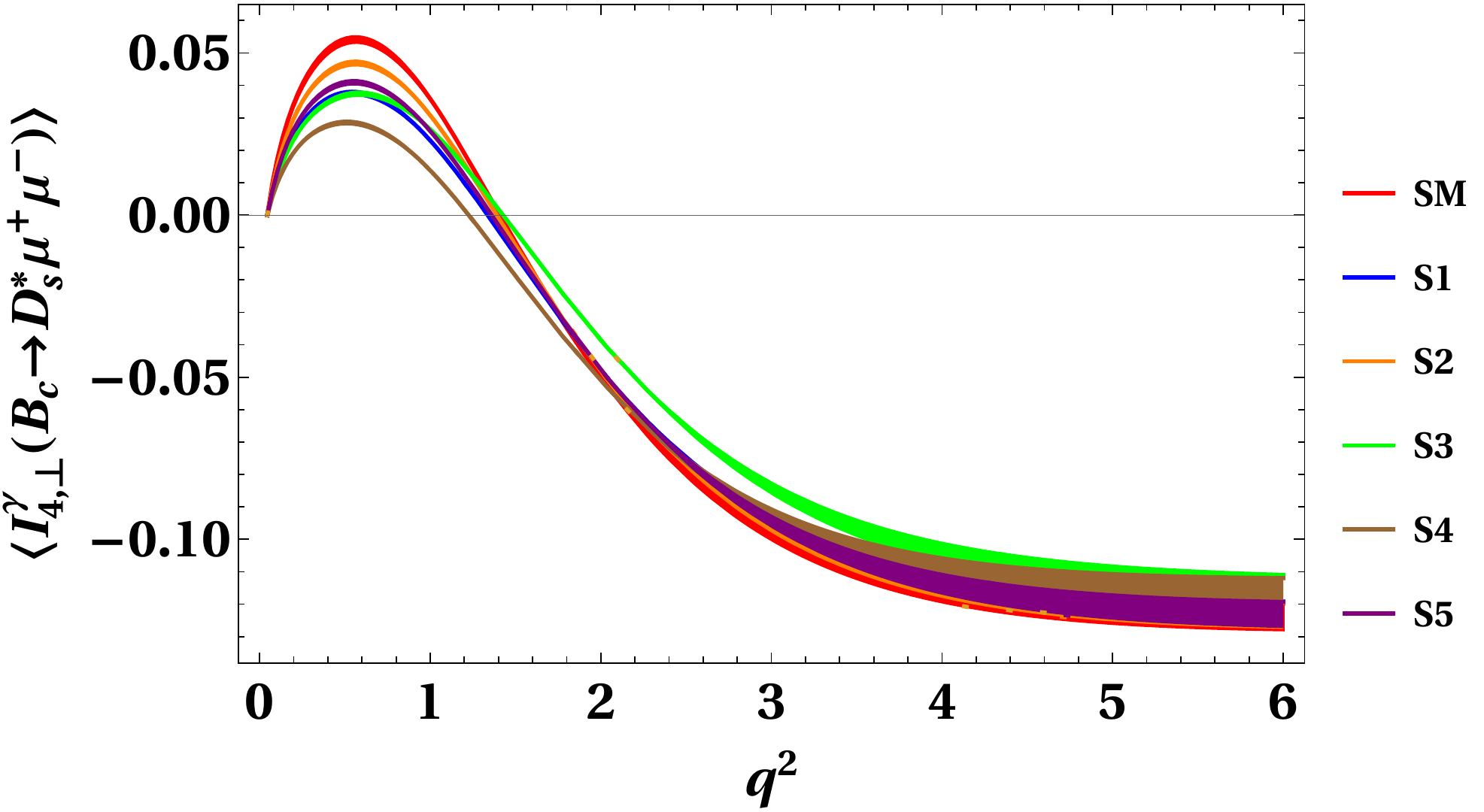}
\end{tabular}
\caption{Angular observables $\langle I^{\gamma}_{1s,\perp}\rangle, \langle I^{\gamma}_{2s,\perp}\rangle, \langle I^{\gamma}_{1c,\perp}\rangle, \langle I^{\gamma}_{2c,\perp}\rangle, \langle I^{\gamma}_{3,\perp}\rangle$, and $\langle I^{\gamma}_{4,\perp}\rangle$ for the decay $B_{c}\to D^{\ast}_{s}(\to D^{\ast}_{s}\gamma)\mu^{+}\mu^{-}$, in the SM and the NP scenarios.}
\label{figIS1}
\end{figure*}
\begin{figure*}[b!]
\begin{tabular}{cc}
\hspace{0.6cm}($\mathbf{a}$)&\hspace{1.2cm}($\mathbf{b}$)\\
\includegraphics[scale=0.40]{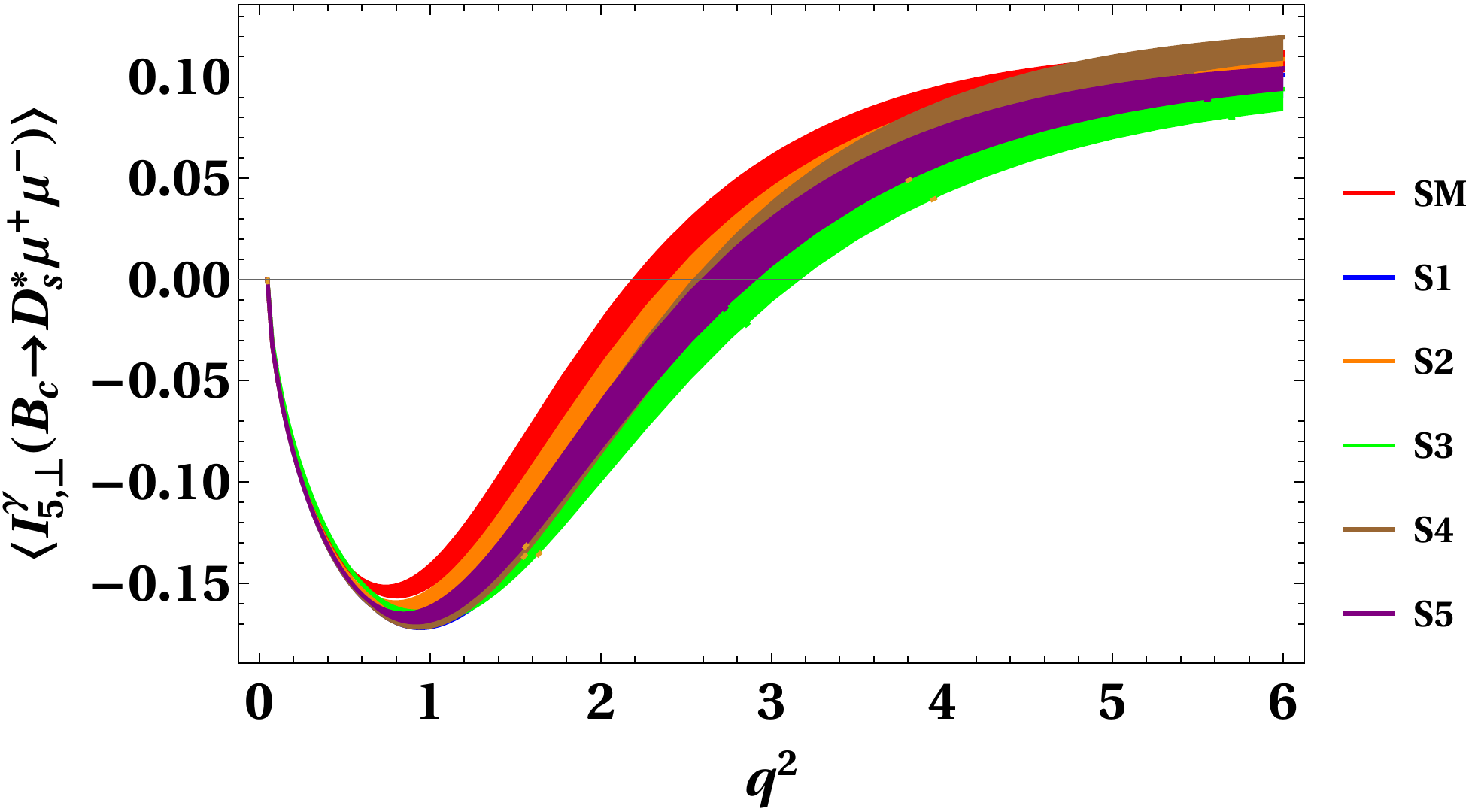}\ \ \
& \ \ \ \includegraphics[scale=0.40]{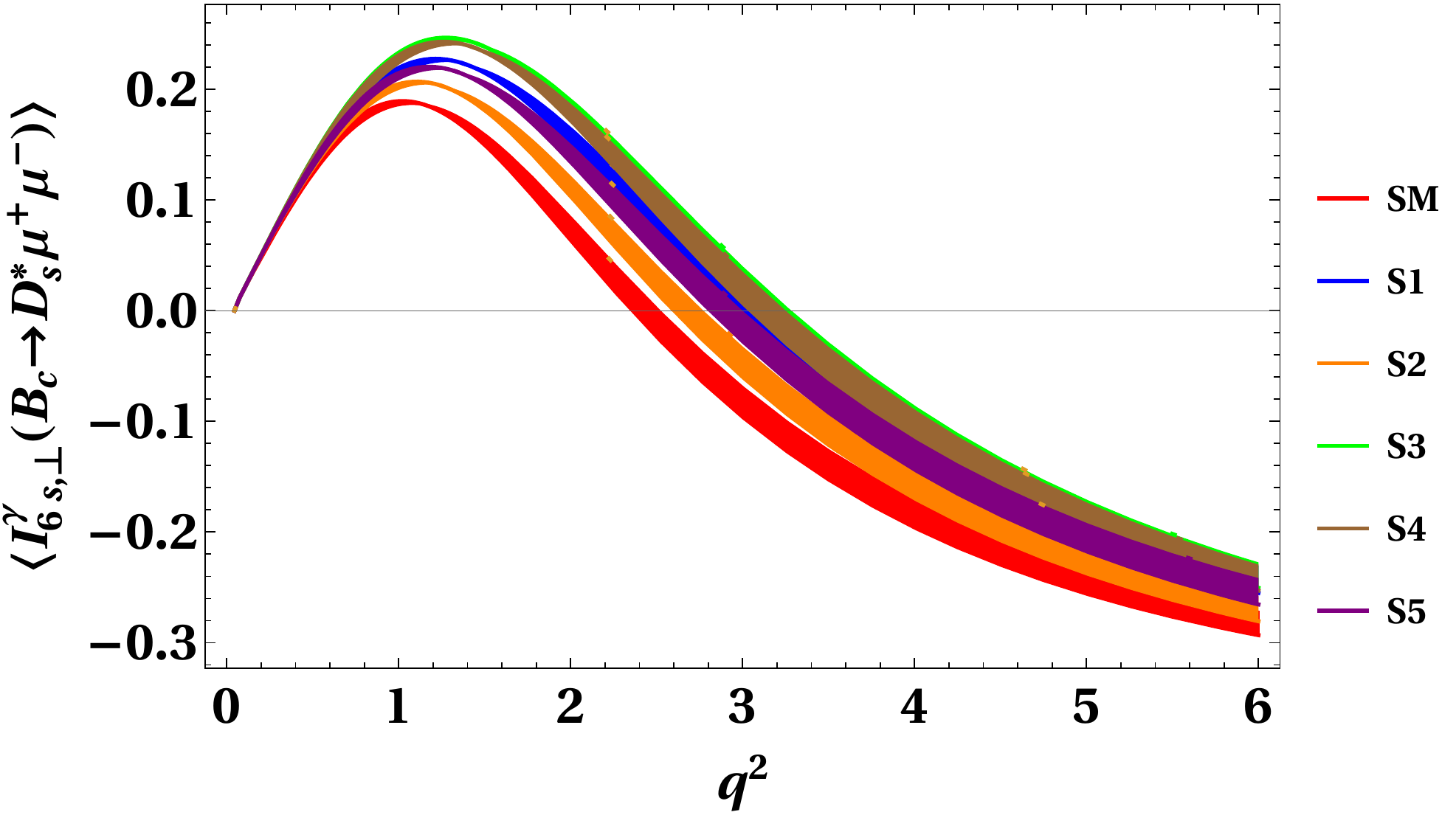}\\\
\hspace{0.6cm}($\mathbf{c}$)&\hspace{1.2cm}\\
\includegraphics[scale=0.40]{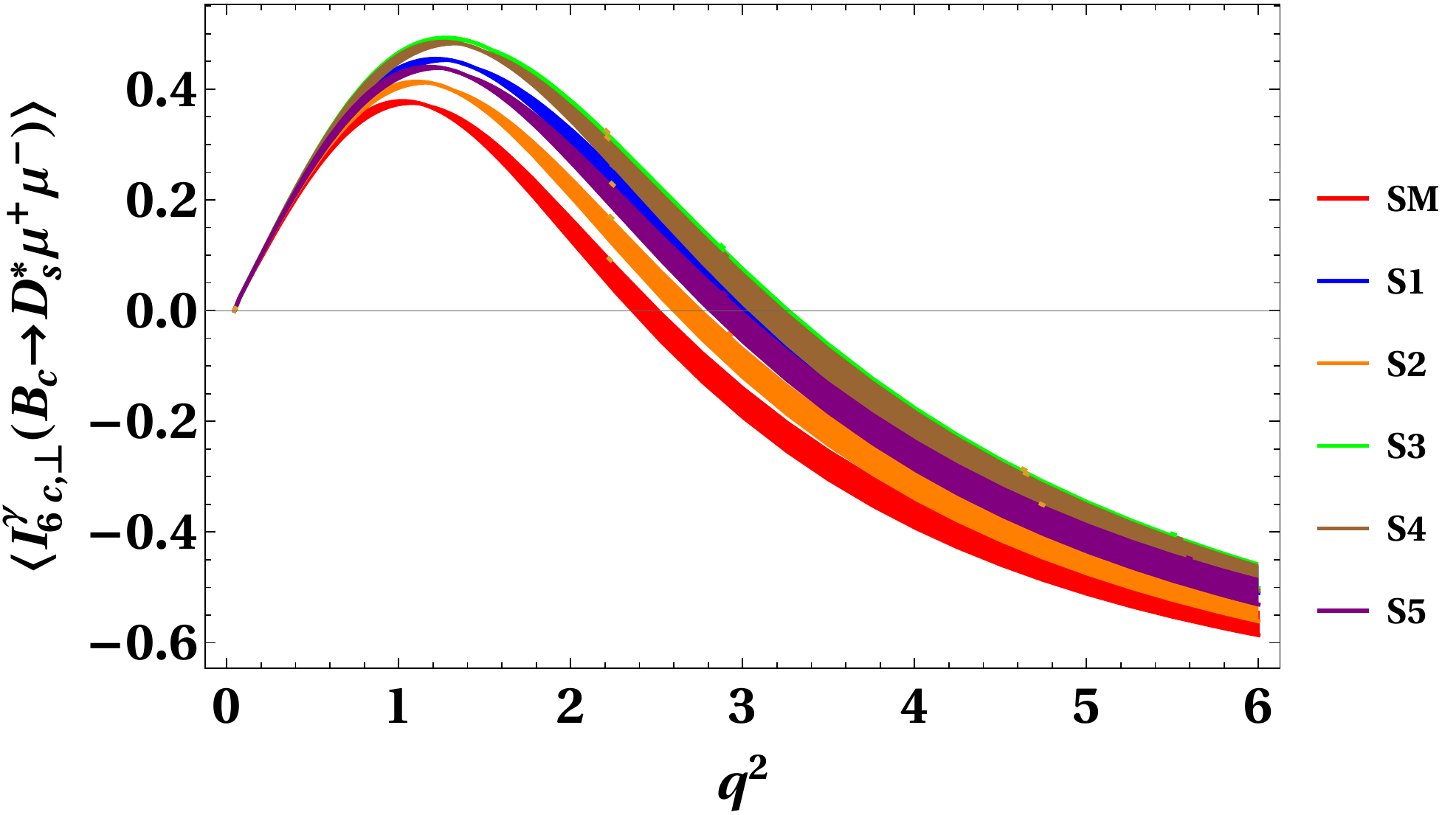}\ \ \
\end{tabular}
\caption{Angular observables $\langle I^{\gamma}_{5,\perp}\rangle, \langle I^{\gamma}_{6s,\perp}\rangle$, and $\langle I^{\gamma}_{6c,\perp}\rangle$ for the decay $B_{c}\to D^{\ast}_{s}(\to D^{\ast}_{s}\gamma)\mu^{+}\mu^{-}$, in the SM and the NP scenarios.}
\label{figIS2 }
\end{figure*}

In this section, we analyze the effects of NP via individual angular coefficients  such as $\langle I^{\gamma}_{1s,\perp}\rangle$$(\langle I_{1s,||}\rangle)$, $\langle I^{\gamma}_{1c,\perp}\rangle $$(\langle I_{1c,||}\rangle)$, $\langle I^{\gamma}_{2s,\perp}\rangle$$(\langle I_{2s,||}\rangle)$,
$\langle I^{\gamma}_{2c,\perp}\rangle$$(\langle I_{2c,||}\rangle)$,
$\langle I^{\gamma}_{3,\perp}\rangle$$(\langle I_{3,||}\rangle)$,
$\langle I^{\gamma}_{4,\perp}\rangle$$(\langle I_{4,||}\rangle)$,
$\langle I^{\gamma}_{5,\perp}\rangle$$(\langle I_{5,||}\rangle)$,
$\langle I^{\gamma}_{6s,\perp}\rangle$
$(\langle I_{6s,||}\rangle)$, and $\langle I^{\gamma}_{6c,\perp}\rangle$$(\langle I_{6c,||}\rangle)$, for $B_{c}\to D_{s}^{\ast}(\to D_{s}\gamma, (D_{s}\pi))\mu^{+}\mu^{-}$ decays. Using the input parameters given in Table \ref{Input}, we estimate the above mentioned angular observables for both $B_{c}\to D^{\ast}_{s}(\to D_{s}\gamma)\mu^{+}\mu^{-}$ and $B_{c}\to D^{\ast}_{s}(\to D_{s}\pi)\mu^{+}\mu^{-}$ decays in the SM as well as in 1D and 2D NP scenarios presented in Table \ref{tab:bestfitWC}. Numerical values of the SM and NP predictions of the averaged angular coefficients, with errors due to the form factors, in different $q^2$ bins, are listed in Tables \ref{tab:Angobs1}-\ref{tab:Angobs7}, of appendix \ref{append1}. We also show the  results of the angular observables as a function of $q^{2}$ in Figs. \ref{figIS1}-\ref{figIS4}. We now discuss these results of angular observables.

\begin{itemize}
\item Fig. \ref{figIS1}, shows the angular coefficients $\langle I^{\gamma}_{1s,\perp}\rangle, \langle I^{\gamma}_{2s,\perp}\rangle, \langle I^{\gamma}_{1c,\perp}\rangle, \langle I^{\gamma}_{2c,\perp}\rangle, \langle I^{\gamma}_{3,\perp}\rangle$, and $\langle I^{\gamma}_{4,\perp}\rangle$ as a function of $q^{2}$ both in the SM and in 1D and 2D NP scenarios. From Figs. \ref{figIS1}(b), \ref{figIS1}(c), \ref{figIS1}(d), and \ref{figIS1}(e), one can see that the angular observables $\langle I^{\gamma}_{2s,\perp}\rangle, \langle I^{\gamma}_{1c,\perp}\rangle, \langle I^{\gamma}_{2c,\perp}\rangle$, and $\langle I^{\gamma}_{3,\perp}\rangle$ deviate significantly from the SM predictions at $q^{2}=[2,3]$ $\text{GeV}^{2}$. Furthermore these observables also discriminate the 1D and 2D NP scenarios at $q^{2}=[2,3]$ $\text{GeV}^{2}$. Similarly, the angular observables $\langle I^{\gamma}_{1s,\perp}\rangle$ and $\langle I^{\gamma}_{4,\perp}\rangle$ presented in Figs. \ref{figIS1}(a), and \ref{figIS1}(e) show a clear departure from the SM predictions and the 1D and 2D NP scenarios are  discriminated around $q^{2}=[1,2]$ $\text{GeV}^{2}$.
\item Fig. \ref{figIS2 }, depicts the angular coefficients $\langle I^{\gamma}_{5,\perp}\rangle, \langle I^{\gamma}_{6s,\perp}\rangle$, and $\langle I^{\gamma}_{6c,\perp}\rangle$ as a function of $q^{2}$ both in the SM and in 1D and 2D NP scenarios. For $\langle I^{\gamma}_{5,\perp}\rangle$ (cf. Fig. \ref{figIS2 }(a)), the value of zero crossing in the SM is $q^{2}\approx 2.5$ $\text{GeV}^{2}$. The deviation of the zero crossing of $\langle I^{\gamma}_{5,\perp}\rangle$ arises in the case of scenarios S2, S3, S4, and S5, but scenario S1 is not distinguishable. For the angular coefficients $\langle I^{\gamma}_{6s,\perp}\rangle$ and $\langle I^{\gamma}_{6c,\perp}\rangle$, there is a shift in zero crossing compared to that of SM, with distinct zero crossing points for scenarios S2, S4, and S5. However, the scenarios S1 and S3 are not much distinguishable.

\begin{figure*}[t!]
\begin{tabular}{cc}
\hspace{0.6cm}($\mathbf{a}$)&\hspace{1.2cm}($\mathbf{b}$)\\
\includegraphics[scale=0.385]{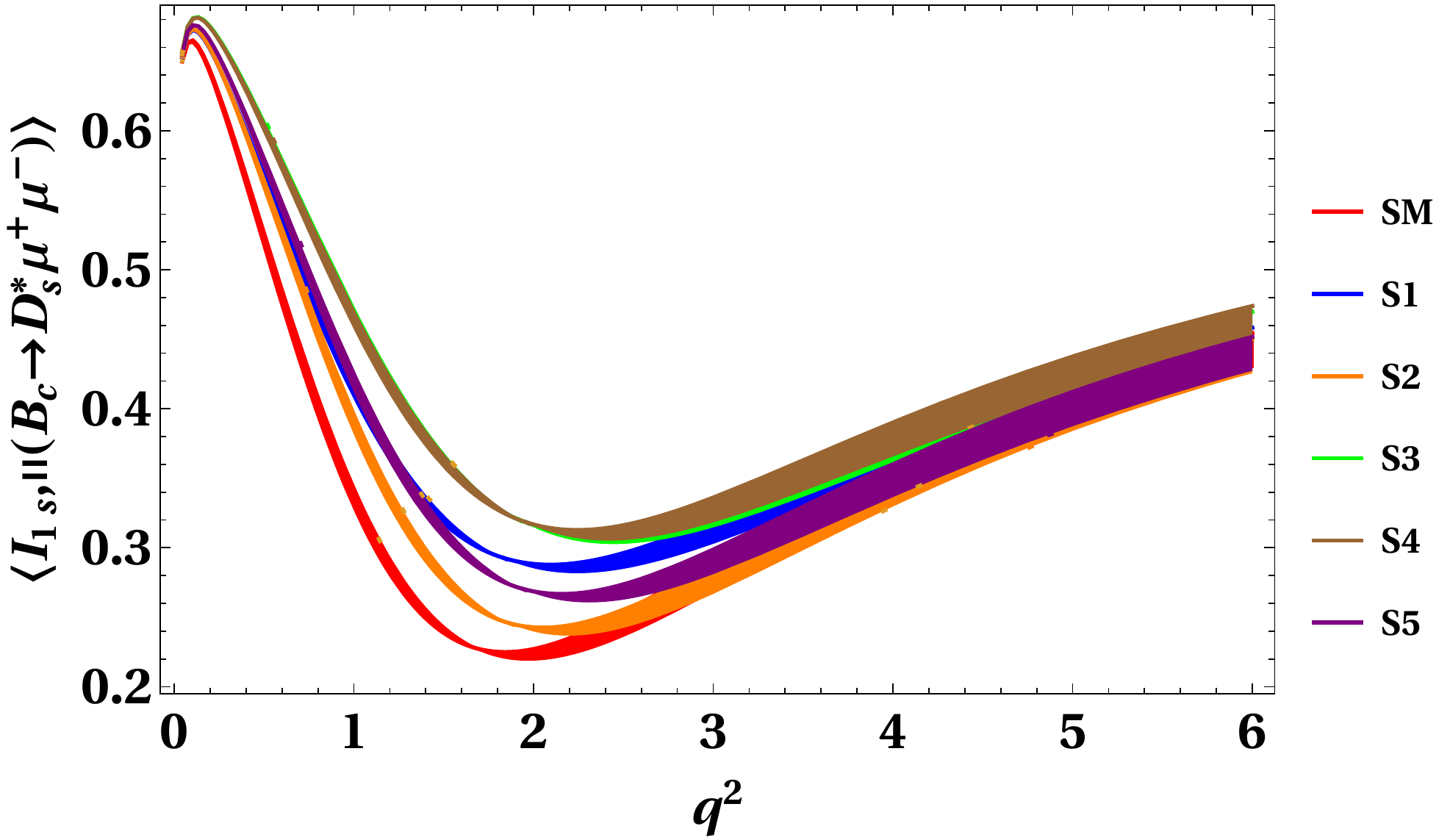}\ \ \
& \ \ \ \includegraphics[scale=0.385]{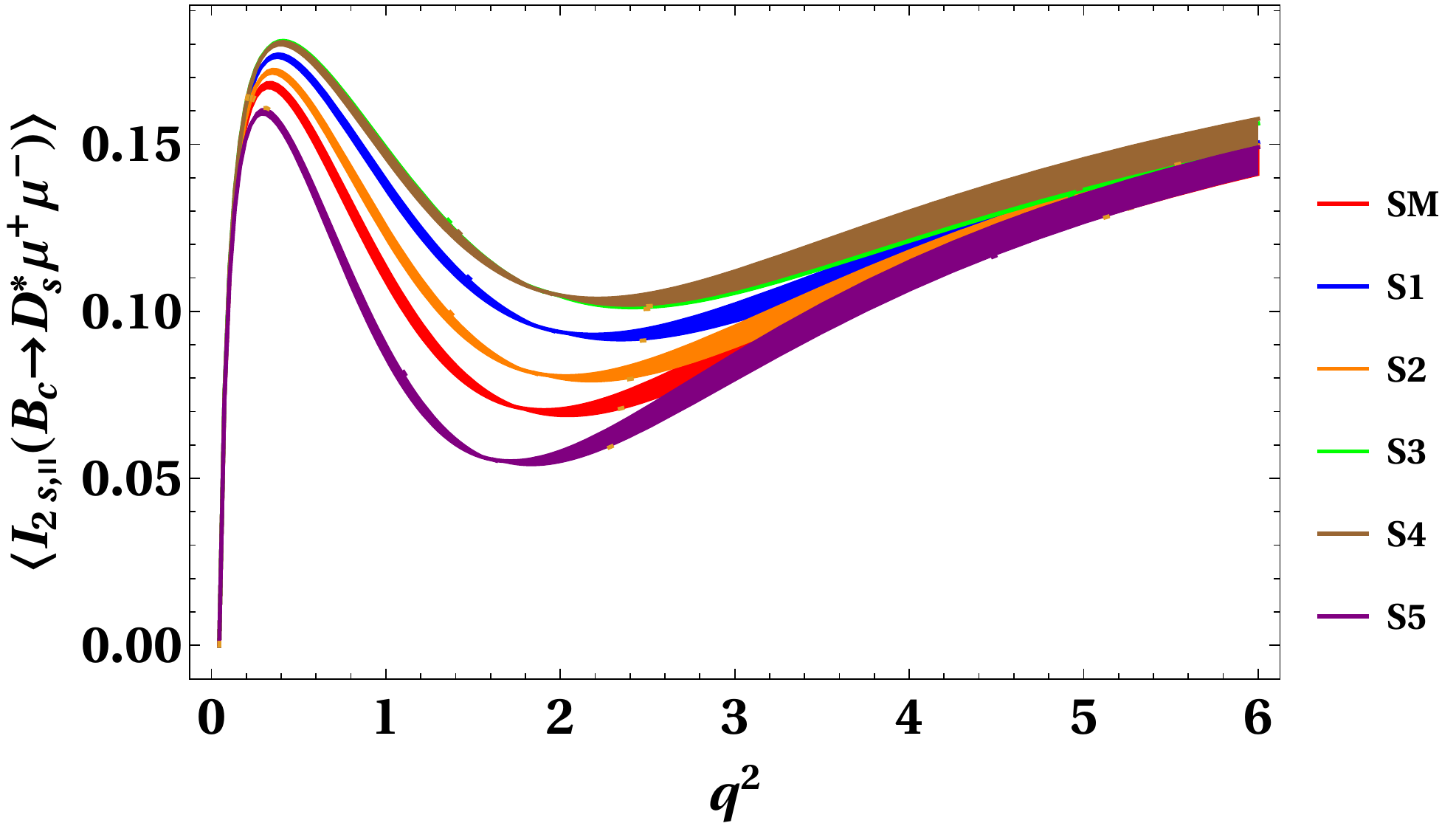}\\\
\hspace{0.6cm}($\mathbf{c}$)&\hspace{1.2cm}($\mathbf{d}$)\\
\includegraphics[scale=0.385]{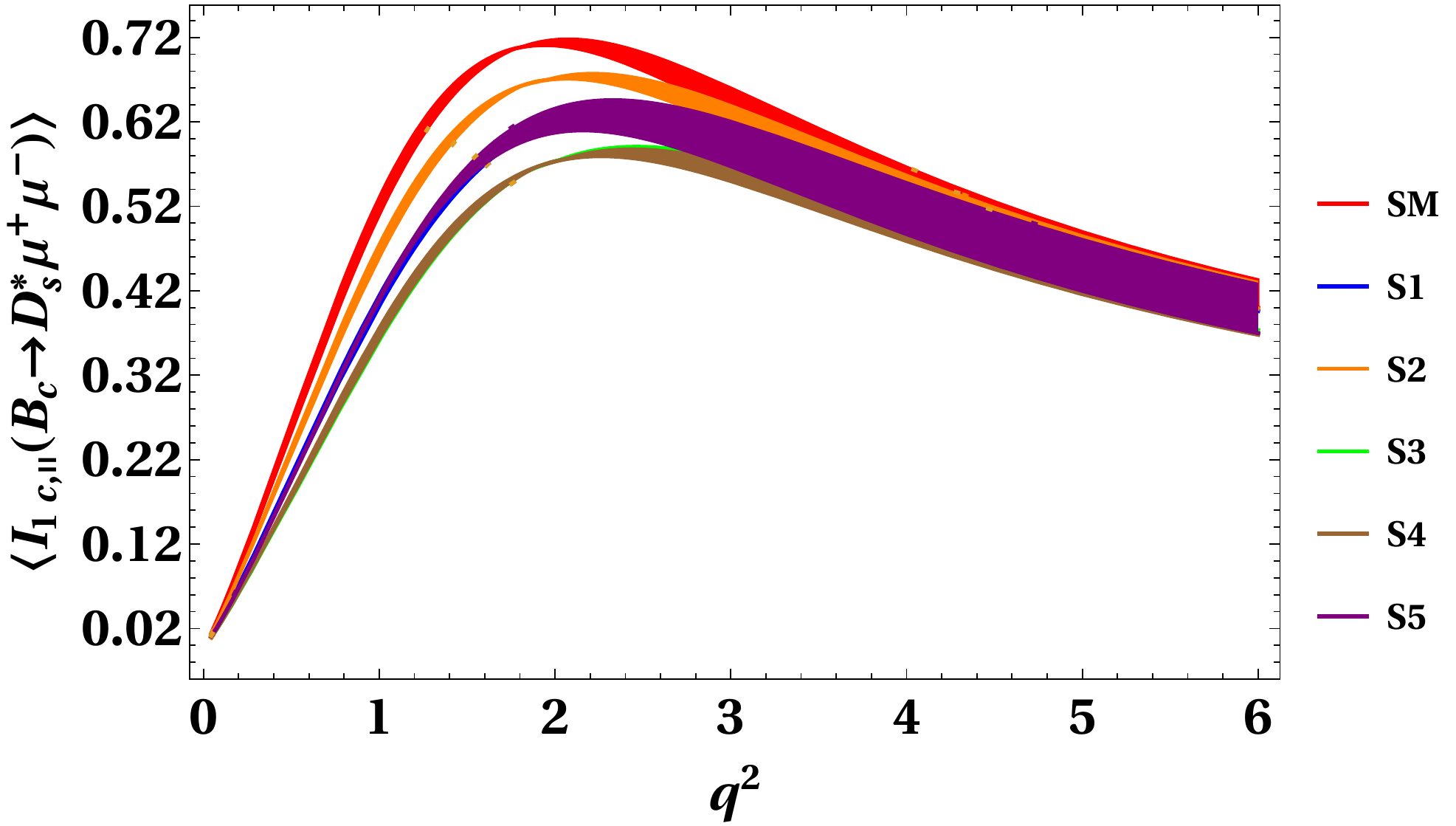}\ \ \
& \ \ \ \includegraphics[scale=0.385]{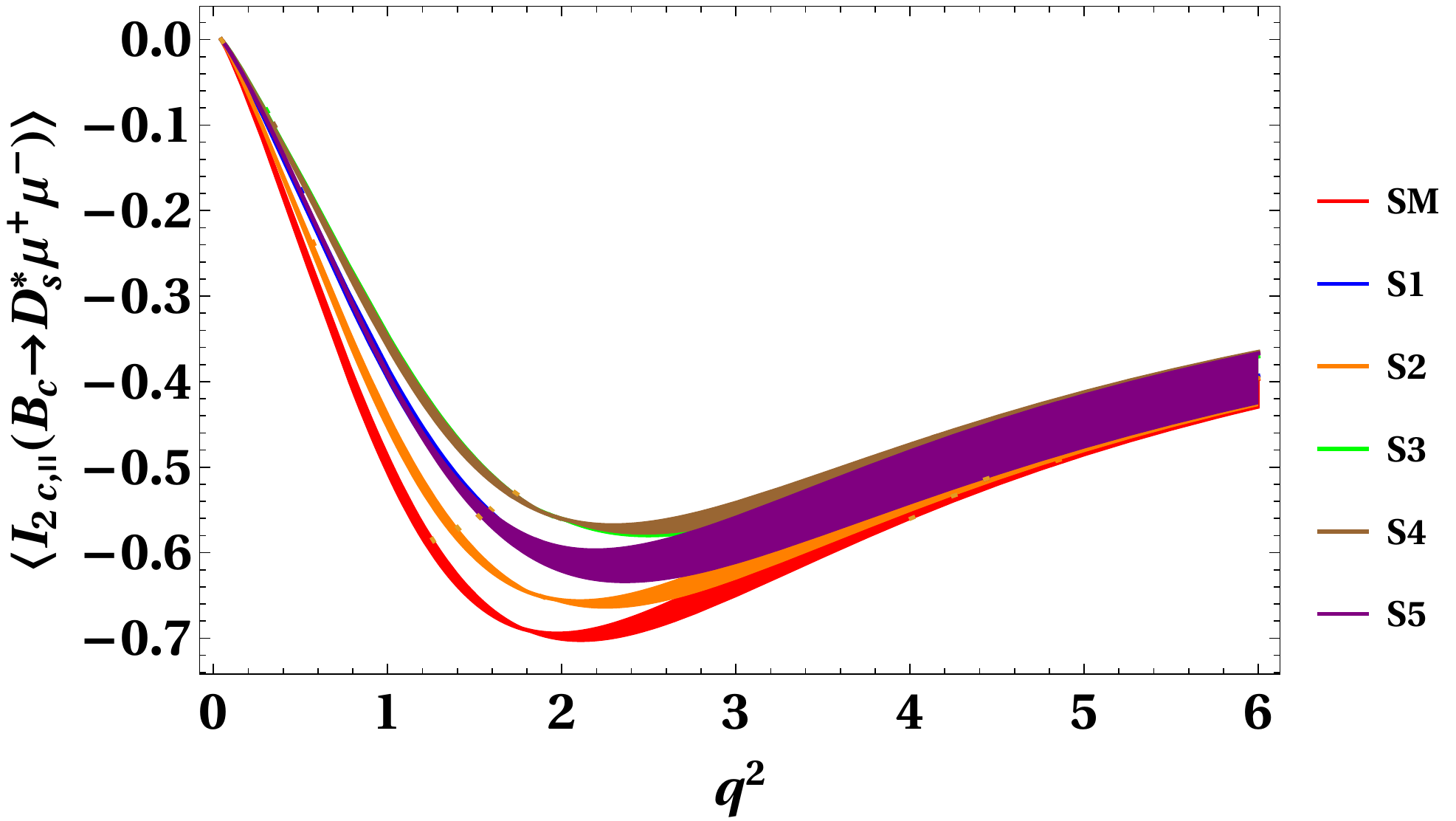}\\\
\end{tabular}
\caption{Angular observables $\langle I_{1s,||}\rangle, \langle I_{2s,||}\rangle, \langle I_{1c,||}\rangle$, and $\langle I_{2c,||}\rangle$ for the decay $B_{c}\to D^{\ast}_{s}(\to D^{\ast}_{s}\pi)\mu^{+}\mu^{-}$, in the SM and the NP scenarios.}
\label{figIS3}
\end{figure*}

\begin{figure*}[b!]
\begin{tabular}{cc}
\hspace{0.6cm}($\mathbf{a}$)&\hspace{1.2cm}($\mathbf{b}$)\\
\includegraphics[scale=0.385]{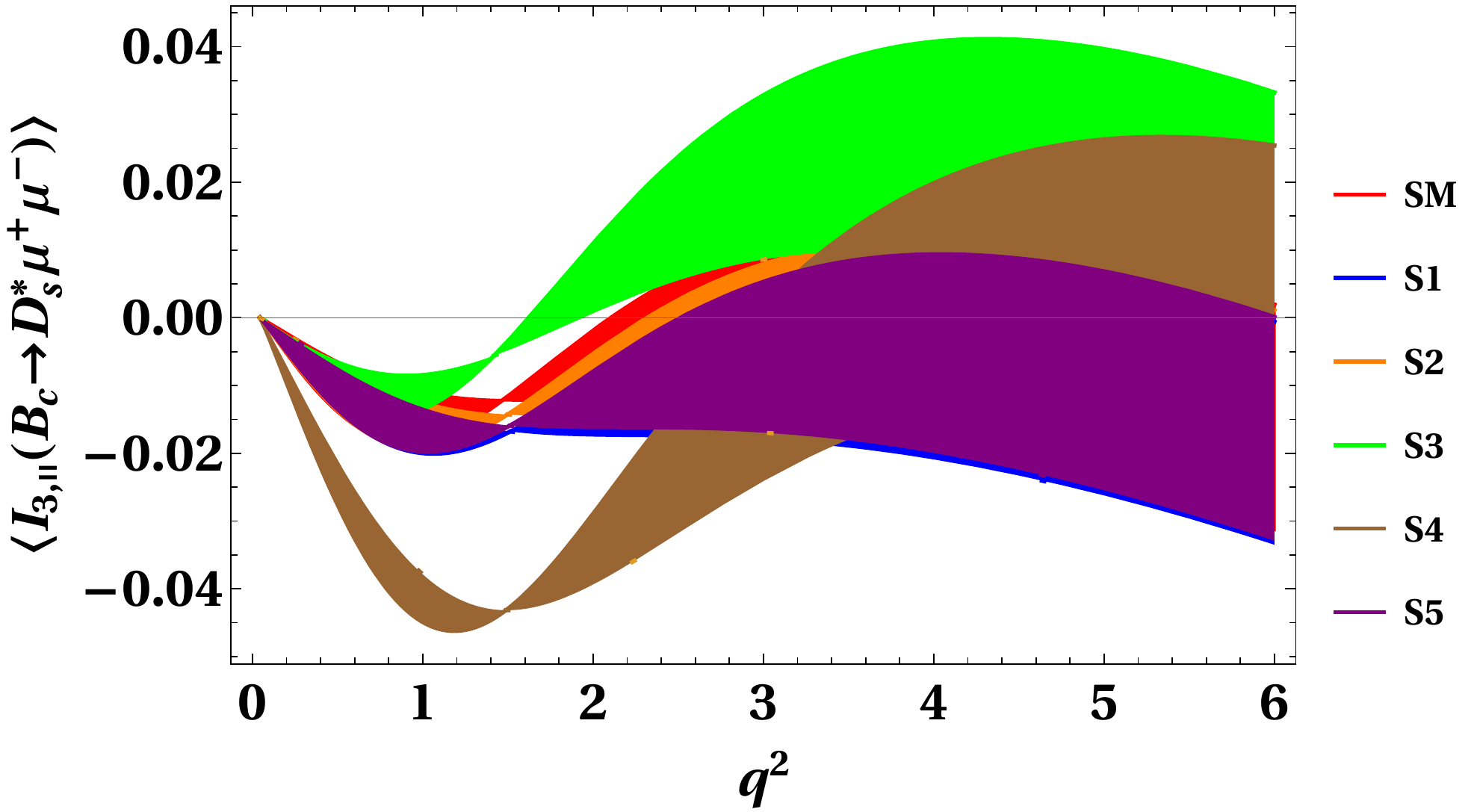}\ \ \
& \ \ \ \includegraphics[scale=0.385]{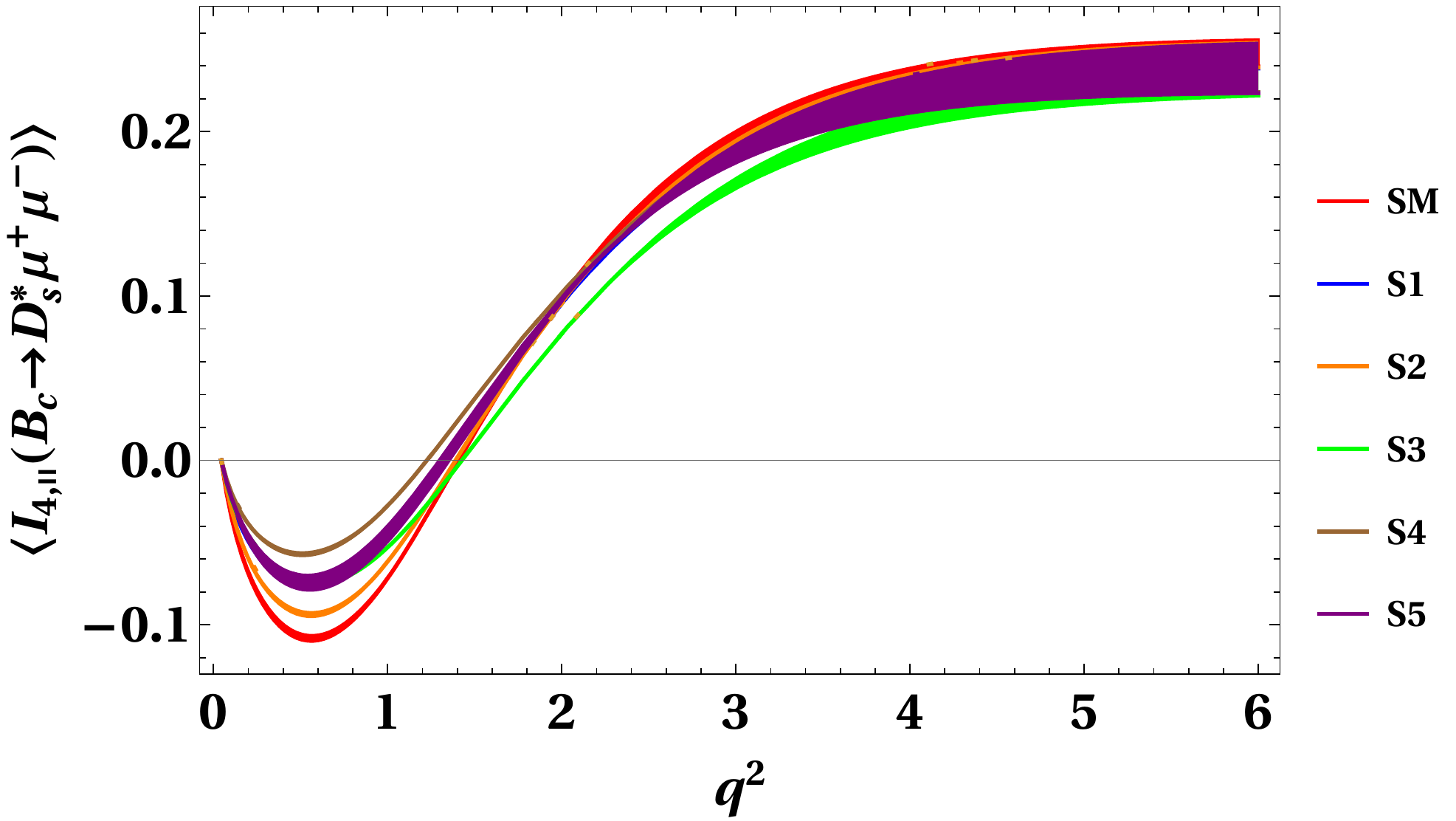}\\\
\hspace{0.6cm}($\mathbf{c}$)&\hspace{1.2cm}($\mathbf{d}$)\\
\includegraphics[scale=0.385]{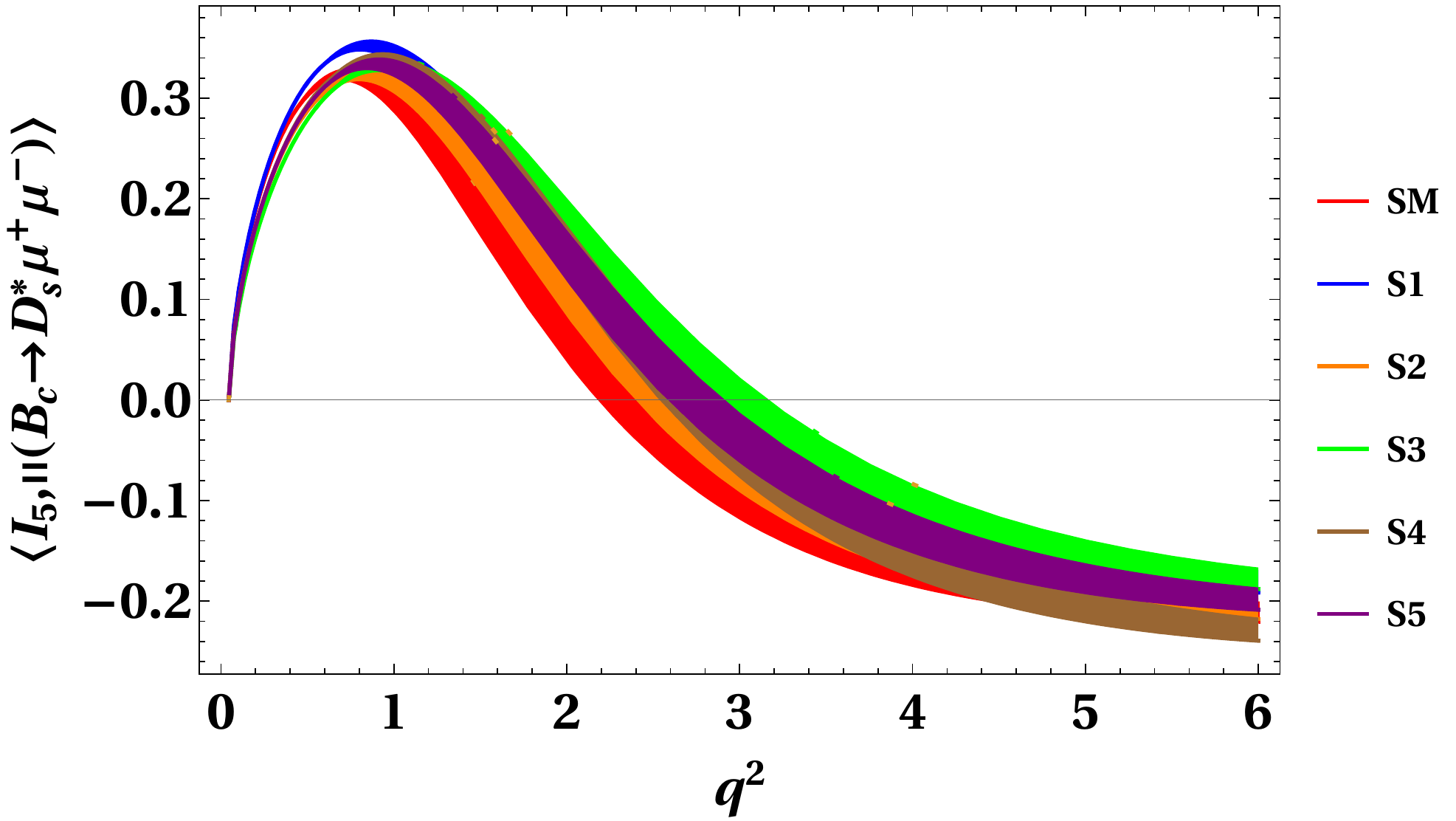}\ \ \
& \ \ \ \includegraphics[scale=0.385]{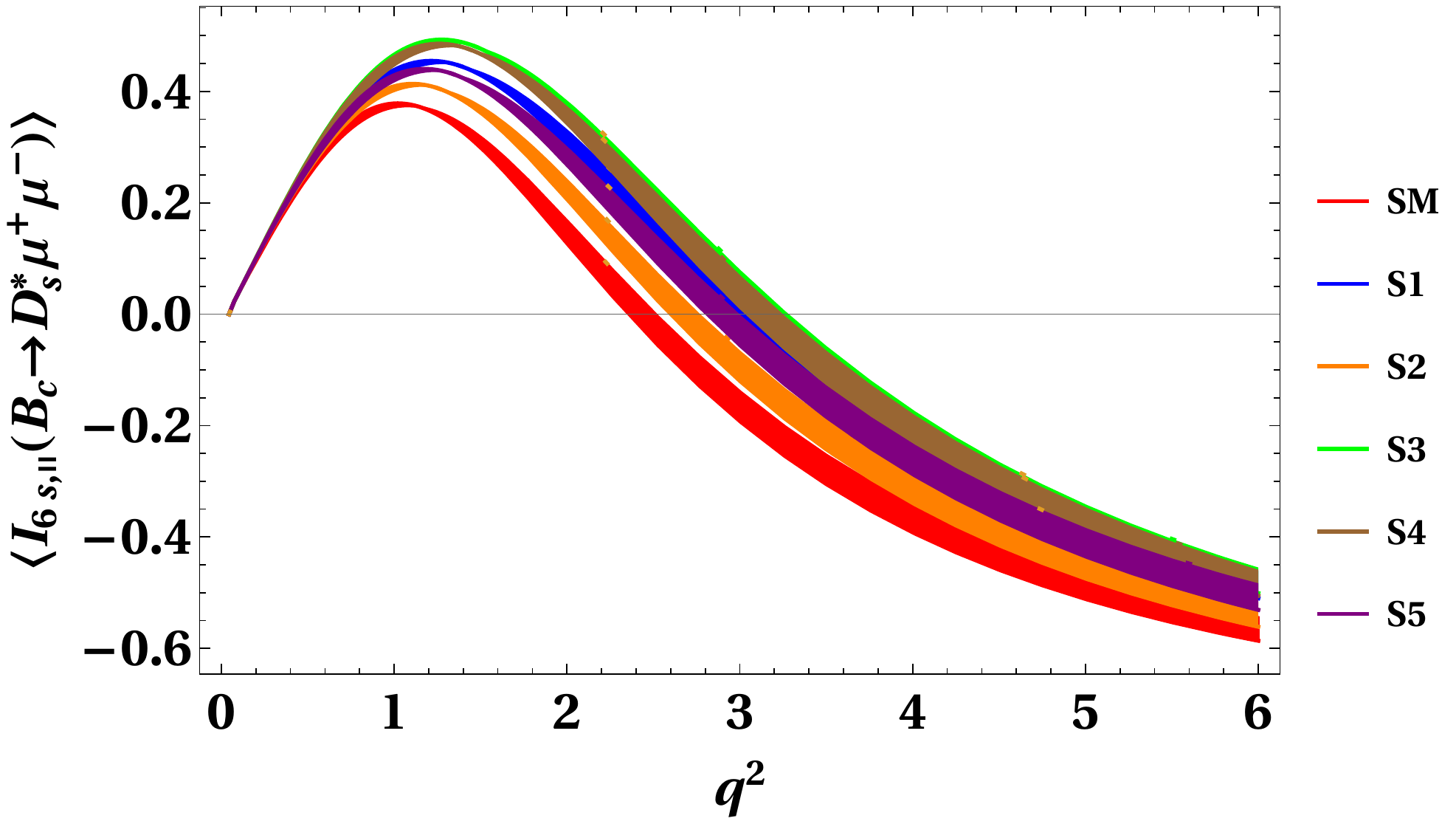}
\end{tabular}
\caption{Angular observables $\langle I_{3,||}\rangle, \langle I_{4,||}\rangle, \langle I_{5,||}\rangle$, and $\langle I_{6s,||}\rangle$ for the decay $B_{c}\to D^{\ast}_{s}(\to D^{\ast}_{s}\pi)\mu^{+}\mu^{-}$, in the SM and the NP scenarios.}
\label{figIS4}
\end{figure*}

\item Fig. \ref{figIS3}, shows the angular coefficients $\langle I_{1s,||}\rangle, \langle I_{2s,||}\rangle, \langle I_{1c,||}\rangle$, and $\langle I_{2c,||}\rangle$ for the decay $B_{c}\to D^{\ast}_{s}(\to D_{s}\pi)\mu^{+}\mu^{-}$ in the SM and in NP scenarios S1, S2, S3, S4 and S5. The effects of NP are distinct compared to that of SM. For angular coefficients $\langle I_{1s,||}\rangle$, and $\langle I_{2s,||}\rangle$ (cf. Fig. \ref{figIS3}(a,b)), the scenarios S1, S2, S4 and S5 are quite distinct in the kinematical region $q^{2}=[1,2]$ $\text{GeV}^{2}$. Similarly for the angular coefficients $\langle I_{1c,||}\rangle$ and $\langle I_{2c,||}\rangle$ (cf. Fig. \ref{figIS3}(c,d)), the NP scenarios S2, S3 and S5 are quite distinct in the region $q^{2}=[1,2]$ $\text{GeV}^{2}$. However in the region $q^{2}=[3,6]$ $\text{GeV}^{2}$ the angular coefficients $\langle I_{1c,||}\rangle$, and $\langle I_{2c,||}\rangle$ for  most of the NP scenarios overlap.
\item In Fig. \ref{figIS4}, we have plotted the angular coefficients $\langle I_{3,||}\rangle, \langle I_{4,||}\rangle, \langle I_{5,||}\rangle$, and $\langle I_{6s,||}\rangle$ for the decay $B_{c}\to D^{\ast}_{s}(\to D_{s}\pi)\mu^{+}\mu^{-}$, in the framework of SM as well as NP scenarios. For the angular coefficients $\langle I_{3,||}\rangle, \langle I_{4,||}\rangle$, and $\langle I_{5,||}\rangle$ shown in Figs. \ref{figIS4}(a), \ref{figIS4}(b), and \ref{figIS4}(c), respectively, the NP scenarios under consideration are discriminated in the region $q^{2}=[1,2]$ $\text{GeV}^{2}$ and $q^{2}=[3,4]$ $\text{GeV}^{2}$, with uncertainties due to form factors negligibly small in the angular coefficient $\langle I_{4,||}\rangle$.
\end{itemize}
\section{Conclusions}\label{concl}
Study of rare semileptonic decays of $B$ meson gives us a path to investigate physics beyond the SM. In literature various exclusive semileptonic decays mediated by the flavor changing neutral current transitions  and flavor changing charged current transitions show reasonable deviations from the SM predictions. As various global fit analyses suggest the presence of NP, in different physical observables of $B\to (K,K^{\ast})\mu^{+}\mu^{-}$ decays, in terms of the fit values of the NP coupling, we analyze the implications of these NP scenarios onto the angular observables of the complementary four-fold $B_{c}\to D^{\ast}_{s}(\to D_{s}\gamma, (D_{s}\pi ))\mu^{+}\mu^{-}$ decays, which are governed by the same quark level transition. Using the effective Hamiltonian by incorporating the vector and axial-vector NP operators $(O_{9}, O_{9^{\prime}}, O_{10}, O_{10^{\prime}})$, we have derived the four-fold angular distributions for $B_{c}\to D^{\ast}_{s}(\to D_{s}\gamma )\mu^{+}\mu^{-}$, and $B_{c}\to D^{\ast}_{s}(\to D_{s}\pi)\mu^{+}\mu^{-}$ decays from which the individual angular coefficients and various physical observables can be extracted. To analyze the NP effects, in these observables, we use the best fit values of the Wilson coefficients coming from the global fit analysis with the assumption of NP present only in the muon sector.

To summarize our work, we have observed sizeable difference between the NP predictions of different physical observables and the angular coefficients for $B_{c}\to D^{\ast}_{s}(\to D_{s}\gamma)\mu^{+}\mu^{-}$ and $B_{c}\to D^{\ast}_{s}(\to D_{s}\pi)\mu^{+}\mu^{-}$ decays, compared to the SM expectations. The NP results of the differential branching ratios for the considered decays indicate decreased values compared to that of the SM, however due to large error bands coming from the errors due to the form factors, NP results remain compatible with the SM estimates. Considering the forward-backward asymmetry and the longitudinal helicity fraction of $D^{\ast}_{s}$ meson, a number of NP scenarios can be distinguished from the SM predictions as well as from each other, in some kinematical ranges. For the unpolarized and polarized LFUV ratios i.e. $R_{D^{\ast}_{s}}$ and $R^{L,T}_{D^{\ast}_{s}}$, our analysis shows that there is no sizeable deviations expected from the SM prediction. Furthermore, the NP analysis considering the individual angular coefficients also shows sizeable deviations from the SM predictions along with distinct predictions for different NP scenarios. Hence the precise measurement of the studied physical observables  for $B_{c}\to D^{\ast}_{s}(\to D_{s}\gamma )\mu^{+}\mu^{-}$, and $B_{c}\to D^{\ast}_{s}(\to D_{s}\pi )\mu^{+}\mu^{-}$ decays at LHCb and the future collider experiments will give useful complementary information, required to clarify the structure of new physics in $b\to s \ell\ell$ decays.
\section*{Acknowledgments}
This work is supported by  Higher Education Commission of Pakistan through Grant no. NRPU/20-15142.
\appendix
\section{SM Wilson Coefficients}\label{append}
The explicit expressions used for the Wilson coefficients are given as follows \cite{Bobeth:1999mk,Beneke:2001at,Asatrian:2001de,Asatryan:2001zw,Greub:2008cy,Du:2015tda},
\begin{eqnarray}
C_{7}^{\text{eff}}(q^2)&=&C_{7}-\frac{1}{3}\left(C_{3}+\frac{4}{3}C_{4}+20C_{5}+\frac{80}{3}C_{6}\right)
-\frac{\alpha_{s}}{4\pi}\left[(C_{1}-6C_{2})F^{(7)}_{1,c}(q^2)+C_{8}F^{(7)}_{8}(q^2)\right],\notag\\
C_{9}^{\text{eff}}(q^2)&=&C_{9}+\frac{4}{3}\left(C_{3}+\frac{16}{3}C_{5}+\frac{16}{9}C_{6}\right)
-h(0, q^2)\left(\frac{1}{2}C_{3}+\frac{2}{3}C_{4}+8C_{5}+\frac{32}{3}C_{6}\right)\notag\\
&-&h(m_{b}^{\text{pole}}, q^2)\big(\frac{7}{2}C_{3}+\frac{2}{3}C_{4}+38C_{5}+\frac{32}{3}C_{6}\big)+h(m_{c}^{\text{pole}}, q^2)
\big(\frac{4}{3}C_{1}+C_{2}+6C_{3}+60C_{5}\big)\notag\\
&-&\frac{\alpha_{s}}{4\pi}\left[C_{1}F^{(9)}_{1,c}(q^2)+C_{2}F^{(9)}_{2,c}(q^2)+C_{8}F^{(9)}_{8}(q^2)\right],\label{WC3}
\end{eqnarray}where the functions $h(m_{q}^{\text{pole}}, q^2)$ with $q=c, b$, and functions $F^{(7,9)}_{8}(q^2)$ are
defined in \cite{Beneke:2001at}, while the functions $F^{(7,9)}_{1,c}(q^2)$, $F^{(7,9)}_{2,c}(q^2)$ are
given in \cite{Asatryan:2001zw} for low $q^{2}$ and in \cite{Greub:2008cy} for high $q^{2}$. The quark masses appearing in all of these functions are defined in the pole scheme.

\section{Binned Predictions of Physical Observables}\label{append1}
In this appendix, we give the SM as well as NP predictions of physical observables in different $q^2$ bins.
\begin{table}[ht!]
\centering
\captionsetup{margin=2cm}
\caption{\small Predictions of averaged values of observables such as differential branching ratios, $d\mathcal{B}(B_{c}\to D^{\ast}_{s}\mu^{+}\mu^{-})/{dq^2}$,
$d\mathcal{B}(B_{c}\to D^{\ast}_{s}(\to D_{s}\pi)\mu^{+}\mu^{-})/{dq^2}$,
$d\mathcal{B}(B_{c}\to D^{\ast}_{s}(\to D_{s}\gamma)\mu^{+}\mu^{-})/{dq^2}$,
longitudinal helicity fraction $f_{L}$,
lepton forward-backward asymmetry $A_{FB}$,
unpolarized LFUV ratio $R_{D^{\ast}_{s}}$, and polarized LFUV ratios $R^{L,T}_{D^{\ast}_{s}}$, in $q^{2}=[0.045-1.0]$ GeV$^2$ bin, for the SM as well as the NP scenarios presented in Table \ref{tab:bestfitWC}. The listed errors arise due to the uncertainties of the form factors.}\label{Obs1}
\renewcommand{\arraystretch}{1.5}
 \scalebox{0.60}{
\begin{tabular}{|M{6.2cm}|M{2cm}|M{2cm}|M{2cm}|M{2cm}|M{2cm}|M{2cm}|}
 \hline
 \multicolumn{7}{|c|}{$q^2=[0.045-1.0]$ GeV$^2$} \\
 \hline
Observables &SM& S1 & S2 & S3 & S4 & S5 \\
 \hline
 $\frac{d\mathcal{B}}{dq^2}(B_{c}\to D^{\ast}_{s}\mu^{+}\mu^{-})\times 10^{8}$& $5.013^{+0.554}_{-0.476}$ &
  $5.041^{+0.480}_{-0.558}$
  &$5.041^{+0.463}_{-0.471}$
  & $5.041^{+0.480}_{-0.458}$
  & $5.040^{+0.476}_{-0.458}$
  & $5.047^{+0.468}_{-0.469}$ \\
  \hline
  $ \frac{d\mathcal{B}}{dq^2}(B_{c}\to D^{\ast}_{s}(\to D_{s}\pi)\mu^{+}\mu^{-})\times 10^{10} $& $2.61^{+0.23}_{-0.24}$ &
 $2.62^{+0.25}_{-0.24}$ &
 $2.61^{+0.25}_{-0.24}$&
 $2.62^{+0.25}_{-0.24}$&
 $2.62^{+0.25}_{-0.24}$&
 $2.62^{+0.25}_{-0.24}$\\
 \hline
  $ \frac{d\mathcal{B}}{dq^2}(B_{c}\to D^{\ast}_{s}(\to D_{s}\gamma)\mu^{+}\mu^{-})\times 10^{8}$
  &$2.760^{+0.290}_{-0.253}$ &
 $2.781^{+0.270}_{-0.250}$ &
 $2.948^{+0.255}_{-0.258}$&
 $2.750^{+0.267}_{-0.253}$&
 $2.752^{+0.270}_{-0.247}$ &
 $2.734^{+0.263}_{-0.251}$\\
 \hline
 ${f}_L$ &$0.175^{+0.004}_{-0.004}$& $0.135^{+0.003}_{-0.004}$ & $0.156^{+0.002}_{-0.004}$ &  $0.118^{+0.003}_{-0.003}$ & $0.119^{+0.003}_{-0.003}$ &  $0.141^{+0.003}_{-0.003}$ \\
 \hline
    $A_{FB}$ &$0.114^{+0.003}_{-0.003}$ &$0.120^{+0.003}_{-0.004}$ & $0.117^{+0.003}_{-0.003}$ &  $0.122^{+0.004}_{-0.004}$& $0.122^{+0.003}_{-0.004}$ &  $0.119^{+0.003}_{-0.004}$ \\
 \hline
 $R_{D^*_s}$& $0.939^{+0.001}_{-0.001}$ &$0.935^{+0.001}_{-0.001}$& $0.934^{+0.001}_{-0.001}$ & $0.934^{+0.001}_{-0.001}$ & $0.934^{+0.001}_{-0.001}$& $0.934^{+0.001}_{-0.001}$\\
 \hline
 $R^L_{D^*_s}$& $0.942^{+0.002}_{-0.002}$ & $0.943^{+0.003}_{-0.003}$ & $0.941^{+0.003}_{-0.004}$ & $0.941^{+0.003}_{-0.003}$ & $0.942^{+0.003}_{-0.002}$ & $0.941^{+0.003}_{-0.002}$ \\
 \hline
 $R^{T}_{D^{\ast}_{s}}$ &$0.933^{+0.001}_{-0.001}$& $0.934^{+0.001}_{-0.001}$ & $0.932^{+0.001}_{-0.001}$ & $0.933^{+0.001}_{-0.001}$& $0.933^{+0.001}_{-0.001}$ & $0.933^{+0.001}_{-0.001}$ \\
 \hline
\end{tabular}}
\end{table}
\begin{table}[ht!]
\centering
\captionsetup{margin=2cm}
\caption{\small Same as in Table \ref{Obs1}, but for $q^{2}=[1.0-2.0]$ GeV$^2$ bin.}\label{Obs2}
\renewcommand{\arraystretch}{1.5}
 \scalebox{0.60}{
\begin{tabular}{|M{6.2cm}|M{2cm}|M{2cm}|M{2cm}|M{2cm}|M{2cm}|M{2cm}|}
 \hline
 \multicolumn{7}{|c|}{$q^2=[1.0-2.0]$ GeV$^2$} \\
 \hline
Observables&SM&S1 & S2 & S3 & S4 & S5\\
 \hline
 $ \frac{d\mathcal{B}}{dq^2}(B_{c}\to D^{\ast}_{s}\mu^{+}\mu^{-})\times 10^{8}$ & $0.938^{+0.114}_{-0.105}$ &$0.887^{+0.103}_{-0.096}$
 &$0.905^{+0.108}_{-0.099}$
 & $0.843^{+0.099}_{-0.089}$ &
 $0.848^{+0.090}_{-0.098} $ &
 $0.893^{+0.103}_{-0.098} $ \\
  \hline
  $ \frac{d\mathcal{B}}{dq^2}(B_{c}\to D^{\ast}_{s}(\to D_{s}\pi)\mu^{+}\mu^{-})\times 10^{10} $ &$4.880^{+0.059}_{-0.050} $&
 $5.150^{+0.032}_{-0.040}$&
 $5.270^{+0.060}_{-0.094}$
 &$3.910^{+0.900}_{-0.800} $&
 $3.940^{+0.050}_{-0.060}$ &
 $4.940^{+0.026}_{-0.080} $ \\
  \hline
 $ \frac{d\mathcal{B}}{dq^2}(B_{c}\to D^{\ast}_{s}(\to D_{s}\gamma)\mu^{+}\mu^{-})\times 10^{8} $&$1.032^{+0.128}_{-0.118}$ &
 $3.980^{+0.118}_{-0.109}$ &
 $0.876^{+0.097}_{-0.106}$&
 $0.932^{+0.113}_{-0.105}$
 &$0.942^{+0.114}_{-0.104}$ &
 $0.918^{+0.110}_{-0.101}$ \\
  \hline
${f}_L$ &$0.661^{+0.004}_{-0.006}$ &$0.542^{+0.005}_{-0.007}$ & $0.606^{+0.005}_{-0.005}$ &  $0.496^{+0.004}_{-0.004}$ & $0.497^{+0.006}_{-0.006} $&  $0.564^{+0.006}_{-0.006}$ \\
  \hline
 $A_{FB}$ &$0.221^{+0.0007}_{-0.008}$ &$0.311^{+0.002}_{-0.002}$ & $0.265^{+0.005}_{-0.006}$ &  $0.343^{+0.001}_{-0.001}$ & $0.337^{+0.002}_{-0.002}$ & $0.296^{+0.003}_{-0.004}$
 \\
  \hline
  $R_{D^*_s}$ & $0.989^{+0.001}_{-0.001} $&  $0.990^{+0.001}_{-0.001}$ &  $0.989^{+0.001}_{-0.001} $& $0.989^{+0.001}_{-0.001}$ &$ 0.988^{+0.001}_{-0.001}$ & $ 0.989^{+0.001}_{-0.001}$ \\
  \hline
 $R^L_{D^*_s}$&$0.997^{+0.001}_{-0.001}$ &$ 0.996^{+0.001}_{-0.001} $& $0.997^{+0.001}_{-0.001}$ & $0.997^{+0.001}_{-0.001}$ & $0.996^{+0.001}_{-0.001}$ & $0.997^{+0.001}_{-0.001}$ \\
  \hline
 $R^T_{D^*_s}$&$0.973^{+0.002}_{-0.002}$ &$0.978^{+0.001}_{-0.001}$ &$ 0.977^{+0.001}_{-0.001}$ & $0.981^{+0.001}_{-0.001}$ &$0.981^{+0.001}_{-0.001}$  & $0.979^{+0.002}_{-0.001}$ \\
 \hline
    \end{tabular}}
\end{table}
\begin{table}[ht!]
\centering
\captionsetup{margin=2cm}
\caption{\small Same as in Table \ref{Obs1}, but for $q^{2}=[2.0-3.0]$ GeV$^2$ bin.}\label{Obs3}
\renewcommand{\arraystretch}{1.5}
 \scalebox{0.60}{
\begin{tabular}{|M{6.2cm}|M{2cm}|M{2cm}|M{2cm}|M{2cm}|M{2cm}|M{2cm}|}
 \hline
 \multicolumn{7}{|c|}{$q^2=[2.0-3.0]$ GeV$^2$} \\
 \hline
 Observables&SM&S1 & S2 & S3 & S4 & S5\\
 \hline
$ \frac{d\mathcal{B}}{dq^2}(B_{c}\to D^{\ast}_{s}\mu^{+}\mu^{-})\times 10^{8} $&$1.091^{+0.172}_{-0.154}$&
 $0.932^{+0.142}_{-0.127} $&
$ 1.004^{+0.156}_{-0.136} $&
 $0.864^{+0.134}_{-0.120}$ &
 $0.875^{+0.134}_{-0.120} $&
$ 0.927^{+0.172}_{-0.104} $ \\
  \hline
   $ \frac{d\mathcal{B}}{dq^2}(B_{c}\to D^{\ast}_{s}(\to D_{s}\pi)\mu^{+}\mu^{-})\times 10^{10} $&$5.607^{+0.090}_{-0.080}$&
 $5.580^{+0.050}_{-0.035}$ &
 $6.030^{+0.050}_{-0.030}$&
 $3.870^{+1.030}_{-0.087}$ &
 $3.930^{+0.050}_{-0.060}$ &
 $5.410^{+0.034}_{-0.098}$  \\
  \hline
  $ \frac{d\mathcal{B}}{dq^2}(B_{c}\to D^{\ast}_{s}(\to D_{s}\gamma)\mu^{+}\mu^{-})\times 10^{8}$ &$1.229^{+0.172}_{-0.193}$&
 $1.079^{+0.164}_{-0.147}$ &
 $0.968^{+0.149}_{-0.133}$&
 $1.007^{+0.156}_{-0.140} $&
 $1.020^{+0.161}_{-0.137}$ &
 $1.003^{+0.153}_{-0.137}$\\
  \hline
 ${f}_L$  &$0.691^{+0.011}_{-0.011} $& $0.622^{+0.006}_{-0.007}$& $0.663^{+0.040}_{-0.054}$ & $ 0.586^{+0.005}_{-0.010}$ & $0.581^{+0.006}_{-0.006}$ &  $0.637^{+0.007}_{-0.007}$ \\
  \hline
 $A_{FB}$ &$-0.020^{+0.018}_{-0.019}$ &$0.116^{+0.014}_{-0.017}$ & $0.039^{+0.020}_{-0.018}$ &$0.156^{+0.016}_{-0.019}$ & $0.160^{+0.020}_{-0.018}$ &  $0.089^{+0.108}_{-0.071}$ \\
  \hline
  $R_{D^*_s}$ & $ 0.989^{+0.001}_{-0.001} $& $ 0.988^{+0.001}_{-0.001} $&  $0.989^{+0.001}_{-0.001}$ & $0.988^{+0.001}_{-0.001}$& $ 0.988^{+0.001}_{-0.001} $&  $0.989^{+0.001}_{-0.001}$ \\
   \hline
 $R^L_{D^*_s}$&$0.997^{+0.001}_{-0.001}$ & $0.996^{+0.001}_{-0.001}$& $0.997^{+0.001}_{-0.001}$ & $0.997^{+0.001}_{-0.001}$ & $0.996^{+0.001}_{-0.001}$ & $0.997^{+0.001}_{-0.001}$ \\
  \hline
 $R^T_{D^*_s}$ &$0.974^{+0.001}_{-0.001}$ &$ 0.974^{+0.001}_{-0.001}$ & $0.974^{+0.001}_{-0.001}$ & $0.976^{+0.001}_{-0.001}$ & $0.976^{+0.001}_{-0.001}$ & $0.975^{+0.001}_{-0.001}$ \\
 \hline
    \end{tabular}}
\end{table}

\begin{table}[ht!]
\centering
\captionsetup{margin=2cm}
\caption{\small Same as in Table \ref{Obs1}, but for $q^{2}=[3.0-4.0]$ GeV$^2$ bin.}\label{Obs4}
\renewcommand{\arraystretch}{1.5}
 \scalebox{0.60}{
\begin{tabular}{|M{6.2cm}| M{2cm}|M{2cm}|M{2cm}|M{2cm}|M{2cm}|M{2cm}|}
 \hline
 \multicolumn{7}{|c|}{$q^2=[3.0-4.0]$ GeV$^2$} \\
 \hline
 Observables&SM&S1 & S2 & S3 & S4 & S5\\
 \hline
 $ \frac{d\mathcal{B}}{dq^2}(B_{c}\to D^{\ast}_{s}\mu^{+}\mu^{-})\times 10^{8}$&$1.537^{+0.281}_{-0.245} $&
 $1.248^{+0.222}_{-0.195}$ &
 $1.248^{+0.386}_{-0.083}$ &
 $1.150^{+0.210}_{-0.184}$ &
 $1.167^{+0.169}_{-0.189}$ &
 $1.522^{+0.580}_{-0.293}$  \\
  \hline
  $ \frac{d\mathcal{B}}{dq^2}(B_{c}\to D^{\ast}_{s}(\to D_{s}\pi)\mu^{+}\mu^{-})\times 10^{10} $ &$7.990^{+0.146}_{-0.127} $&
 $7.650^{+0.087}_{-0.201}$ &
 $8.500^{+0.156}_{-0.178}$ &
 $5.020^{+0.200}_{-0.198}$&
 $5.100^{+0.156}_{-0.189}$ &
 $7.570^{+0.345}_{-0.321}$ \\
  \hline
  $ \frac{d\mathcal{B}}{dq^2}(B_{c}\to D^{\ast}_{s}(\to D_{s}\gamma)\mu^{+}\mu^{-})\times 10^{8}$&$1.723^{+0.272}_{-0.310}$ &
 $1.448^{+0.220}_{-0.255}$ &
 $1.330^{+0.239}_{-0.210}$ &
 $1.134^{+0.342}_{-0.140}$&
 $1.370^{+0.246}_{-0.216}$ &
 $1.355^{+0.241}_{-0.211}$\\
  \hline
 ${f}_L$  &$0.595^{+0.018}_{-0.016}$ &$0.565^{+0.025}_{-0.015}$ & $0.584^{+0.030}_{-0.015}$ & $ 0.538^{+0.040}_{-0.019} $& $0.528^{+0.016}_{-0.016}$ &  $0.572^{+0.016}_{-0.014}$ \\
  \hline
 $A_{FB}$ &$-0.212^{+0.008}_{-0.022}$ &$-0.093^{+0.026}_{-0.024}$ &$ -0.164^{+0.017}_{-0.020}$ &$-0.063^{+0.024}_{-0.019}$ &$-0.145^{+0.025}_{-0.015}$ &  $-0.120^{+0.020}_{-0.023}$
 \\
  \hline
  $R_{D^*_s}$& $0.991^{+0.001}_{-0.001}$ & $0.988^{+0.001}_{-0.001} $ &  $0.991^{+0.001}_{-0.001} $&  $0.990^{+0.001}_{-0.001}$ &  $0.989^{+0.001}_{-0.001}$ & $0.990^{+0.001}_{-0.001}$ \\
  \hline
 $R^L_{D^*_s}$& $0.997^{+0.001}_{-0.001}$& $0.996^{+0.002}_{-0.002}$ & $0.997^{+0.002}_{-0.001}$ & $0.997^{+0.001}_{-0.001}$ & $0.996^{+0.001}_{-0.001}$ & $0.997^{+0.002}_{-0.001}$ \\
  \hline
 $R^T_{D^*_s}$&$0.983^{+0.002}_{-0.002}$ & $0.981^{+0.002}_{-0.003}$ & $0.982^{+0.001}_{-0.002}$ & $0.981^{+0.001}_{-0.001}$ & $0.981^{+0.001}_{-0.001}$ & $0.982^{+0.001}_{-0.001}$ \\
 \hline
\end{tabular}}
\end{table}

\begin{table}[ht!]
\centering
\captionsetup{margin=2cm}
\caption{\small Same as in Table \ref{Obs1}, but for $q^{2}=[4.0-5.0]$ GeV$^2$ bin.}\label{Obs5}
\renewcommand{\arraystretch}{1.5}
 \scalebox{0.60}{
\begin{tabular}{|M{6.2cm}| M{2cm}|M{2cm}|M{2cm}|M{2cm}|M{2cm}|M{2cm}|}
 \hline
 \multicolumn{7}{|c|}{$q^2=[4.0-5.0]$ GeV$^2$} \\
 \hline
 Observables&SM&S1 & S2 & S3 & S4 & S5\\
 \hline
$ \frac{d\mathcal{B}}{dq^2}(B_{c}\to D^{\ast}_{s}\mu^{+}\mu^{-})\times 10^{8}$ &$2.187^{+0.445}_{-0.381}$&
 $1.733^{+0.347}_{-0.298}$ &
 $1.733^{+0.612}_{-0.120}$ &
 $1.596^{+0.328}_{-0.281}$&
 $1.621^{+0.333}_{-0.285}$  &
 $1.670^{+0.494}_{-0.179}$  \\
  \hline
  $\frac{d\mathcal{B}}{dq^2}(B_{c}\to D^{\ast}_{s}(\to D_{s}\pi)\mu^{+}\mu^{-})\times 10^{10}$ &$1.137^{+0.232}_{-0.201} $&
 $1.080^{+0.340}_{-0.321}$ &
 $1.220^{+0.296}_{-0.281}$ &
 $0.684^{+0.316}_{-0.350}$ &
 $0.695^{+0.300}_{-0.234}$ &
 $1.023^{+0.212}_{-0.345}$  \\
  \hline
  $\frac{d\mathcal{B}}{dq^2}(B_{c}\to D^{\ast}_{s}(\to D_{s}\gamma)\mu^{+}\mu^{-})\times 10^{8} $&$2.434^{+0.490}_{-0.418}$ &
 $1.999^{+0.341}_{-0.395} $&
 $1.875^{+0.376}_{-0.324}$ &
 $1.858^{+0.376}_{-0.323}$ &
 $1.892^{+0.0.381}_{-0.328}$ &
 $1.887^{+0.376}_{-0.323}$\\
  \hline
 ${f}_L$  &$0.508^{+0.018}_{-0.017}$ &$0.494^{+0.016}_{-0.016} $ & $0.503^{+0.023}_{-0.420}$  &  $0.474^{+0.013}_{-0.020}$  & $0.461^{+0.018}_{-0.018}$  &  $0.497^{+0.017}_{-0.016}$ \\
  \hline
 $A_{FB}$ &$-0.329^{+0.016}_{-0.020}$&$-0.237^{+0.019}_{-0.019}$ &$-0.294^{+0.018}_{-0.020}$ &  $-0.215^{+0.025}_{-0.015}$ & $-0.225^{+0.020}_{-0.021}$ &  $-0.260^{+0.018}_{-0.022}$ \\
  \hline
 $R_{D^*_s}$& $0.993^{+0.001}_{-0.001} $&  $0.989^{+0.001}_{-0.001}$ &  $0.992^{+0.001}_{-0.001}$ &  $0.991^{+0.001}_{-0.001}$ & $0.991^{+0.002}_{-0.002}$ & $0.992^{+0.002}_{-0.002}$ \\
  \hline
 $R^L_{D^*_s}$&$0.997^{+0.001}_{-0.001}$ & $0.996^{+0.002}_{-0.002}$ & $0.997^{+0.001}_{-0.001}$ & $0.997^{+0.001}_{-0.002}$ & $0.996^{+0.001}_{-0.001}$ & $0.997^{+0.001}_{-0.001}$ \\
  \hline
 $R^T_{D^*_s}$&$0.988^{+0.001}_{-0.001}$ &$0.986^{+0.001}_{-0.001}$ & $0.988^{+0.001}_{-0.001}$ & $0.986^{+0.001}_{-0.001}$ & $0.987^{+0.001}_{-0.001}$ & $0.987^{+0.001}_{-0.001}$ \\
 \hline
\end{tabular}}
\end{table}
\begin{table}[ht!]
\centering
\captionsetup{margin=2cm}
\caption{\small Same as in Table \ref{Obs1}, but for $q^{2}=[5.0-6.0]$ GeV$^2$ bin.}\label{Obs6}
\renewcommand{\arraystretch}{1.5}
 \scalebox{0.60}{
\begin{tabular}{|M{6.2cm}| M{2cm}|M{2cm}|M{2cm}|M{2cm}|M{2cm}|M{2cm}|}
 \hline
 \multicolumn{7}{|c|}{$q^2=[5.0-6.0]$ GeV$^2$} \\
 \hline
 Observables&SM&S1 & S2 & S3 & S4 & S5\\
 \hline
 $\frac{d\mathcal{B}}{dq^2}(B_{c}\to D^{\ast}_{s}\mu^{+}\mu^{-})\times 10^{8}$&$3.056^{+0.683}_{-0.574}$&
 $2.391^{+0.529}_{-0.446}$ &
 $2.391^{+0.990}_{-0.186}$ &
 $2.204^{+0.500}_{-0.419}$ &
 $2.237^{+0.508}_{-0.427} $ &
 $2.341^{+0.708}_{-0.312}$  \\
  \hline
  $\frac{d\mathcal{B}}{dq^2}(B_{c}\to D^{\ast}_{s}(\to D_{s}\pi)\mu^{+}\mu^{-})\times 10^{10}$ &$1.589^{+0.355}_{-0.298}$ &
$1.519^{+0.476}_{-0.507}$ &
$1.785^{+0.500}_{-0.494}$ &
$9.281^{+0.480}_{-0.478}$&
$9.422^{+0.450}_{-0.480}$ &
$1.407^{+0.389}_{-0.283}$ \\
 \hline
 $ \frac{d\mathcal{B}}{dq^2}(B_{c}\to D^{\ast}_{s}(\to D_{s}\gamma)\mu^{+}\mu^{-})\times 10^{8} $&$3.379^{+0.745}_{-0.628} $&
$2.740^{+0.599}_{-0.505}$ &
$2.610^{+0.578}_{-0.487}$ &
$2.548^{+0.569}_{-0.479}$&
$2.593^{+0.577}_{-0.486}$ &
$2.608^{+0.559}_{-0.487}$\\
 \hline
 ${f}_L$  &$0.442^{+0.060}_{-0.025} $ &$0.435^{+0.052}_{-0.019}$  &$0.440^{+0.060}_{-0.024} $  & $ 0.421^{+0.430}_{-0.395}$  & $0.406^{+0.424}_{-0.424} $ &  $0.437^{+0.054}_{-0.021} $ \\
  \hline
 $A_{FB}$ &$-0.400^{+0.015}_{-0.017}$&$-0.332^{+0.017}_{-0.019}$ &$-0.376^{+0.016}_{-0.019}$ & $-0.316^{+0.024}_{-0.013}$ & $-0.327^{+0.017}_{-0.020}$ & $-0.350^{+0.017}_{-0.021}$ \\
  \hline
 $R_{D^*_s}$&$0.992^{+0.002}_{-0.002} $& $0.991^{+0.002}_{-0.002} $& $0.994^{+0.002}_{-0.002}$ & $0.993^{+0.002}_{-0.002}$ & $0.993^{+0.002}_{-0.002}$ & $0.993^{+0.002}_{-0.002}$ \\
  \hline
 $R^L_{D^*_s}$&$ 0.997^{+0.001}_{-0.001}$& $0.996^{+0.001}_{-0.001} $&$0.997^{+0.001}_{-0.001}$ & $0.997^{+0.002}_{-0.002}$ & $0.996^{+0.002}_{-0.002} $& $0.997^{+0.002}_{-0.002}$ \\
  \hline
 $R^T_{D^*_s}$&$0.991^{+0.001}_{-0.001}$ &$0.990^{+0.002}_{-0.002} $ & $0.991^{+0.001}_{-0.002}$&$ 0.990^{+0.001}_{-0.001}$ & $0.990^{+0.002}_{-0.002}$ & $0.990^{+0.001}_{-0.001}$ \\
 \hline
\end{tabular}}
\end{table}

\begin{table}[ht!]
\centering
\captionsetup{margin=2cm}
\caption{\small Same as in Table \ref{Obs1}, but for $q^{2}=[1.0-6.0]$ GeV$^2$ bin.}\label{Obs7}
\renewcommand{\arraystretch}{1.5}
 \scalebox{0.60}{
\begin{tabular}{|M{6.2cm}| M{2cm}|M{2cm}|M{2cm}|M{2cm}|M{2cm}|M{2cm}|}
 \hline
 \multicolumn{7}{|c|}{$q^2=[1.0-6.0]$ GeV$^2$} \\
 \hline
 Observables&SM&S1 & S2 & S3 & S4 & S5\\
 \hline
$\frac{d\mathcal{B}}{dq^2}(B_{c}\to D^{\ast}_{s}\mu^{+}\mu^{-})\times 10^{8} $&$2.106^{+0.455}_{-0.384}$ &
 $1.659^{+0.349}_{-0.294}$ &
 $1.659^{+0.616}_{-0.124}$ &
 $1.529^{+0.330}_{-0.277}$ &
 $1.552^{+0.335}_{-0.280}$ &
 $1.626^{+0.468}_{-0.208}$  \\
  \hline
  $ \frac{d\mathcal{B}}{dq^2}(B_{c}\to D^{\ast}_{s}(\to D_{s}\pi)\mu^{+}\mu^{-})\times 10^{10} $ &$1.095^{+0.237}_{-0.199}$ &
 $1.040^{+0.303}_{-0.331} $&
 $9.740^{+0.209}_{-0.300}$&
$7.806^{+0.860}_{-0.296}$ &
 $8.073^{+0.740}_{-0.461}$&
 $9.759^{+0.131}_{-0.384}$\\
  \hline
  $ \frac{d\mathcal{B}}{dq^2}(B_{c}\to D^{\ast}_{s}(\to D_{s}\gamma)\mu^{+}\mu^{-})\times 10^{8} $&$9.797^{+0.863}_{-0.613}$ &
 $8.246^{+0.532}_{-0.520}$ &
 $7.657^{+0.448}_{-0.249}$&
$7.691^{+0.457}_{-0.260}$ &
 $7.521^{+0.475}_{-0.276}$&
 $7.821^{+0.452}_{-0.256}$  \\
  \hline
${f}_L$  &$0.539^{+0.056}_{-0.023}$  &$0.509^{+0.523}_{-0.496}$  & $0.528^{+0.056}_{-0.013}$ &$0.485^{+0.093}_{-0.065}$ & $0.475^{+0.089}_{-0.061}$  &  $0.516^{+0.030}_{-0.002}$ \\
 \hline
 $A_{FB}$ &$-0.236^{+0.020}_{-0.026}$ &$-0.130^{+0.023}_{-0.026}$ &$-0.194^{+0.023}_{-0.024}$&  $-0.101^{+0.029}_{-0.020}$ &$-0.171^{+0.022}_{-0.25}$& $-0.154^{+0.023}_{-0.024}$ \\
  \hline
 $R_{D^*_s}$&$0.993^{+0.001}_{-0.001}$ & $0.993^{+0.001}_{-0.001} $& $0.992^{+0.001}_{-0.001}$ &$ 0.991^{+0.001}_{-0.001}$ & $0.991^{+0.001}_{-0.001}$ & $0.991^{+0.002}_{-0.002}$ \\
  \hline
 $R^L_{D^*_s}$&$0.997^{+0.002}_{-0.002}$ & $0.997^{+0.001}_{-0.002}$& $0.997^{+0.001}_{-0.002}$ & $0.997^{+0.001}_{-0.001}$ & $0.996^{+0.001}_{-0.001}$ & $0.997^{+0.001}_{-0.001}$ \\
  \hline
 $R^T_{D^*_s}$&$0.985^{+0.001}_{-0.001}$ &$ 0.984^{+0.001}_{-0.001}$ &$ 0.985^{+0.001}_{-0.001}$ & $0.984^{+0.001}_{-0.001}$ & $0.985^{+0.002}_{-0.002}$ & $0.985^{+0.002}_{-0.002}$ \\
  \hline
\end{tabular}}
\end{table}

\begin{table}
\centering
\caption{\small Predictions of averaged values of the angular coefficients for the $B_{c}\to D^{\ast}_{s}(\to D_{s}\pi)\mu^{+}\mu^{-}$, and $B_{c}\to D^{\ast}_{s}(\to D_{s}\gamma)\mu^{+}\mu^{-}$ decay, in $q^2=[0.045-1]$ GeV$^2$ bin, for the SM as well as the NP scenarios presented in Table-\ref{tab:bestfitWC}. The listed errors arise due to the uncertainties of the form factors.}\label{tab:Angobs1}
    \renewcommand{\arraystretch}{1.5}
    \scalebox{0.50}{
    \begin{tabular}{|c||c|c||c|c||c|c||c|c||c|c||c|c|}
        \hline
        \multicolumn{13}{|c|}{$q^2=[0.045-1]$ GeV$^2$}\\
        \hline
      Angular   &\multicolumn{2}{|c|}{SM} &\multicolumn{2}{|c|}{S1} &\multicolumn{2}{|c|}{S2} &\multicolumn{2}{|c|}{S3} &\multicolumn{2}{|c|}{S4} &\multicolumn{2}{|c|}{S5} \\
         \cline{2-13}
       observables   & $D_s\pi$ & $D_s\gamma$ &  $D_s\pi$ & $D_s\gamma$ &  $D_s\pi$ & $D_s\gamma$&  $D_s\pi$ & $D_s\gamma$ &  $D_s\pi$ & $D_s\gamma$ &  $D_s\pi$ & $D_s\gamma$ \\
         \hline
         $\langle I_{1s,^{\|}}\rangle,\langle I^{\gamma}_{1s,\perp}\rangle$ &
          $0.580^{+0.003}_{-0.03}$ &$0.391^{+0.001}_{-0.001}$  &$0.610^{+0.003}_{-0.003}$&$0.386^{+0.001}_{-0.001}$ &$0.608^{+0.003}_{-0.002}$& $0.387^{+0.001}_{-0.001}$ &$0.635^{+0.002}_{-0.001}$ & $0.383^{+0.002}_{-0.001}$ &$0.635^{+0.002}_{-0.002}$ & $0.384^{+0.001}_{-0.000}$ &$0.619^{+0.003}_{-0.003}$& $0.386^{+0.001}_{-0.000}$ \\
          \hline
         $\langle I_{2s,^{\|}}\rangle,\langle I^{\gamma}_{2s,\perp}\rangle $&
         $0.129^{+0.001}_{-0.001}$ &  $-0.015^{+0.003}_{-0.002}$ &
         $0.137^{+0.002}_{-0.001}$ & $0.005^{+0.002}_{-0.002}$ &
         $0.135^{+0.001}_{-0.001}$ &  $0.004^{+0.002}_{-0.002}$ &
         $0.141^{+0.002}_{-0.001}$& $0.0014^{+0.002}_{-0.002}$ &
         $0.142^{+0.001}_{-0.002}$ &  $0.013^{+0.001}_{-0.001}$&
         $0.138^{+0.001}_{-0.001}$ &  $0.007^{+0.001}_{-0.001}$\\
         \hline
         $\langle I_{1c,^{\|}}\rangle,\langle I^{\gamma}_{1c,\perp}\rangle$&
         $0.205^{+0.004}_{-0.003}$ &$0.580^{+0.003}_{-0.03}$&
         $0.163^{+0.003}_{-0.004}$ & $0.610^{+0.003}_{-0.003}$ &
         $0.164^{+0.004}_{-0.004}$ & $0.608^{+0.003}_{-0.002}$  &
         $0.124^{+0.023}_{-0.010}$ & $0.635^{+0.002}_{-0.001}$&
         $0.0129^{+0.015}_{-0.003}$ & $0.635^{+0.002}_{-0.002}$&
         $0.155^{+0.007}_{-0.005}$ & $0.619^{+0.003}_{-0.003}$ \\
\hline
    \end{tabular}}
\end{table}
\begin{table}
    \centering
    \caption{\small Same as in Table \ref{tab:Angobs1}, but for $q^{2}=[1.0-2.0]$ GeV$^2$ bin.}\label{tab:Angobs2}
    \renewcommand{\arraystretch}{1.5}
    \scalebox{0.50}{
    \begin{tabular}{|c||c|c||c|c||c|c||c|c||c|c||c|c|}
     \hline
        \multicolumn{13}{|c|}{$q^2=[1-2]$ GeV$^2$}\\
        \hline
       Angular &\multicolumn{2}{|c|}{SM} &\multicolumn{2}{|c|}{S1} &\multicolumn{2}{|c|}{S2} &\multicolumn{2}{|c|}{S3} &\multicolumn{2}{|c|}{S4} &\multicolumn{2}{|c|}{S5} \\
         \cline{2-13}
        observables & $D_s\pi$ & $D_s\gamma$ &  $D_s\pi$ & $D_s\gamma$ &  $D_s\pi$ & $D_s\gamma$&  $D_s\pi$ & $D_s\gamma$ &  $D_s\pi$ & $D_s\gamma$ &  $D_s\pi$ & $D_s\gamma$ \\
         \hline
          $\langle I_{1s,^{\parallel}}\rangle,\langle I^{\gamma}_{1s,\perp}\rangle$
          & $0.253^{+0.005}_{-0.003}$ &  $0.461^{+0.001}_{-0.001}$  &
          $0.343^{+0.004}_{-0.006}$ & $0.448^{+0.001}_{-0.001}$ &
          $0.294^{+0.005}_{-0.004}$ & $0.455^{+0.001}_{-0.001}$ &
          $0.377^{+0.004}_{-0.003}$ & $0.442^{+0.001}_{-0.000}$ &
          $0.376^{+0.004}_{-0.005}$ & $0.443^{+0.001}_{-0.000}$ &
          $0.326^{+0.005}_{-0.004}$ &  $0.450^{+0.001}_{-0.000}$
          \\
          \hline
         $\langle  I_{2s,^{\parallel}}\rangle,\langle I^{\gamma}_{2s,\perp}\rangle $&
         $0.083^{+0.004}_{-0.002}$ & $-0.274^{+0.002}_{-0.002}$ &
         $0.111^{+0.013}_{-0.010}$ & $-0.212^{+0.002}_{-0.003}$ &
         $0.096^{+0.007}_{-0.005}$ & $-0.242^{+0.003}_{-0.004}$ &
         $0.122^{+0.002}_{-0.001}$ & $-0.187^{+0.002}_{-0.003}$ &
         $0.119^{+0.004}_{-0.008}$ & $-0.189^{+0.002}_{-0.003}$ &
         $0.106^{+0.001}_{-0.001}$ &$-0.220^{+0.003}_{-0.003}$ \\
         \hline
         $\langle  I_{1c,^{\parallel}}\rangle,\langle I^{\gamma}_{1c,\perp}\rangle $&
         $0.671^{+0.004}_{-0.006}$ & $0.253^{+0.005}_{-0.003}$ &
         $0.615^{+0.005}_{-0.006}$ & $0.343^{+0.004}_{-0.006}$ &
         $0.549^{+0.006}_{-0.004}$ & $0.294^{+0.005}_{-0.004}$ &
         $0.503^{+0.004}_{-0.006}$ & $0.377^{+0.004}_{-0.003}$ &
         $0.504^{+0.005}_{-0.006}$ & $0.376^{+0.004}_{-0.005}$ &
         $0.572^{+0.006}_{-0.007}$ &$0.326^{+0.005}_{-0.004}$ \\
\hline
          \end{tabular}}
\end{table}
\begin{table}
    \centering
    \caption{\small Same as in Table \ref{tab:Angobs1}, but for $q^{2}=[2.0-3.0]$ GeV$^2$ bin.}\label{tab:Angobs3}
    \renewcommand{\arraystretch}{1.5}
    \scalebox{0.50}{
    \begin{tabular}{|c||c|c||c|c||c|c||c|c||c|c||c|c|}
    \hline
    \multicolumn{13}{|c|}{$q^2=[2-3]$ GeV$^2$}\\
    \hline
  Angular &\multicolumn{2}{|c|}{SM} &\multicolumn{2}{|c|}{S1} &\multicolumn{2}{|c|}{S2} &\multicolumn{2}{|c|}{S3} &\multicolumn{2}{|c|}{S4} &\multicolumn{2}{|c|}{S5} \\
        \cline{2-13}
    observables    & $D_s\pi$ & $D_s\gamma$ &  $D_s\pi$ & $D_s\gamma$ &  $D_s\pi$ & $D_s\gamma$&  $D_s\pi$ & $D_s\gamma$ &  $D_s\pi$ & $D_s\gamma$ &  $D_s\pi$ & $D_s\gamma$\\
           \hline
          $\langle  I_{1s,^{\parallel}}\rangle,\langle I^{\gamma}_{1s,\perp}\rangle$ &
           $0.232^{+0.008}_{-0.008}$ & $0.462^{+0.001}_{-0.001}$ &
           $0.283^{+0.005}_{-0.003}$ & $0.453^{+0.001}_{-0.001}$ &
           $0.253^{+0.007}_{-0.007}$& $0.460^{+0.002}_{-0.001}$  &
           $0.311^{+0.007}_{-0.004}$& $0.449^{+0.001}_{-0.002}$ &
           $0.315^{+0.005}_{-0.006}$& $0.449^{+0.001}_{-0.002}$ &
           $0.272^{+0.006}_{-0.005}$& $0.456^{+0.001}_{-0.001}$
          \\ \hline
         $\langle  I_{2s,^{\parallel}}\rangle,\langle I^{\gamma}_{2s,\perp}\rangle$ &
         $0.077^{+0.003}_{-0.002}$ &  $-0.285^{+0.006}_{-0.007}$ &
         $0.094^{+0.002}_{-0.001}$ &  $-0.248^{+0.005}_{-0.005}$ &
         $0.084^{+0.001}_{-0.002}$ &  $-0.281^{+0.005}_{-0.005}$  &
         $0.103^{+0.003}_{-0.001}$  &  $-0.226^{+0.006}_{-0.005}$ &
         $0.105^{+0.001}_{-0.001}$ &  $-0.227^{+0.006}_{-0.007}$&
         $0.091^{+0.001}_{-0.002}$&  $-0.261^{+0.004}_{-0.005}$  \\
         \hline
        $\langle  I_{1c,^{\parallel}}\rangle,\langle I^{\gamma}_{1c,\perp}\rangle$&
         $0.685^{+0.022}_{-0.009}$ & $0.232^{+0.008}_{-0.008}$ &
         $0.668^{+0.009}_{-0.009}$& $0.283^{+0.005}_{-0.003}$ &
         $0.627^{+0.006}_{-0.007}$& $0.253^{+0.007}_{-0.007}$ &
         $0.590^{+0.005}_{-0.010}$& $0.311^{+0.007}_{-0.004}$ &
         $0.585^{+0.007}_{-0.007}$&  $0.315^{+0.005}_{-0.006}$&
         $0.643^{+0.006}_{-0.008}$& $0.272^{+0.006}_{-0.005}$  \\
\hline
 \end{tabular}}
\end{table}

\begin{table}
    \centering
    \caption{\small Same as in Table \ref{tab:Angobs1}, but for $q^{2}=[3.0-4.0]$ GeV$^2$ bin.}\label{tab:Angobs4}
    \renewcommand{\arraystretch}{1.5}
    \scalebox{0.50}{
    \begin{tabular}{|c||c|c||c|c||c|c||c|c||c|c||c|c|}
     \hline
        \multicolumn{13}{|c|}{$q^2=[3-4]$ GeV$^2$}\\
        \hline
       Angular  &\multicolumn{2}{|c|}{SM} &\multicolumn{2}{|c|}{S1} &\multicolumn{2}{|c|}{S2} &\multicolumn{2}{|c|}{S3} &\multicolumn{2}{|c|}{S4} &\multicolumn{2}{|c|}{S5} \\
         \cline{2-13}
         observables & $D_s\pi$ & $D_s\gamma$ &  $D_s\pi$ & $D_s\gamma$ &  $D_s\pi$ & $D_s\gamma$&  $D_s\pi$ & $D_s\gamma$ &  $D_s\pi$ & $D_s\gamma$ &  $D_s\pi$ & $D_s\gamma$  \\
         \hline
          $\langle I_{1s,^{\parallel}}\rangle,\langle I^{\gamma}_{1s,\perp}\rangle$&
          $0.303^{+0.012}_{-0.013}$&$0.449^{+0.002}_{-0.002}$&
          $0.326^{+0.011}_{-0.003}$&$0.445^{+0.002}_{-0.001}$&
         $0.312^{+0.011}_{-0.008}$&$0.450^{+0.002}_{-0.002}$&
          $0.346^{+0.014}_{-0.008}$ &$0.439^{+0.002}_{-0.004}$&
          $0.354^{+0.015}_{-0.013}$& $0.439^{+0.002}_{-0.005}$&
          $0.321^{+0.010}_{-0.011}$&$0.447^{+0.002}_{-0.002}$ \\
          \hline
         $\langle  I_{2s,^{\parallel}}\rangle,\langle I^{\gamma}_{2s,\perp}\rangle$
         &$0.101^{+0.005}_{-0.004}$& $-0.231^{+0.010}_{-0.010}$&
         $0.109^{+0.003}_{-0.004}$& $-0.213^{+0.010}_{-0.008}$&
         $0.104^{+0.005}_{-0.001}$& $-0.235^{+0.010}_{-0.009}$&
         $0.115^{+0.005}_{-0.002}$ & $-0.195^{+0.009}_{-0.009}$&
         $0.118^{+0.003}_{-0.004}$ & $-0.194^{+0.009}_{-0.008}$&
         $0.107^{+0.003}_{-0.004}$ & $-0.222^{+0.010}_{-0.008}$ \\
         \hline
         $\langle  I_{1c,^{\parallel}}\rangle,\langle I^{\gamma}_{1c,\perp}\rangle$&
         $0.599^{+0.017}_{-0.018}$ &$0.303^{+0.012}_{-0.013}$&
         $0.587^{+0.026}_{-0.016} $&$0.326^{+0.011}_{-0.003}$&
         $0.568^{+0.014}_{-0.014}  $& $0.312^{+0.011}_{-0.008}$&
         $0.541^{+0.014}_{-0.028} $  &$0.346^{+0.014}_{-0.008}$&
         $0.531^{+0.015}_{-0.014}  $&$0.354^{+0.015}_{-0.013}$&
         $0.575^{+0.015}_{-0.014} $&$0.321^{+0.010}_{-0.011}$\\
\hline
          \end{tabular}}
\end{table}
\begin{table}
    \centering
    \caption{\small Same as in Table \ref{tab:Angobs1}, but for $q^{2}=[4.0-5.0]$ GeV$^2$ bin.}\label{tab:Angobs5}
    \renewcommand{\arraystretch}{1.5}
    \scalebox{0.50}{
    \begin{tabular}{|c||c|c||c|c||c|c||c|c||c|c||c|c|}
     \hline
        \multicolumn{13}{|c|}{$q^2=[4-5]$ GeV$^2$}\\
        \hline
       Angular  &\multicolumn{2}{|c|}{SM} &\multicolumn{2}{|c|}{S1} &\multicolumn{2}{|c|}{S2} &\multicolumn{2}{|c|}{S3} &\multicolumn{2}{|c|}{S4} &\multicolumn{2}{|c|}{S5} \\
         \cline{2-13}
        observables  & $D_s\pi$ & $D_s\gamma$ &  $D_s\pi$ & $D_s\gamma$ &  $D_s\pi$ & $D_s\gamma$&  $D_s\pi$ & $D_s\gamma$ &  $D_s\pi$ & $D_s\gamma$ &  $D_s\pi$ & $D_s\gamma$   \\
         \hline
         $\langle I_{1s,^{\parallel}}\rangle,\langle I^{\gamma}_{1s,\perp}\rangle$ &
          $0.369^{+0.013}_{-0.014}$&$0.438^{+0.002}_{-0.002}$&
          $0.380^{+0.011}_{-0.002}$&$0.436^{+0.002}_{-0.002}$&
          $0.375^{+0.010}_{-0.013}$& $0.439^{+0.002}_{-0.002}$&
          $0.386^{+0.008}_{-0.008}$&$0.431^{+0.002}_{-0.002}$&
          $0.404^{+0.012}_{-0.013}$ &$0.430^{+0.002}_{-0.002}$&
          $0.377^{+0.012}_{-0.013}$&$0.437^{+0.002}_{-0.002}$\\
          \hline
          $\langle  I_{2s,^{\parallel}}\rangle,\langle I^{\gamma}_{2s,\perp}\rangle$&
          $0.123^{+0.005}_{-0.005}$& $-0.181^{+0.011}_{-0.010}$&
          $0.130^{+0.007}_{-0.000}$&$-0.173^{+0.011}_{-0.010}$&
          $0.129^{+0.000}_{-0.009}$ & $-0.186^{+0.011}_{-0.010}$&
          $0.131^{+0.006}_{-0.000}$ & $-0.157^{+0.010}_{-0.010}$&
          $0.134^{+0.005}_{-0.004}$ &$-0.148^{+0.010}_{-0.010}$&
          $0.125^{+0.004}_{-0.004}$&$-0.179^{+0.010}_{-0.010}$  \\
         \hline
         $\langle  I_{c,^{\parallel}}\rangle,\langle I^{\gamma}_{1c,\perp}\rangle$&
         $0.510^{+0.018}_{-0.017}$& $0.369^{+0.013}_{-0.014}$&
         $0.505^{+0.017}_{-0.016}$&$0.380^{+0.011}_{-0.002}$&
         $0.496^{+0.016}_{-0.016}$&$0.375^{+0.010}_{-0.013}$&
         $0.476^{+0.011}_{-0.021}$&$0.386^{+0.008}_{-0.008}$&
         $0.463^{+0.021}_{-0.026}$&$0.404^{+0.012}_{-0.013}$&
         $0.499^{+0.017}_{-0.022}$&$0.377^{+0.012}_{-0.013}$   \\
\hline
          \end{tabular}}
\end{table}
\begin{table}
    \centering
    \caption{\small Same as in Table \ref{tab:Angobs1}, but for $q^{2}=[5.0-6.0]$ GeV$^2$ bin.}\label{tab:Angobs6}
    \renewcommand{\arraystretch}{1.5}
    \scalebox{0.50}{
    \begin{tabular}{|c||c|c||c|c||c|c||c|c||c|c||c|c|}
     \hline
        \multicolumn{13}{|c|}{$q^2=[5-6]$ GeV$^2$}\\
        \hline
       Angular &\multicolumn{2}{|c|}{SM} &\multicolumn{2}{|c|}{S1} &\multicolumn{2}{|c|}{S2} &\multicolumn{2}{|c|}{S3} &\multicolumn{2}{|c|}{S4} &\multicolumn{2}{|c|}{S5} \\
         \cline{2-13}
         observables & $D_s\pi$ & $D_s\gamma$ &  $D_s\pi$ & $D_s\gamma$ &  $D_s\pi$ & $D_s\gamma$&  $D_s\pi$ & $D_s\gamma$ &  $D_s\pi$ & $D_s\gamma$ &  $D_s\pi$ & $D_s\gamma$  \\
         \hline
         $\langle I_{1s,^{\parallel}}\rangle,\langle I^{\gamma}_{1s,\perp}\rangle$&
          $0.418^{+0.013}_{-0.013}$& $0.365^{+0.007}_{-0.006}$&
         $0.424^{+0.011}_{-0.005}$ &$0.395^{+0.005}_{-0.005}$&
          $0.420^{+0.012}_{-0.013} $& $0.372^{+0.007}_{-0.006}$&
          $0.435^{+0.018}_{-0.007}$ &$0.396^{+0.004}_{-0.005}$&
          $0.445^{+0.013}_{-0.013}$& $0.396^{+0.004}_{-0.005}$&
          $0.422^{+0.012}_{-0.013}$&$0.387^{+0.006}_{-0.006}$ \\
          \hline
         $\langle  I_{2s,^{\parallel}}\rangle,\langle I^{\gamma}_{2s,\perp}\rangle$ &
         $0.139^{+0.004}_{-0.004}$&$-0.144^{+0.011}_{-0.010}$&
         $0.141^{+0.004}_{-0.004}$& $-0.139^{+0.011}_{-0.009}$&
         $0.140^{+0.005}_{-0.005}$& $-0.148^{+0.011}_{-0.010}$&
         $0.145^{+0.006}_{-0.003}$ & $-0.125^{+0.011}_{-0.009}$&
         $0.148^{+0.005}_{-0.004}$ & $-0.124^{+0.011}_{-0.010}$&
         $0.140^{+0.004}_{-0.005}$&$-0.143^{+0.011}_{-0.010}$ \\
         \hline
          $\langle  I_{1c,^{\parallel}}\rangle,\langle I^{\gamma}_{1c,\perp}\rangle$&
         $0.443^{+0.024}_{-0.016}$&$0.418^{+0.013}_{-0.013}$&
         $0.436^{+0.017}_{-0.015}$  &$0.424^{+0.011}_{-0.005}$&
         $0.441^{+0.018}_{-0.016}$ &$0.420^{+0.012}_{-0.013}$&
         $0.422^{+0.009}_{-0.024}$&$0.435^{+0.018}_{-0.007}$&
         $0.407^{+0.018}_{-0.016} $&$0.445^{+0.013}_{-0.013}$&
         $0.438^{+0.018}_{-0.016}$&$0.422^{+0.012}_{-0.013}$   \\
\hline
          \end{tabular}}
\end{table}
\begin{table}[htbp!]
    \centering
    \caption{\small Same as in Table \ref{tab:Angobs1}, but for $q^{2}=[1.0-6.0]$ GeV$^2$ bin.}\label{tab:Angobs7}
    \renewcommand{\arraystretch}{1.5}
    \scalebox{0.50}{
    \begin{tabular}{|c||c|c||c|c||c|c||c|c||c|c||c|c|}
     \hline
        \multicolumn{13}{|c|}{$q^2=[1-6]$ GeV$^2$}\\
        \hline
     Angular  &\multicolumn{2}{|c|}{SM} &\multicolumn{2}{|c|}{S1} &\multicolumn{2}{|c|}{S2} &\multicolumn{2}{|c|}{S3} &\multicolumn{2}{|c|}{S4} &\multicolumn{2}{|c|}{S5} \\
         \cline{2-13}
      observables & $D_s\pi$ & $D_s\gamma$ &  $D_s\pi$ & $D_s\gamma$ &  $D_s\pi$ & $D_s\gamma$&  $D_s\pi$ & $D_s\gamma$ &  $D_s\pi$ & $D_s\gamma$ &  $D_s\pi$ & $D_s\gamma$ \\
         \hline
        $\langle I_{1s,^{\parallel}}\rangle,\langle I^{\gamma}_{1s,\perp}\rangle$ &
          $0.354^{+0.012}_{-0.012}$& $0.442^{+0.003}_{-0.002}$&
          $0.375^{+0.01}_{-0.010}$ &$0.439^{+0.001}_{-0.001}$&
          $0.354^{+0.011}_{-0.011}$ & $0.442^{+0.002}_{-0.002}$&
          $0.385^{+0.005}_{-0.006}$ &$0.435^{+0.002}_{-0.002}$&
          $0.393^{+0.011}_{-0.010}$ &$0.436^{+0.002}_{-0.002}$&
          $0.362^{+0.014}_{-0.006}$&$0.440^{+0.002}_{-0.002}$ \\
          \hline
         $\langle  I_{2s,^{\parallel}}\rangle,\langle I^{\gamma}_{2s,\perp}\rangle$ &
         $0.117^{+0.004}_{-0.004}$& $-0.199^{+0.011}_{-0.010}$&
         $0.124^{+0.004}_{-0.003}$& $-0.183^{+0.009}_{-0.008}$&
         $0.118^{+0.002}_{-0.004}$ &$-0.200^{+0.010}_{-0.009}$&
         $0.128^{+0.007}_{-0.001}$ &$-0.166^{+0.008}_{-0.010}$&
         $0.130^{+0.003}_{-0.002}$ &$-0.165^{+0.009}_{-0.011}$&
         $0.120^{+0.005}_{-0.002}$&$-0.181^{+0.011}_{-0.010}$ \\
         \hline
         $\langle  I_{1c,^{\parallel}}\rangle,\langle I^{\gamma}_{1c,\perp}\rangle$&
         $0.530^{+0.003}_{-0.002} $&$0.354^{+0.012}_{-0.012}$&
         $0.503^{+0.013}_{-0.014} $&$0.375^{+0.010}_{-0.010}$&
         $0.531^{+0.017}_{-0.015} $&$0.354^{+0.011}_{-0.011}$&
         $0.488^{+0.010}_{-0.015} $&$0.385^{+0.005}_{-0.006}$&
         $0.477^{+0.010}_{-0.014} $&$0.393^{+0.011}_{-0.010}$&
         $0.519^{+0.014}_{-0.019}$&$ 0.362^{+0.014}_{-0.006}$  \\
\hline
          \end{tabular}}
    \label{AngC}
\end{table}

\clearpage
\bibliographystyle{refstyle}
\bibliography{references}

\providecommand{\href}[2]{#2}\begingroup\raggedright\begin{thebibliography}{10}

\bibitem{LHCb:2015svh}
{\bf LHCb} Collaboration, R.~Aaij et~al., {\it {Angular analysis of the $B^{0}
  \to K^{*0} \mu^{+} \mu^{-}$ decay using 3 fb$^{-1}$ of integrated
  luminosity}},  {\em JHEP} {\bf 02} (2016) 104,
  [\href{https://arxiv.org/abs/1512.04442}{{\tt arXiv:1512.04442}}].

\bibitem{LHCb:2020lmf}
{\bf LHCb} Collaboration, R.~Aaij et~al., {\it {Measurement of $CP$-Averaged
  Observables in the $B^{0}\rightarrow K^{*0}\mu^{+}\mu^{-}$ Decay}},  {\em
  Phys. Rev. Lett.} {\bf 125} (2020), no.~1 011802,
  [\href{https://arxiv.org/abs/2003.04831}{{\tt arXiv:2003.04831}}].

\bibitem{LHCb:2020gog}
{\bf LHCb} Collaboration, R.~Aaij et~al., {\it {Angular Analysis of the
  $B^{+}\rightarrow K^{\ast+}\mu^{+}\mu^{-}$ Decay}},  {\em Phys. Rev. Lett.}
  {\bf 126} (2021), no.~16 161802,
  [\href{https://arxiv.org/abs/2012.13241}{{\tt arXiv:2012.13241}}].

\bibitem{CMS:2020oqb}
{\bf CMS} Collaboration, A.~M. Sirunyan et~al., {\it {Angular analysis of the
  decay B$^+$ $\to$ K$^*$(892)$^+\mu^+\mu^-$ in proton-proton collisions at
  $\sqrt{s} =$ 8 TeV}},  {\em JHEP} {\bf 04} (2021) 124,
  [\href{https://arxiv.org/abs/2010.13968}{{\tt arXiv:2010.13968}}].

\bibitem{LHCb:2014auh}
{\bf LHCb} Collaboration, R.~Aaij et~al., {\it {Angular analysis of charged and
  neutral $B \to K \mu^+\mu^-$ decays}},  {\em JHEP} {\bf 05} (2014) 082,
  [\href{https://arxiv.org/abs/1403.8045}{{\tt arXiv:1403.8045}}].

\bibitem{LHCb:2016due}
{\bf LHCb} Collaboration, R.~Aaij et~al., {\it {Measurement of the phase
  difference between short- and long-distance amplitudes in the $B^{+}\to
  K^{+}\mu^{+}\mu^{-}$ decay}},  {\em Eur. Phys. J. C} {\bf 77} (2017), no.~3
  161, [\href{https://arxiv.org/abs/1612.06764}{{\tt arXiv:1612.06764}}].

\bibitem{CMS:2018qih}
{\bf CMS} Collaboration, A.~M. Sirunyan et~al., {\it {Angular analysis of the
  decay B$^+\to$ K$^+\mu^+\mu^-$ in proton-proton collisions at $\sqrt{s} =$ 8
  TeV}},  {\em Phys. Rev. D} {\bf 98} (2018), no.~11 112011,
  [\href{https://arxiv.org/abs/1806.00636}{{\tt arXiv:1806.00636}}].

\bibitem{Becirevic:2011bp}
D.~Becirevic and E.~Schneider, {\it {On transverse asymmetries in $B \to
  K^{\ast} \ell^+\ell^-$}},  {\em Nucl. Phys. B} {\bf 854} (2012) 321--339,
  [\href{https://arxiv.org/abs/1106.3283}{{\tt arXiv:1106.3283}}].

\bibitem{Ciuchini:2015qxb}
M.~Ciuchini, M.~Fedele, E.~Franco, S.~Mishima, A.~Paul, L.~Silvestrini, and
  M.~Valli, {\it {$B\to K^* \ell^+ \ell^-$ decays at large recoil in the
  Standard Model: a theoretical reappraisal}},  {\em JHEP} {\bf 06} (2016) 116,
  [\href{https://arxiv.org/abs/1512.07157}{{\tt arXiv:1512.07157}}].

\bibitem{Hiller:2003js}
G.~Hiller and F.~Kruger, {\it {More model-independent analysis of $b \to s$
  processes}},  {\em Phys. Rev. D} {\bf 69} (2004) 074020,
  [\href{https://arxiv.org/abs/hep-ph/0310219}{{\tt hep-ph/0310219}}].

\bibitem{LHCb:2022qnv}
{\bf LHCb} Collaboration, {\it {Test of lepton universality in $b \rightarrow s
  \ell^+ \ell^-$ decays}},  [\href{https://arxiv.org/abs/2212.09152}{{\tt
  arXiv:2212.09152}}].

\bibitem{LHCb:2022zom}
{\bf LHCb} Collaboration, {\it {Measurement of lepton universality parameters
  in $B^+\to K^+\ell^+\ell^-$ and $B^0\to K^{*0}\ell^+\ell^-$ decays}},
  [\href{https://arxiv.org/abs/2212.09153}{{\tt arXiv:2212.09153}}].

\bibitem{LHCb:2014cxe}
{\bf LHCb} Collaboration, R.~Aaij et~al., {\it {Differential branching
  fractions and isospin asymmetries of $B \to K^{(*)} \mu^+ \mu^-$ decays}},
  {\em JHEP} {\bf 06} (2014) 133, [\href{https://arxiv.org/abs/1403.8044}{{\tt
  arXiv:1403.8044}}].

\bibitem{LHCb:2013zuf}
{\bf LHCb} Collaboration, R.~Aaij et~al., {\it {Differential branching fraction
  and angular analysis of the decay $B^{0} \to K^{*0} \mu^{+}\mu^{-}$}},  {\em
  JHEP} {\bf 08} (2013) 131, [\href{https://arxiv.org/abs/1304.6325}{{\tt
  arXiv:1304.6325}}].

\bibitem{LHCb:2016ykl}
{\bf LHCb} Collaboration, R.~Aaij et~al., {\it {Measurements of the S-wave
  fraction in $B^{0}\rightarrow K^{+}\pi^{-}\mu^{+}\mu^{-}$ decays and the
  $B^{0}\rightarrow K^{\ast}(892)^{0}\mu^{+}\mu^{-}$ differential branching
  fraction}},  {\em JHEP} {\bf 11} (2016) 047,
  [\href{https://arxiv.org/abs/1606.04731}{{\tt arXiv:1606.04731}}]. [Erratum:
  JHEP 04, 142 (2017)].

\bibitem{LHCb:2013tgx}
{\bf LHCb} Collaboration, R.~Aaij et~al., {\it {Differential branching fraction
  and angular analysis of the decay $B_s^0\to\phi\mu^{+}\mu^{-}$}},  {\em JHEP}
  {\bf 07} (2013) 084, [\href{https://arxiv.org/abs/1305.2168}{{\tt
  arXiv:1305.2168}}].

\bibitem{LHCb:2015wdu}
{\bf LHCb} Collaboration, R.~Aaij et~al., {\it {Angular analysis and
  differential branching fraction of the decay $B^0_s\to\phi\mu^+\mu^-$}},
  {\em JHEP} {\bf 09} (2015) 179, [\href{https://arxiv.org/abs/1506.08777}{{\tt
  arXiv:1506.08777}}].

\bibitem{Descotes-Genon:2012isb}
S.~Descotes-Genon, J.~Matias, M.~Ramon, and J.~Virto, {\it {Implications from
  clean observables for the binned analysis of $B \to K^{\ast}\mu^+\mu^-$ at
  large recoil}},  {\em JHEP} {\bf 01} (2013) 048,
  [\href{https://arxiv.org/abs/1207.2753}{{\tt arXiv:1207.2753}}].

\bibitem{Descotes-Genon:2013vna}
S.~Descotes-Genon, T.~Hurth, J.~Matias, and J.~Virto, {\it {Optimizing the
  basis of $B\to K^*ll$ observables in the full kinematic range}},  {\em JHEP}
  {\bf 05} (2013) 137, [\href{https://arxiv.org/abs/1303.5794}{{\tt
  arXiv:1303.5794}}].

\bibitem{ATLAS:2018gqc}
{\bf ATLAS} Collaboration, M.~Aaboud et~al., {\it {Angular analysis of $B^0_d
  \rightarrow K^{*}\mu^+\mu^-$ decays in $pp$ collisions at $\sqrt{s}= 8$ TeV
  with the ATLAS detector}},  {\em JHEP} {\bf 10} (2018) 047,
  [\href{https://arxiv.org/abs/1805.04000}{{\tt arXiv:1805.04000}}].

\bibitem{Aebischer:2018iyb}
J.~Aebischer, J.~Kumar, P.~Stangl, and D.~M. Straub, {\it {A Global Likelihood
  for Precision Constraints and Flavour Anomalies}},  {\em Eur. Phys. J. C}
  {\bf 79} (2019), no.~6 509, [\href{https://arxiv.org/abs/1810.07698}{{\tt
  arXiv:1810.07698}}].

\bibitem{Belle:2016xuo}
{\bf Belle} Collaboration, A.~Abdesselam et~al., {\it {Angular analysis of $B^0
  \to K^\ast(892)^0 \ell^+ \ell^-$}},  in {\em {LHC Ski 2016}: {A First
  Discussion of 13 TeV Results}}, 4, 2016.
\newblock \href{https://arxiv.org/abs/1604.04042}{{\tt arXiv:1604.04042}}.

\bibitem{Belle:2016fev}
{\bf Belle} Collaboration, S.~Wehle et~al., {\it {Lepton-Flavor-Dependent
  Angular Analysis of $B\to K^\ast \ell^+\ell^-$}},  {\em Phys. Rev. Lett.}
  {\bf 118} (2017), no.~11 111801,
  [\href{https://arxiv.org/abs/1612.05014}{{\tt arXiv:1612.05014}}].

\bibitem{CMS:2017rzx}
{\bf CMS} Collaboration, A.~M. Sirunyan et~al., {\it {Measurement of angular
  parameters from the decay $\mathrm{B}^0 \to \mathrm{K}^{*0} \mu^+ \mu^-$ in
  proton-proton collisions at $\sqrt{s} = $ 8 TeV}},  {\em Phys. Lett. B} {\bf
  781} (2018) 517--541, [\href{https://arxiv.org/abs/1710.02846}{{\tt
  arXiv:1710.02846}}].

\bibitem{Alguero:2021anc}
M.~Alguer\'o, B.~Capdevila, S.~Descotes-Genon, J.~Matias, and M.~Novoa-Brunet,
  {\it {$\boldsymbol{b\to s\ell\ell}$ global fits after Moriond 2021 results}},
   in {\em {55th Rencontres de Moriond on QCD and High Energy Interactions}},
  4, 2021.
\newblock \href{https://arxiv.org/abs/2104.08921}{{\tt arXiv:2104.08921}}.

\bibitem{Descotes-Genon:2015uva}
S.~Descotes-Genon, L.~Hofer, J.~Matias, and J.~Virto, {\it {Global analysis of
  $b\to s\ell\ell$ anomalies}},  {\em JHEP} {\bf 06} (2016) 092,
  [\href{https://arxiv.org/abs/1510.04239}{{\tt arXiv:1510.04239}}].

\bibitem{Altmannshofer:2017fio}
W.~Altmannshofer, C.~Niehoff, P.~Stangl, and D.~M. Straub, {\it {Status of the
  $B\rightarrow K^*\mu ^+\mu ^-$ anomaly after Moriond 2017}},  {\em Eur. Phys.
  J.} {\bf C77} (2017), no.~6 377,
  [\href{https://arxiv.org/abs/1703.09189}{{\tt arXiv:1703.09189}}].

\bibitem{Alok:2017sui}
A.~K. Alok, B.~Bhattacharya, A.~Datta, D.~Kumar, J.~Kumar, and D.~London, {\it
  {New Physics in $b \to s \mu^+ \mu^-$ after the Measurement of $R_{K^*}$}},
  {\em Phys. Rev.} {\bf D96} (2017), no.~9 095009,
  [\href{https://arxiv.org/abs/1704.07397}{{\tt arXiv:1704.07397}}].

\bibitem{Altmannshofer:2017yso}
W.~Altmannshofer, P.~Stangl, and D.~M. Straub, {\it {Interpreting Hints for
  Lepton Flavor Universality Violation}},  {\em Phys. Rev.} {\bf D96} (2017),
  no.~5 055008, [\href{https://arxiv.org/abs/1704.05435}{{\tt
  arXiv:1704.05435}}].

\bibitem{Geng:2017svp}
L.-S. Geng, B.~Grinstein, S.~Jäger, J.~Martin~Camalich, X.-L. Ren, and R.-X.
  Shi, {\it {Towards the discovery of new physics with lepton-universality
  ratios of $b\to s\ell\ell$ decays}},  {\em Phys. Rev.} {\bf D96} (2017),
  no.~9 093006, [\href{https://arxiv.org/abs/1704.05446}{{\tt
  arXiv:1704.05446}}].

\bibitem{Ciuchini:2017mik}
M.~Ciuchini, A.~M. Coutinho, M.~Fedele, E.~Franco, A.~Paul, L.~Silvestrini, and
  M.~Valli, {\it {On Flavourful Easter eggs for New Physics hunger and Lepton
  Flavour Universality violation}},  {\em Eur. Phys. J.} {\bf C77} (2017),
  no.~10 688, [\href{https://arxiv.org/abs/1704.05447}{{\tt
  arXiv:1704.05447}}].

\bibitem{Capdevila:2017bsm}
B.~Capdevila, A.~Crivellin, S.~Descotes-Genon, J.~Matias, and J.~Virto, {\it
  {Patterns of New Physics in $b\to s\ell^+\ell^-$ transitions in the light of
  recent data}},  {\em JHEP} {\bf 01} (2018) 093,
  [\href{https://arxiv.org/abs/1704.05340}{{\tt arXiv:1704.05340}}].

\bibitem{Alguero:2019ptt}
M.~Algueró, B.~Capdevila, A.~Crivellin, S.~Descotes-Genon, P.~Masjuan,
  J.~Matias, M.~Novoa~Brunet, and J.~Virto, {\it {Emerging patterns of New
  Physics with and without Lepton Flavour Universal contributions}},  {\em Eur.
  Phys. J.} {\bf C79} (2019), no.~8 714,
  [\href{https://arxiv.org/abs/1903.09578}{{\tt arXiv:1903.09578}}]. [Addendum:
  Eur.Phys.J.C 80, 511 (2020)].

\bibitem{Alok:2019ufo}
A.~K. Alok, A.~Dighe, S.~Gangal, and D.~Kumar, {\it {Continuing search for new
  physics in $b \to s \mu \mu$ decays: two operators at a time}},  {\em JHEP}
  {\bf 06} (2019) 089, [\href{https://arxiv.org/abs/1903.09617}{{\tt
  arXiv:1903.09617}}].

\bibitem{Ciuchini:2019usw}
M.~Ciuchini, A.~M. Coutinho, M.~Fedele, E.~Franco, A.~Paul, L.~Silvestrini, and
  M.~Valli, {\it {New Physics in $b \to s \ell^+ \ell^-$ confronts new data on
  Lepton Universality}},  {\em Eur. Phys. J.} {\bf C79} (2019), no.~8 719,
  [\href{https://arxiv.org/abs/1903.09632}{{\tt arXiv:1903.09632}}].

\bibitem{Datta:2019zca}
A.~Datta, J.~Kumar, and D.~London, {\it {The $B$ anomalies and new physics in
  $b \to s e^+ e^-$}},  {\em Phys. Lett.} {\bf B797} (2019) 134858,
  [\href{https://arxiv.org/abs/1903.10086}{{\tt arXiv:1903.10086}}].

\bibitem{Aebischer:2019mlg}
J.~Aebischer, W.~Altmannshofer, D.~Guadagnoli, M.~Reboud, P.~Stangl, and D.~M.
  Straub, {\it {$B$-decay discrepancies after Moriond 2019}},  {\em Eur. Phys.
  J. C} {\bf 80} (2020), no.~3 252,
  [\href{https://arxiv.org/abs/1903.10434}{{\tt arXiv:1903.10434}}].

\bibitem{Kowalska:2019ley}
K.~Kowalska, D.~Kumar, and E.~M. Sessolo, {\it {Implications for new physics in
  $b\rightarrow s \mu \mu $ transitions after recent measurements by Belle and
  LHCb}},  {\em Eur. Phys. J.} {\bf C79} (2019), no.~10 840,
  [\href{https://arxiv.org/abs/1903.10932}{{\tt arXiv:1903.10932}}].

\bibitem{Arbey:2019duh}
A.~Arbey, T.~Hurth, F.~Mahmoudi, D.~M. Santos, and S.~Neshatpour, {\it {Update
  on the $b\to s$ anomalies}},  {\em Phys. Rev.} {\bf D100} (2019), no.~1
  015045, [\href{https://arxiv.org/abs/1904.08399}{{\tt arXiv:1904.08399}}].

\bibitem{Bhattacharya:2019dot}
S.~Bhattacharya, A.~Biswas, S.~Nandi, and S.~K. Patra, {\it {Exhaustive model
  selection in $b \to s \ell \ell$ decays: Pitting cross-validation against the
  Akaike information criterion}},  {\em Phys. Rev. D} {\bf 101} (2020), no.~5
  055025, [\href{https://arxiv.org/abs/1908.04835}{{\tt arXiv:1908.04835}}].

\bibitem{Biswas:2020uaq}
A.~Biswas, S.~Nandi, S.~K. Patra, and I.~Ray, {\it {New physics in $b\to s
  \ell\ell$ decays with complex Wilson coefficients}},  {\em Nucl. Phys. B}
  {\bf 969} (2021) 115479, [\href{https://arxiv.org/abs/2004.14687}{{\tt
  arXiv:2004.14687}}].

\bibitem{Alok:2022pjb}
A.~K. Alok, N.~R. Singh~Chundawat, S.~Gangal, and D.~Kumar, {\it {A global
  analysis of $b \rightarrow s \ell \ell $ data in heavy and light $Z'$
  models}},  {\em Eur. Phys. J. C} {\bf 82} (2022), no.~10 967,
  [\href{https://arxiv.org/abs/2203.13217}{{\tt arXiv:2203.13217}}].

\bibitem{Huang:2018rys}
Z.-R. Huang, M.~A. Paracha, I.~Ahmed, and C.-D. L\"u, {\it {Testing Leptoquark
  and $Z^{\prime}$ Models via $B\to K_{1}(1270,1400)\mu^{+}\mu^{-}$ Decays}},
  {\em Phys. Rev. D} {\bf 100} (2019), no.~5 055038,
  [\href{https://arxiv.org/abs/1812.03491}{{\tt arXiv:1812.03491}}].

\bibitem{MunirBhutta:2020ber}
F.~Munir~Bhutta, Z.-R. Huang, C.-D. L\"u, M.~A. Paracha, and W.~Wang, {\it {New
  physics in $b\to s\ell\ell$ anomalies and its implications for the
  complementary neutral current decays}},  {\em Nucl. Phys. B} {\bf 979} (2022)
  115763, [\href{https://arxiv.org/abs/2009.03588}{{\tt arXiv:2009.03588}}].

\bibitem{Das:2018orb}
D.~Das, B.~Kindra, G.~Kumar, and N.~Mahajan, {\it {$B\to
  K^\ast_2(1430)\ell^+\ell^-$ distributions at large recoil in the Standard
  Model and beyond}},  {\em Phys. Rev. D} {\bf 99} (2019), no.~9 093012,
  [\href{https://arxiv.org/abs/1812.11803}{{\tt arXiv:1812.11803}}].

\bibitem{Mohapatra:2021izl}
M.~K. Mohapatra and A.~Giri, {\it {Implications of light $Z^{\prime}$ on
  semileptonic $B(B_s)\to
  T\{K_2^{\ast}(1430)(f_2^{\prime}(1525))\}\ensuremath{\ell}^+\ensuremath{\ell}^-$
  decays at large recoil}},  {\em Phys. Rev. D} {\bf 104} (2021), no.~9 095012,
  [\href{https://arxiv.org/abs/2109.12382}{{\tt arXiv:2109.12382}}].

\bibitem{Rajeev:2020aut}
N.~Rajeev, N.~Sahoo, and R.~Dutta, {\it {Angular analysis of $B_s\, \to\,
  f_{2}'\,(1525)\,(\to K^+\,K^-)\,\mu^+ \,\mu^-$ decays as a probe to lepton
  flavor universality violation}},  {\em Phys. Rev. D} {\bf 103} (2021), no.~9
  095007, [\href{https://arxiv.org/abs/2009.06213}{{\tt arXiv:2009.06213}}].

\bibitem{Dutta:2019wxo}
R.~Dutta, {\it {Model independent analysis of new physics effects on $B_c \to
  (D_s,\,D^{\ast}_s)\,\mu^+\mu^-$ decay observables}},  {\em Phys. Rev. D} {\bf
  100} (2019), no.~7 075025, [\href{https://arxiv.org/abs/1906.02412}{{\tt
  arXiv:1906.02412}}].

\bibitem{Mohapatra:2021ynn}
M.~K. Mohapatra, N.~Rajeev, and R.~Dutta, {\it {Combined analysis of $B_c\to
  D_s^{(\ast)}\ensuremath{\mu}^+\ensuremath{\mu}^-$ and $B_c\to
  D_s^{(\ast)}\ensuremath{\nu}\ensuremath{\bar\nu}$ decays within $Z^{\prime}$
  and leptoquark new physics models}},  {\em Phys. Rev. D} {\bf 105} (2022),
  no.~11 115022, [\href{https://arxiv.org/abs/2108.10106}{{\tt
  arXiv:2108.10106}}].

\bibitem{Paul:2016urs}
A.~Paul and D.~M. Straub, {\it {Constraints on new physics from radiative $B$
  decays}},  {\em JHEP} {\bf 04} (2017) 027,
  [\href{https://arxiv.org/abs/1608.02556}{{\tt arXiv:1608.02556}}].

\bibitem{Bobeth:1999mk}
C.~Bobeth, M.~Misiak, and J.~Urban, {\it {Photonic penguins at two loops and
  $m_t$ dependence of $BR[B \to X_s l^+ l^-]$}},  {\em Nucl. Phys. B} {\bf 574}
  (2000) 291--330, [\href{https://arxiv.org/abs/hep-ph/9910220}{{\tt
  hep-ph/9910220}}].

\bibitem{Beneke:2001at}
M.~Beneke, T.~Feldmann, and D.~Seidel, {\it {Systematic approach to exclusive
  $B \to V l^+ l^-$, $V \gamma$ decays}},  {\em Nucl. Phys.} {\bf B612} (2001)
  25--58, [\href{https://arxiv.org/abs/hep-ph/0106067}{{\tt hep-ph/0106067}}].

\bibitem{Asatrian:2001de}
H.~H. Asatrian, H.~M. Asatrian, C.~Greub, and M.~Walker, {\it {Two loop virtual
  corrections to $B \to X_s l^+ l^-$ in the standard model}},  {\em Phys. Lett.
  B} {\bf 507} (2001) 162--172,
  [\href{https://arxiv.org/abs/hep-ph/0103087}{{\tt hep-ph/0103087}}].

\bibitem{Asatryan:2001zw}
H.~H. Asatryan, H.~M. Asatrian, C.~Greub, and M.~Walker, {\it {Calculation of
  two loop virtual corrections to $b \to s l^+ l^-$ in the standard model}},
  {\em Phys. Rev.} {\bf D65} (2002) 074004,
  [\href{https://arxiv.org/abs/hep-ph/0109140}{{\tt hep-ph/0109140}}].

\bibitem{Greub:2008cy}
C.~Greub, V.~Pilipp, and C.~Schupbach, {\it {Analytic calculation of two-loop
  QCD corrections to $b \to sl^+ l^-$ in the high $q^2$ region}},  {\em JHEP}
  {\bf 12} (2008) 040, [\href{https://arxiv.org/abs/0810.4077}{{\tt
  arXiv:0810.4077}}].

\bibitem{Du:2015tda}
D.~Du, A.~X. El-Khadra, S.~Gottlieb, A.~S. Kronfeld, J.~Laiho, E.~Lunghi, R.~S.
  Van~de Water, and R.~Zhou, {\it {Phenomenology of semileptonic B-meson decays
  with form factors from lattice QCD}},  {\em Phys. Rev.} {\bf D93} (2016),
  no.~3 034005, [\href{https://arxiv.org/abs/1510.02349}{{\tt
  arXiv:1510.02349}}].

\bibitem{Faessler:2002ut}
A.~Faessler, T.~Gutsche, M.~A. Ivanov, J.~G. Korner, and V.~E. Lyubovitskij,
  {\it {The Exclusive rare decays $B \to$ K(K*) $\bar{\ell} \ell$ and $B_c \to$
  D(D*) $\bar{\ell} \ell$ in a relativistic quark model}},  {\em Eur. Phys. J.
  direct} {\bf 4} (2002), no.~1 18,
  [\href{https://arxiv.org/abs/hep-ph/0205287}{{\tt hep-ph/0205287}}].

\bibitem{Ebert:2010dv}
D.~Ebert, R.~N. Faustov, and V.~O. Galkin, {\it {Rare Semileptonic Decays of
  $B$ and $B_c$ Mesons in the Relativistic Quark Model}},  {\em Phys. Rev. D}
  {\bf 82} (2010) 034032, [\href{https://arxiv.org/abs/1006.4231}{{\tt
  arXiv:1006.4231}}].

\bibitem{Ebert:2001pc}
D.~Ebert, R.~N. Faustov, and V.~O. Galkin, {\it {Form-factors of heavy to light
  B decays at large recoil}},  {\em Phys. Rev. D} {\bf 64} (2001) 094022,
  [\href{https://arxiv.org/abs/hep-ph/0107065}{{\tt hep-ph/0107065}}].

\bibitem{ParticleDataGroup:2022pth}
{\bf Particle Data Group} Collaboration, R.~L. Workman et~al., {\it {Review of
  Particle Physics}},  {\em PTEP} {\bf 2022} (2022) 083C01.

\bibitem{Blake:2016olu}
T.~Blake, G.~Lanfranchi, and D.~M. Straub, {\it {Rare $B$ Decays as Tests of
  the Standard Model}},  {\em Prog. Part. Nucl. Phys.} {\bf 92} (2017) 50--91,
  [\href{https://arxiv.org/abs/1606.00916}{{\tt arXiv:1606.00916}}].

\end{thebibliography}\endgroup
\end{document}